\documentclass{article}
\usepackage{graphicx}
\usepackage{tikz}
\usetikzlibrary{arrows}
\usetikzlibrary{arrows.meta}
\usetikzlibrary{shapes}
\usetikzlibrary{backgrounds}
\usetikzlibrary{positioning}
\usetikzlibrary{decorations.markings}
\usetikzlibrary{patterns}
\usetikzlibrary{calc}
\usetikzlibrary{fit}
\usetikzlibrary{petri,decorations.markings}

\newcommand{\OPT}{\mathtt{OPT}}

\newcommand{\VS}{\mathtt{VS}}

\usepackage{fancybox}
\usepackage{enumitem}

\usepackage[noend]{algpseudocode}
\usepackage[linesnumbered,ruled]{algorithm2e}
\setlistdepth{9}

\newlist{myEnumerate}{enumerate}{5}
\setlist[myEnumerate,1]{label=\arabic*.}
\setlist[myEnumerate,2]{label=(\alph*)}
\setlist[myEnumerate,3]{label=\roman*.}
\setlist[myEnumerate,4]{label=\Alph*.}
\setlist[myEnumerate,5]{label=\Roman*.}

\usepackage{amsmath,amsfonts,amssymb,bm,xparse, amsthm}
\usepackage{bm,xspace}
\usepackage{color,xcolor}
\usepackage[colorlinks=true,citecolor=blue]{hyperref}
\usepackage{nameref}
\definecolor{ForestGreen}{rgb}{0.1333,0.5451,0.1333}
\definecolor{Red}{rgb}{0.9,0,0}
\usepackage[noabbrev,nameinlink, capitalise]{cleveref}
\crefname{property}{property}{Property}
\creflabelformat{property}{(#1)#2#3}
\crefname{equation}{eq}{Eq}
\creflabelformat{equation}{(#1)#2#3}

 \usepackage{thm-restate}
 \usepackage{thmtools}

\let\oldnl\nl
\newcommand{\nonl}{\renewcommand{\nl}{\let\nl\oldnl}}%

\newtheorem{theorem}{Theorem}[section]
\newtheorem{corollary}[theorem]{Corollary}
\newtheorem{lemma}[theorem]{Lemma}
\newtheorem{proposition}[theorem]{Proposition}

\theoremstyle{definition}
\newtheorem{claim}[theorem]{Claim}
\newtheorem{definition}{Definition}[section]

\newtheorem{remark}{Remark}[section]

\newenvironment{fminipage}%
  {\begin{Sbox}\begin{minipage}}%
  {\end{minipage}\end{Sbox}\fbox{\TheSbox}}

 \SetKwInOut{Input}{Input}
 \SetKwInOut{Output}{Output}

\def\pr#1{\mathrm{Pr}\left[ #1 \right]}

\def\ip#1#2{\left\langle #1, #2 \right\rangle}

\def\defeq{\stackrel{\mathrm{def}}{=}}
\def\setof#1{\left\{#1  \right\}}
\def\sizeof#1{\left|#1  \right|}

\def\ceil#1{\left\lceil #1 \right\rceil}

\def\union{\cup}
\def\intersect{\cap}
\def\Union{\bigcup}

\def\eps{\epsilon}

\def\stars#1{\sizeof{\mathtt{stars}\kh{#1}}}
\def\levels#1{\sizeof{\mathtt{levels}\kh{#1}}}

\def\R{\mathbb{R}}

\def\Scal{\mathcal{S}}
\def\Ecal{\mathcal{E}}

\def\Wcal{\mathcal{W}}

\def\Gcal{\mathcal{G}}

\def\Scal{\mathcal{S}}

\newcommand\EE{\boldsymbol{\mathit{E}}}

\newcommand\ve{\vec{\mathcal{E}}}

\renewcommand{\tilde}{\wildetilde}

\newcommand\Otil{\widetilde{O}}

\DeclareMathOperator*{\argmin}{arg\,min}
\DeclareMathOperator*{\argmax}{arg\,max}

\newcommand{\kh}[1]{\left(#1\right)}

\renewcommand{\ip}[2]{\left\langle#1,#2\right\rangle}

\newcommand{\poly}{\mathrm{poly}}
\DeclareMathOperator{\polylog}{polylog}

\NewDocumentCommand{\ot}{g}{
    \IfNoValueTF{#1}{\tilde{O}}{\tilde{O}(#1)}
}
\definecolor{RURed}{rgb}{0.95, 0.1, 0.1} 

\usepackage{tcolorbox}
\tcbuselibrary{skins,breakable}
\tcbset{enhanced jigsaw}

\usepackage{mdframed}

\newtheorem{mdresult}{Result}

\usepackage{bbm}

\makeatletter
\def\thmt@innercounters{equation,algocf}
\makeatother
\pdfoutput=1

\title{
Near-optimal Linear Sketches and Fully-Dynamic Algorithms for Hypergraph Spectral Sparsification
}
\author{Sanjeev Khanna\thanks{School of Engineering and Applied Sciences, University of Pennsylvania, Philadelphia, PA. Email: \texttt{sanjeev@cis.upenn.edu}.
Supported in part by NSF awards CCF-2008305 and CCF-2402284.} \and Huan Li\thanks{School of Engineering and Applied Sciences, University of Pennsylvania, Philadelphia, PA. Email: \texttt{huanli@cis.upenn.edu}. Supported in part by NSF awards CCF-2008305 and CCF-2402284.} \and Aaron (Louie) Putterman\thanks{School of Engineering and Applied Sciences, Harvard University, Cambridge, Massachusetts, USA. Supported in part by the Simons Investigator Awards of Madhu Sudan and Salil Vadhan, NSF Award CCF 2152413 and a Hudson River Trading PhD Research Scholarship. Email: \texttt{aputterman@g.harvard.edu}.
}}
\date{\today}

\begin{document}

\maketitle

\begin{abstract}
A hypergraph spectral sparsifier of a hypergraph $G$ is a weighted subgraph $H$ that approximates the Laplacian of $G$ to a specified precision. Recent work has shown that similar to ordinary graphs, there exist $\widetilde{O}(n)$-size hypergraph spectral sparsifiers. 
However, the task of computing such sparsifiers turns out to be much more involved, and all known algorithms rely on the notion of balanced weight assignments, whose computation inherently relies on repeated, complete access to the underlying hypergraph. We introduce a significantly simpler framework for hypergraph spectral sparsification which bypasses the need to compute such weight assignments, essentially reducing hypergraph sparsification to repeated effective resistance sampling in \textit{ordinary graphs}, which are obtained by \textit{oblivious vertex-sampling} of the original hypergraph.

Our framework immediately yields a simple, new
nearly-linear time algorithm for nearly-linear size spectral hypergraph sparsification. 
Furthermore, as a direct consequence of our framework, we obtain the first nearly-optimal algorithms in several other models of computation:
\begin{enumerate}
    \item The first nearly-optimal size \textit{linear sketches} for spectral hypergraph sparsification. For hypergraphs on $n$ vertices, with hyperedges of arity $\leq r$ and with $\leq m$ hyperedges, these sketches require only $\widetilde{O}(n r \polylog(m) / \eps^2)$ bits and recover a $(1 \pm \eps)$ spectral-hypergraph sparsifier with high probability. It is known that linear sketches require $\Omega(nr \log(m))$ bits even for the easier task of cut sparsification (Khanna-Putterman-Sudan FOCS 2024).
    \item The first nearly-optimal \emph{fully dynamic} $(1 \pm \eps)$ spectral (and cut) hypergraph sparsification algorithm.
    Our algorithm has an amortized, expected update time of $\widetilde{O}(r \polylog(m) / \eps^2)$, and produces sparsifiers with $\widetilde{O}(n \polylog(m) / \eps^2)$ hyperedges. This is nearly-optimal as even to read a single hyperedge takes time $\Omega(r)$.
    \item The first nearly-optimal algorithm for \textit{online} hypergraph spectral sparsification. On a sequence of $m$ (unweighted) hyperedges, our algorithm creates a $(1 \pm \eps)$ hypergraph spectral sparsifier with $\widetilde{O}(n \polylog(m) / \eps^2)$ hyperedges in an online manner.
    When $m\leq \poly(n)$,
    this improves upon the work of Soma, Tung, and Yoshida (IPCO 2024) by a factor of $r$, who created online sparsifiers with $\widetilde{O}(n (r + \log(m)) / \eps^2)$ hyperedges.
    We complement this result with an $\Omega(n \log(m))$ lower-bound for any online sparsifier, thus provably separating the classical and online settings.  
\end{enumerate}

Our main conceptual and technical contributions are introduction of (a) the \textit{vertex sampling} framework to reduce spectral sparsification in hypergraphs to ordinary graphs, and (b)   
a notion of \textit{collective energy} in hypergraphs that may be seen
as a continuous generalization of $k$-cuts. 

\end{abstract}
\pagenumbering{gobble}

\pagebreak

\setcounter{tocdepth}{2}
\tableofcontents

\pagebreak

\pagenumbering{arabic}

\section{Introduction}

Ever since its conception in the work of Karger \cite{Kar93}, \emph{graph sparsification} has been a powerful algorithmic tool in the design of efficient graph algorithms. Roughly speaking, given a graph $G = (V, E)$, sparsification is the process of choosing a re-weighted subgraph $G' \subseteq G$ such that certain properties of $G$ are preserved. In the works of Karger \cite{Kar93} and Bencz\'ur and Karger \cite{BK96}, these properties took the form of the \emph{cut sizes in} the graph, and this research culminated in the design of an $\widetilde{O}(n + m)$ time algorithm that produces a ``nearly-linear size cut-sparsifier.'' Specifically, they showed that there exists a re-weighted subgraph of $G$ with only $O(n \log (n) / \eps^2)$ edges that preserves the weight of every cut to a $(1 \pm \eps)$-factor. This result continues to play a central role in obtaining significantly more space-efficient and time-efficient algorithms for cut and flow problems on graphs across many models of computation.

Subsequently, other works extended the notion of cut sparsification in different directions. Perhaps most notably, the landmark
works of Spielman and Teng \cite{ST11},
Spielman and Srivastava \cite{SS11},
and Batson, Spielman, and Srivastava \cite{BatsonSS12} studied \emph{spectral} graph sparsification. Here, instead of simply preserving the cuts in the graph, their sparsifiers preserve the entire spectrum of the graph Laplacian. Recall that for a graph $G$, the Laplacian $L_G = D - A$, where $D$ is a diagonal matrix of vertex degrees, and $A$ is the adjacency matrix. One can check that for $S \subseteq [n]$, $\mathbf{1}_S^T L_G \mathbf{1}_S = |\mathrm{cut}_G(S)|$. Thus, preserving the spectrum of $L_G$ to a $(1 \pm \eps)$ factor is a \emph{stronger} notion than $(1 \pm \eps)$ cut-sparsification.
Yet, \cite{ST11,SS11,BatsonSS12} still showed the existence of spectral sparsifiers of (nearly) linear size.

Paralleling the work in graphs, recent efforts in sparsification have focused heavily on the \emph{hypergraph} setting. Originally proposed in the cut setting by Kogan and Krauthgamer \cite{KK15}, here one is tasked with selecting a re-weighted \emph{sub-hypergraph}, such that all cuts have their weight preserved to a $(1 \pm \eps)$-factor. As in graphs, a cut is given by a subset $S \subseteq V$ of vertices, and a hyperedge $e \subseteq V$ is said to be crossing the cut if $e \cap S \neq \emptyset$ and $e \cap (V-S) \neq \emptyset$. Likewise, there is a notion of a hypergraph Laplacian (see \cite{CLTZ18}),  where for a hypergraph $H = (V, E)$ and a vector $x \in \R^n$, we say that 
\[
Q_H(x) = \sum_{e \in E} w_e \cdot \max_{(u,v) \in e}(x[u] - x[v])^2.
\]
Initially studied by Soma and  Yoshida \cite{SomaY19} and Bansal, Svensson, and Trevisan \cite{BansalST19},
a $(1 \pm \eps)$ hypergraph spectral sparsifier $H'$ of a hypergraph $H$ is thus a re-weighted sub-hypergraph such that $\forall x \in \R^n, (1 - \eps)Q_H(x) \leq Q_{H'}(x) \leq (1 + \eps)Q_H(x)$.

As with graphs, for a set $S \subseteq V$, it is also the case that $Q_H(\mathbf{1}_S) = |\mathrm{cut}_H(S)|$, and hence hypergraph spectral sparsification generalizes cut sparsification. In a long line of works \cite{CKN20, KKTY21a, KKTY21b, Lee23, JambulapatiLS23, JLLS23, OST23, Qua23, KPS24} the existence of nearly-optimal size $\widetilde{O}(n / \eps^2)$ cut and spectral sparsifiers was established, along with a host of efficient algorithms for computing such sparsifiers.

\subsection{Computing Sparsifiers in Modern Computational Settings}

However, the extensive research into efficient algorithms for hypergraph sparsifiers has so far primarily focused on the classical model of computation where the underlying hypergraph is static and the algorithm has unrestricted random access to the hypergraph. In contrast, one highlight of the extensive literature on graph sparsification algorithms has been the deployment of these algorithms in modern computational settings where either the underlying graph is not static or the algorithm has only a restricted access to the underlying graph, say, via {\em linear measurements} only.
For instance, the work of \cite{AGM12}, and \cite{KLMMS14} established algorithms for creating cut and spectral sparsifiers of graphs in the \emph{linear sketching} model of computation which require only $\widetilde{O}(n / \eps^2)$ bits of space. These algorithms are then immediately amenable to other computational settings such as the massively parallel computation (MPC) model (see e.g., \cite{AssadiCLMW22,AgarwalKLP22}) and the \emph{dynamic streaming} setting (where edges in the stream can be both inserted and deleted).

In a different vein, the work of \cite{ADKKP16} studied algorithms for \emph{dynamically} maintaining sparsifiers of ordinary graphs. In this model, the graph is no longer static but is undergoing edge insertions and deletions, and the goal is to maintain a sparsifier without recomputing it from scratch after each update. The work of \cite{ADKKP16} showed that one can maintain a $(1 \pm \eps)$-cut sparsifier of a graph with worst-case update time $\poly(\log(n), 1 / \eps)$, and spectral sparsifiers with an amortized update time of $\poly(\log(n), 1 / \eps)$. Beyond this, there has also been work on graph sparsification in the \emph{small space} regime by \cite{DMVZ20} (i.e., designing spectral sparsification algorithms that use a small work-tape but have an unrestricted access to the input graph) and development of {\em deterministic} algorithms for computing spectral sparsifiers of graphs by Batson, Spielman, and Srivastava \cite{BSS09}.

In contrast, hypergraph sparsification is well-understood only in a small number of settings. Guha, McGregor, and Tench \cite{GMT15} gave the first {\em linear sketches} for hypergraph cut-sparsifiers. However, the linear sketches of \cite{GMT15} require $\widetilde{O}(n r^2 \polylog(m) / \eps^2)$\footnote{Throughout the paper, we use $\Otil$ to hide $\polylog(n)$ factors.} bits to recover a $(1 \pm \eps)$  cut-sparsifier for hypergraphs on $n$ vertices, with at most $m$ hyperedges of arity bounded by $r$. A very recent work of Khanna, Putterman, and Sudan \cite{KPS24c} constructed linear sketches for hypergraph cut-sparsifiers that achieve a nearly-optimal size of $\widetilde{O}(n r \log(m) / \eps^2)$ bits. As mentioned above, this immediately yields space-efficient dynamic streaming algorithms and communication-efficient MPC algorithms. Finally, the very recent work of Soma, Tung, and Yoshida \cite{STY24} studied hypergraph spectral sparsification in the \emph{online} model. Here, hyperedges arrive one at a time, and the sparsifier must decide, upon each arrival, whether to keep the hyperedge or not: if the hyperedge is kept, the algorithm must decide its weight immediately. The work of~\cite{STY24} shows that one can compute $(1 \pm \eps)$ hypergraph spectral sparsifiers with $\widetilde{O}(n(r + \log(m))/\eps^2)$ hyperedges, where $r$ is the maximum arity of any hyperedge.
However, to the best of our knowledge, neither hypergraph cut-sparsification nor spectral-sparsification has been studied in the fully dynamic setting.

To summarize, there are many computational models where our understanding of hypergraph sparsification \emph{significantly} lags behind our understanding of the graph setting, with either no non-trivial results known or there is a striking gap between upper bounds achieved by algorithmic results and known lower bound results.

 Because spectral hypergraph sparsification has found numerous applications
in e.g., semi-supervised learning
\cite{ZhangHTC20}, clustering \cite{LVSLG21, ZLS22}, and practical compression \cite{BestaWGGIOH19}, designing efficient sparsifiers is ever more important.
Indeed, in these modern applications, one is often facing a staggering volume of data,
which is itself evolving in real-time.
Thus, these applications vastly benefit from the ability to process this kind of data in a space- and time-efficient
manner.

Motivated by this, our work introduces a new unifying framework that significantly narrows these gaps in our understanding by providing nearly-tight algorithms for hypergraph sparsification (both cut and spectral) in the settings of linear sketching, fully-dynamic algorithms, and online algorithms. In the following section, we describe in detail our contributions to hypergraph sparsification in these settings.

\subsection{Our Contributions}

As a step towards overcoming the gap in our understanding of computing hypergraph sparsifiers in restricted settings, we introduce a general framework for reducing hypergraph sparsification to an inherently graph-theoretic task. Our starting point, as in prior works on spectral hypergraph sparsification, is the notion of a \emph{weight assignment}. Vaguely speaking, hypergraph sparsification relies on computing appropriate estimates of the \emph{importance} of each hyperedge. After computing a probability distribution $p_e$ over the hyperedges $e \in E$, one then samples every hyperedge $e$ with probability $p_e$, and reweighs the hyperedge to have weight $\frac{1}{p_e}$. Weight assignments provide a mechanism for computing these probabilities by carefully studying an associated ordinary \emph{multi-graph} for the hypergraph at hand. Unfortunately, these algorithms \emph{iteratively} refine the weight assignment to the underlying multi-graph, and therefore rely on repeated, complete access to the entire hypergraph,
which we do \textit{not} have in restricted models of computation such as linear sketching.
Our first contribution is to provide a simple algorithm for static hypergraph sparsification which removes the need for computing these weight assignments to identify important hyperedges, and instead relegates them entirely to the analysis. 

\paragraph{Vertex-Sampling Framework for Hypergraph Sparsification.} We introduce a notion of \emph{vertex-sampling} whereby we repeatedly, independently, sample subsets of the vertex set $V$. On these vertex-sampled hypergraphs, we then associate a canonical \emph{multi-graph} which replaces each hyperedge with a clique on the corresponding vertices. We show that sampling the multi-edges in proportion to their effective resistance (and then recovering the corresponding hyperedges for whichever multi-edges survive) suffices for recovering the hyperedges with ``high importance''. Further, when we build sparsifiers using this framework, the sparsity of the returned hypergraphs is \emph{nearly-optimal}. 
While the technique of sampling vertices and looking at the resulting induced subgraphs
has been used previously for ordinary graph sparsification \cite{FiltserKN21,ChenKL22},
to the best of our knowledge,
we are the first to use it for sparsification of \textit{hypergraphs}.

Note that we only present our results for unweighted hypergraphs,
but one can extend them to weighted graphs by standard geometric weight grouping tricks (see \cite{KLMMS14} or \cite{ADKKP16}).
Specifically, we show the following result:

\begin{theorem}[Informal]\label{thm:vertexSamplingMainintro}
    There is a randomized algorithm that reduces the task of creating a $(1 \pm \eps)$ spectral-sparsifier for a hypergraph $H$ with at most $m$ hyperedges to the task of sampling multi-edges at rate $\polylog(m,n) / \eps^2$ times their effective resistance in a collection $\Gcal$ of ordinary graphs.
    The graphs in $\Gcal$ are created in an oblivious manner, and moreover, the total number of vertices in graphs in $\Gcal$ is bounded by $n \polylog(m,n)$, thus ensuring that the sparsifier has only $\widetilde{O}(n \polylog(n,m) / \eps^2)$ hyperedges.
\end{theorem}

Since our algorithm is conceptually fairly simple, we present its full pseudocode
in \cref{sec:metaalgoov}.
Our framework immediately yields a novel and simple
nearly-linear time algorithm for computing
hypergraph spectral sparsifiers of nearly-linear size (see \cref{sec:nltime}).

\paragraph{Linear Sketches for Hypergraph Spectral Sparsification.} Furthermore,
due to the simplicity of our framework, we are able to directly take advantage of many well-developed techniques for estimating effective resistances in ordinary graphs and extend these to the hypergraph setting. Our first such extension is in designing \emph{linear sketches} for hypergraph spectral sparsification, where we take advantage of the prior work of \cite{KLMMS14} which designed linear sketches for effective resistance sampling in ordinary graphs. A linear sketch for a hypergraph $H$ is specified by a matrix $ P \in \R^{s \times 2^n}$, where a hypergraph is represented by an indicator vector in $\{0,1\}^{2^n}$. The sketch itself is then $P \cdot H$, and the number of bits required to store the sketch is typically $\widetilde{O}(s)$ (there are only $s$ entries in the sketch, but they may require more precision to represent). In this regime, we show the following:

\begin{theorem}\label{thm:linearSketchintro}
    There is a linear sketch for hypergraphs on $n$ vertices, $\leq m$ hyperedges, and arity $\leq r$ which uses $\widetilde{O}(n r \polylog(m) / \eps^2)$ bits of space, and can be used to recover a $(1 \pm \eps)$ spectral-sparsifier with probability $1 - 1 / \poly(n,m)$.
\end{theorem}

Recall that the work of \cite{KPS24c} showed a lower bound of $\Omega(nr \log(m))$ bits even for the simpler task of computing a $(1 \pm \eps)$ \emph{cut-sparsifier} of a hypergraph using a linear sketch.
Thus, a dependence on $\log m$ is unavoidable in the linear sketching setting.
Our linear sketch has nearly the same size and succeeds even in computing $(1 \pm \eps)$ \emph{spectral}-sparsifiers. As an immediate corollary of the above theorem, we also get the first nearly-optimal space complexity algorithm for recovering hypergraph spectral sparsifiers from dynamic streams, and show that its complexity nearly-matches that of the cut-sparsification setting:

\begin{corollary}\label{cor:dynamicStreamingintro}
    For any $\eps \in (0,1)$, there is a randomized dynamic streaming algorithm using $\widetilde{O}(nr \polylog(m) / \eps^2)$ bits of space that, for any sequence of insertions / deletions of hyperedges in an $n$-vertex unweighted hypergraph $H$ with at most $m$ edges of arity bounded by $r$, allows recovery of a $(1 \pm \eps)$ \emph{spectral}-sparsifier of $H$ with probability $1 - 1 / \poly(n,m)$ at the end of the stream.
\end{corollary}

\paragraph{Fully Dynamic Hypergraph Spectral Sparsification.} We also present the first construction of \emph{fully-dynamic} hypergraph sparsifiers with nearly-optimal update time and sparsity (for both the cut and spectral settings). Recall that in the fully-dynamic setting, there is a sequence of insertion / deletion operations of hyperedges, and after each such operation, the algorithm must output a list of changes that must be made to the existing hypergraph sparsifier such that it maintains a $(1 \pm \eps)$-approximation to the spectrum (or cut-sizes). Along these lines, we show the following:

\begin{theorem}\label{thm:fullyDynamicintro}
    There is a fully dynamic data structure that for a hypergraph $H$ on $n$ vertices with $\leq m$ hyperedges, and arity $\leq r$ undergoing a sequence of updates maintains a $(1 \pm \eps)$-\emph{spectral}-sparsifier (and thus cut-sparsifier too) with probability $1 - 1 / \mathrm{poly}(n,m)$. The expected, amortized update time of the data structure is $O(r\polylog(m,n) \cdot (1/\eps)^2\log(1/\eps))$, and it maintains a sparsifier with $\widetilde{O}(n \polylog(m) / \eps^2)$ hyperedges.
\end{theorem}

Note that even to maintain a $(1\pm\eps)$ \textit{ordinary graph} spectral sparsifier in the
fully dynamic setting, the best known result is an \textit{amortized} update time
of $\poly(\log n, \eps^{-1})$ by \cite{ADKKP16}. 
We also observe that our update-time is essentially the best possible, as simply to read each hyperedge requires $\Omega(r)$ time.
This result also resolves in the affirmative the open question 6.3
of \cite{STY24} about the existence of efficient fully dynamic hypergraph spectral sparsification algorithms.

In real-world applications, fully-dynamic sparsifiers allow for maintaining a sparse representation of the hypergraph, even as it undergoes hyperedge insertions / deletions, without re-computing the sparsifier from scratch. Because of the broad applicability of hypergraph spectral sparsifiers for faster clustering algorithms \cite{LM17, LM18, VBK20, VBK21,LVSLG21, ZLS22}, this immediately yields faster clustering as the underlying hypergraph evolves. We note that our results hold only for the oblivious adversary fully-dynamic model, where the hyperedge updates are independent of the current sparsifier (as opposed to adaptive adversary models, such as the one considered in the work of \cite{BBGNSS022} for ordinary graphs).

\paragraph{Online Hypergraph Spectral Sparsification.} Finally, we also present a new nearly-optimal algorithm for online hypergraph spectral sparsification. Here, the algorithm is presented with a stream of hyperedges to be inserted. After seeing each hyperedge, the algorithm must immediately decide whether or not to keep the hyperedge (as well as what weight should be assigned if kept). Once a hyperedge is kept, the algorithm cannot remove it and cannot change its weight. In the context of hypergraphs, this problem was first studied in \cite{STY24}, who showed the existence of online hypergraph spectral sparsification algorithms which require only $\widetilde{O}(n^2 \log(m))$ bits of space, and produce sparsifiers with $\widetilde{O}(n(r + \log(m)) / \eps^2)$ hyperedges. Using our framework, we are able to improve the sparsity
by a factor of $r$ in the case when $m\leq \poly(n)$:

\begin{theorem}\label{thm:onlineintro}
    There is an online hypergraph spectral sparsification algorithm, which for hypergraphs $H$ on $n$ vertices with $\leq m$ hyperedges, and arity $\leq r$ undergoing a sequence of insertions maintains a $(1 \pm \eps)$-\emph{spectral}-sparsifier (and thus cut-sparsifier too) with probability $1 - 1 / \mathrm{poly}(n,m)$. The space complexity of the algorithm is $\widetilde{O}(nr\polylog(m) / \eps^2)$ bits and the sparsifier contains $\widetilde{O}(n \polylog(m) / \eps^2)$ hyperedges.
\end{theorem}
The space complexity of $\widetilde{O}(nr\polylog(m) / \eps^2)$ bits of our online algorithm matches
\textit{exactly}
the bound proposed in the open question 6.2 of \cite{STY24}, thus resolves it affirmatively.

We complement the algorithmic result of Theorem~\ref{thm:onlineintro} with an $\Omega(n \log(m))$ lower-bound on the number of hyperedges that must be kept by any online sparsifier in the worst-case (\cref{sec:ollb}). Hence, unlike the static setting, in the online setting a dependence on $\log m$ is unavoidable.

The following table depicts the landscape of hypergraph sparsification in different models of computation,
with our contributions highlighted in \textcolor{red}{red} which includes as a reference a linear sketching algorithm for hypergraph spectral sparsification that can be easily derived from known results (see \cref{sec:lsnaive}).

\renewcommand{\arraystretch}{1.5}
\begin{center}
\begin{table}
\begin{tabular}{||c c c c c||}
 \hline
 Work & Setting & Sparsification & Space Complexity & Time Complexity \\ [0.5ex] 
 \hline\hline
 \cite{KPS24c} & Linear Sketching & Cut & $\widetilde{O}(nr\log(m)/\eps^2)$ bits & N/A \\
 \hline
  Naive ({\hypersetup{linkcolor=black}\cref{sec:lsnaive}}) & Linear Sketching & Spectral & $\widetilde{O}(n r^2 \polylog(m)/\eps^2)$ bits & N/A \\
 \hline
 \textcolor{red}{This work} & \textcolor{red}{Linear Sketching} & \textcolor{red}{Spectral} & \textcolor{red}{$\widetilde{O}(nr\polylog(m) / \eps^2)$ bits} & \textcolor{red}{N/A} \\
 \hline
 \textcolor{red}{This work} & \textcolor{red}{Fully-Dynamic} & \textcolor{red}{Cut} &
 \textcolor{red}{$\widetilde{O}(n\polylog(m) / \eps^2)$ edges} &
 \textcolor{red}{$\widetilde{O}({r \polylog(m)/\eps^{2}}{})$} \\ 
 \hline
 \textcolor{red}{This work} & \textcolor{red}{Fully-Dynamic} &
 \textcolor{red}{Spectral} & \textcolor{red}{$\widetilde{O}(n\polylog(m) / \eps^2)$ edges} & \textcolor{red}{$\widetilde{O}({r\polylog(m)/\eps^{2}}{})$} \\
 \hline
 
 \cite{STY24} & Online & Spectral & $O(n (r + \log(m)) / \eps^2)$ edges & N/A \\ 
 \hline
 \textcolor{red}{This work} & \textcolor{red}{Online} & \textcolor{red}{Spectral} &
 \textcolor{red}{$\widetilde{O}(n\polylog(m) / \eps^2)$ edges} & \textcolor{red}{N/A} \\[1ex] 
 \hline
\end{tabular}
\caption{Summary of our results.}
\end{table}
\end{center}

\section{Technical Overview}

In this section, we give a high level overview of our techniques.
Throughout this section,
we focus on
unweighted
hypergraphs that only have hyperedges of arity (cardinality) between $[r,2r]$,
as one can extend our results to general weighted hypergraphs of arbitrary arity by geometrically grouping hyperedges by arity and weights.

\subsection{A New Hypergraph Spectral Sparsification Framework}\label{sec:metaalgoov}

We now present our new unified framework for spectral hypergraph sparsification which as we will see later, lends itself easily to efficient implementation in a variety of different computational models. At a high-level, our meta-algorithm is based on a very natural strategy for sparsification. Given a hypergraph $H(V,E)$, identify a set $F \subseteq E$ of {\em critical} hyperedges that almost certainly needs to be included in any sparsifier. The remaining hyperedges in $E - F$ are not critical, and we can afford to subsample them with probability $1/2$ to create a new hypergraph $H'(V,E')$. 
One can then focus on the task of recursively building a sparsifier for the graph $H'$ which contains only half as many hyperedges as the starting hypergraph $H$. 
The main challenge then is in efficiently identifying the set of critical hyperedges.

Our meta-algorithm is completely specified by a pair of subroutines that implement the strategy above. The first subroutine, given by \cref{alg:VS}, reduces the problem of identifying critical hyperedges to effective resistance sampling in {\em ordinary} graphs. The second subroutine, given by \cref{alg:sparsifyov}, then recursively builds a sparsifier by invoking \cref{alg:VS} to identify critical hyperedges at each level of recursion. The heart of our approach is \cref{alg:VS} for recognizing critical hyperedges, and we describe it next.

As mentioned before, our key new insight is that the task of identifying critical hyperedges can be reduced to effective resistance sampling in a suitable collection of ordinary graphs, generated by sub-sampling vertices of the hypergraph.
Specifically, for a hypergraph $H = (V, E)$, \textit{vertex-sampling} at rate $p$ refers to sampling each vertex in $V$ independently with probability $p$. If we denote the resulting vertex set by $V'$, then it defines a new hypergraph $H' = (V', E')$, where $E' = \{e \cap V': e \in E \}$. Another key notion that we will utilize is 
the notion of a \textit{multi-graph} of a hypergraph $H$, denoted by $\Phi(H)$, which is an {\em ordinary graph} obtained by replacing each hyperedge with a clique on the vertices of the hyperedge. Note that since a pair of vertices may be contained in multiple hyperedges, this gives rise to a multi-graph.
Notationally, we also write $R^G(u,v)$ for an ordinary (multi-)graph $G$ to denote the effective
resistance between $u,v$ in $G$.

\quad

\begin{restatable}[H]{algorithm}{vsov}
    \caption{$\VS(H,\lambda)$}
    \label{alg:VS}
    \Input{A hypergraph $H$ with $\leq m$ hyperedges of arity $[r, 2r]$,
    and an oversampling rate $\lambda \geq 1$.}
    \Output{A set of hyperedges $F$.}
    Initialize $F\gets \emptyset$. \\
    \For{$r \polylog(n,m)$ {\rm rounds}}{
        Vertex sample $H$ at rate $1/r$ to get $H'$, with $\Phi(H')$ being its multi-graph. \\
        Independently sample each multi-edge $(u,v)$ of $\Phi(H')$ with
        probability $\lambda\cdot R^{\Phi(H')}(u,v) $. \\
        Let $F'$ contain all hyperedges of $H$
        for which at least one associated multi-edge
        got sampled. \\
        Let $F\gets F\cup F'$ and delete $F'$ from $H$.
    }
    \Return{$F$.}
\end{restatable}

\quad

\begin{restatable}[H]{algorithm}{sparsifyov}
\Input{A hypergraph $H$ with $\leq m$ hyperedges of arity $[r, 2r]$, and a parameter $\epsilon$.}
\Output{A reweighted subgraph of $H$.}
    \caption{HypergraphSpectralSparsify$(H, \eps, r, m)$} \label{alg:sparsifyov}\label{alg:sparsifyintro}
    \label{alg:sparsify}
    Let $H_0 = H$. \\
    \For{$i = 0, \dots, \log(m)$}{
    Let $F_i = \VS(H_i, \polylog(n,m) / \eps^2)$. \\
    Let $H_{i+1} = H_i - F_i$ with its hyperedges sampled at rate $1/2$.
    }
    \Return{$\bigcup_{i = 0}^{\log(m)} 2^i \cdot F_i$,
    \rm{i.e. put weight $2^i$ on hyperedges in $F_i$.}}
\end{restatable}

\subsection{Analysis of Our Meta-Algorithm}

In spectral sparsification of ordinary graphs, critical/important edges are defined to be edges whose effective resistance is $\Omega(1/\log n)$ \cite{SS11}. For spectral sparsification of hypergraphs, previous work \cite{KKTY21b,JambulapatiLS23,Lee23} determined important edges based on the notion
of a weight assignment which are computed by an iterative refinement approach. In contrast, Algorithm~\ref{alg:VS}
above identifies important hyperedges by implementing effective resistance sampling in ordinary graphs that are obtained by \textit{oblivious} vertex sampling of the original hypergraph,
completely bypassing the need for computing a weight assignment. The analysis of Algorithm~\ref{alg:VS}, however, will rely on the notion of weight assignments, to argue correctness of our approach.

\paragraph{Weight Assignments.}
Recall that given a hypergraph $H = (V, E)$, we can associate with it a \emph{multi-graph}, denoted by $G = \Phi(H)$, where for each hyperedge $e \in E$, we replace it with a clique $K_e$ on the corresponding vertices. A \textit{weight assignment} $W$ assigns a weight $w_f$ to every multi-edge in $G$ such that for every hyperedge $e \in H$, the sum of the weights of the corresponding multi-edges in $G$ is $1$.

As shown in \cite{KKTY21b,JambulapatiLS23,Lee23}, for any weight assignment $W$ of $G$, sampling each hyperedge $e$
with probability given by $p_e = \eps^{-2} \polylog(n)\cdot \max_{(u,v) \in e} R^{G(W)}(u,v)$ suffices to yield a $(1\pm\eps)$-spectral sparsifier,
where we write $G(W)$ to denote $G$ with edges weighted by $W$,
and write $R^{G(W)}(u,v)$ to denote the effective resistance between $u,v$
in $G(W)$.

That is,
the sampling rate for each hyperedge $e$ is proportional to the \emph{maximum} over all multi-edges $f = (u,v) \in e$ of the effective resistance between $u,v$ in $G(W)$.
Consequently, the important hyperedges are the ones that contain some $u,v$ between which
the effective resistance is at least $1/\polylog(n,m)$ in the graph $G(W)$. Naturally then, the difficulty comes in showing that one can compute weight assignments such that $\sum_{e \in E} p_e$ is not too large (in particular, the number of important hyperedges is $\Otil(n)$), as this then immediately yields small spectral hypergraph sparsifiers. Unfortunately, finding such weight assignments requires an iterative refinement procedure that needs
repeated, complete access to the entire hypergraph,
which we do not have in modern models of computation such as linear sketching.

\paragraph{Bypassing Weight Assignments Algorithmically.}
Instead, we present a sampling procedure that operates on the underlying (unweighted) multi-graph, i.e. $\VS(H,\lambda)$ in \cref{alg:VS},
to recover the important hyperedges.
This procedure samples multi-edges without needing to compute a weight assignment. Then, for each sampled multi-edge $f$, we recover the corresponding hyperedge (i.e., the $e \in H$ such that $f \in e$). In order to show the correctness of our procedure, the key is to show that \emph{there exists} a single weight assignment $W^*$ to the multi-graph $\Phi(H)$ such that every hyperedge $e$ with a large sampling rate under $W^*$ is recovered by our procedure. For the remaining hyperedges which are not recovered, this weight assignment $W^*$ is a certificate to the fact that their sampling rates are small, and therefore they do not yet need to be recovered. Crucially, we do \textit{not} need to explicitly
compute such a $W^*$, but only need its existence for our analysis.

There is a delicate and complex argument needed to show the existence
of a weight assignment $W^*$ to which we can couple the success of the scheme.
In particular, our key technical theorem is the following:

\begin{restatable}[Key Technical Theorem]{theorem}{keythmov}
\label{prop:goalintro}
\label{prop:goal}
    For any $\theta \in (0,1)$, given a hypergraph
    $H$ and its multi-graph $G = \Phi(H)$,
    there exists a single weight assignment $W^*$ of $G$
    such that for \textit{every} hyperedge
    $e\in H$, one of the following statements holds:
    \begin{enumerate}[label=(\alph*)]
        \item either $\max_{(u,v)\in e} R^{G(W^*)}(u,v) \leq
            \theta$, or
        \item we have that $\pr{e\in \VS(H, \lambda = \theta^{-1}\polylog(n,m))}
            \geq 1 - 1/\poly(n,m) $,
        where the $\poly(n,m)$ in the success probability depends
        on the $\polylog(n,m)$ in the parameter $\lambda$ of \cref{alg:VS}.
    \end{enumerate}
\end{restatable}

By setting $\theta = \eps^{2}/\polylog(n,m)$ in
\cref{prop:goalintro},
we get that there exists a weight assignment $W^*$ such that
each hyperedge either (i) can be recovered by our vertex sampling algorithm
$\VS(H,\polylog(n,m))$ with high probability, or (ii) can be sampled with probability
$1/2$ while preserving the entire spectrum to within $1\pm\eps/\polylog(n,m)$.
We thus classify the hyperedges satisfying (i) as important ones,
and sample the rest with probability $1/2$ in \cref{alg:sparsifyintro}.

At a high level, the proof of \cref{prop:goalintro} consists of two parts.
First, we connect important hyperedges
to a new notion of multi-way energy that we call
\textit{collective energy}\footnote{This may be seen as a continuous generalization
of $k$-cuts, just as ordinary energy vs. 2-cuts;
see \cref{rmk:kcut} below.}, defined for
a set of potential vectors $x_1,\ldots,x_k\in\mathbb{R}^{n}$ as
\begin{align}\label{eq:collectiveenergyearlyearly}
    \Ecal^H(x_1,\ldots,x_k) \defeq
    \sum_{e\in H}
    \max_{(a,b)\in e}
    \kh{
    \sum_{i=1}^{k} (x_i[a]-x_i[b])^2
    }.
\end{align}
Notably, $\Ecal^H(x_1,\ldots,x_k)$
is solely determined by the hypergraph $H$ and the potential vectors,
without reference to any weight assignments.
We derive that
each important hyperedge is ``witnessed'' by a set of $k$ potential
vectors with small collective energy (in particular, $\leq k\polylog(n,m)$).
Then in the second part,
we show that the existence of such a set of potential vectors
witnessing important hyperedges
can be in turn used to show that the witnessed
hyperedges can be recovered
by our vertex sampling algorithm with high probability.
Crucially, we do \textit{not} ever need to compute these potential vectors explicitly -
they are only used in our analysis.

The formal statement of the first part is summarized in our
\textit{Collective Energy Lemma} (\cref{lem:medium}),
and the formal statement of the second part is summarized in our
 \textit{Vertex Sampling Lemma} (\cref{lem:vse}).
We also refer the reader to our \hyperref[proof:thm1]{Proof of \cref*{prop:goal}}
for how we combine the two parts
to prove the above theorem.

Due to space limitations,
we defer a detailed discussion of the proof of \cref{prop:goalintro} to \cref{sec:techOverviewAnalysis}, with the first part discussed in \cref{sec:collectiveenergyintro} and the second part discussed in
\cref{sec:vsintro}.

For now, we proceed by
understanding the correctness of \cref{alg:sparsifyintro}, as well as
describing the implementation of our framework in modern models
of computation.

\paragraph{Correctness of \cref{alg:sparsifyintro}.}
 We now show how \cref{alg:sparsifyintro} leverages \cref{prop:goalintro} into a simple algorithm for sparsifying the overall hypergraph. This type of framework is relatively standard, appearing in many works (\cite{KoutisX16, GMT15, ADKKP16, ChenKL22, KPS24c} to name a few).

To understand the above algorithm, let us focus on the first iteration, as it turns out the reasoning extends to future iterations as well. When we run the vertex-sampling algorithm on $H_0 = H$, we recover some set $F_0 \subseteq H_0$ of hyperedges. By \cref{prop:goalintro}, we can then argue that with high probability, there exists a single weight assignment $W^*$ for which \emph{all remaining non-recovered hyperedges} $H_0 - F_0$ would have maximum pairwise resistance $\leq \frac{\eps^2}{\polylog(m,n)}$ in the multi-graph weighted by $W^*$. Per \cite{JambulapatiLS23},
by choosing an appropriate $\polylog(n,m)$,
when we now sample at rate $1/2$ to get $H_{1}$, it will be the case that $F_0 \cup 2 \cdot H_1$ is a $(1 \pm \eps)$ spectral-sparsifier for $H_0$ with high probability. 

In general, we can extend this reasoning beyond the first iteration: in the $i$th iteration $F_i \cup 2 \cdot H_{i+1}$ is a $(1 \pm \eps)$-spectral sparsifier for $H_i$ with high probability. By composing the sparsifiers,
if $\eps \leq O(1/\log m)$,
we see that the final returned result is a $(1 \pm O(\eps \log(m)))$-sparsifier to $H$ with high probability. By instead running the sparsification algorithm with $\eps' = \eps / \log(m)$, we then get our desired result.

To see the sparsity of the sparsifier, we see that each hyperedge which is recovered must correspond to a multi-edge that is recovered. But, recall that we are doing effective resistance sampling on $r \polylog(m,n)$ multi-graphs, each on $n \polylog(n,m) / r$ vertices, with an oversampling factor of $\polylog(m,n) / \eps^2$. On each multi-graph, we therefore recover $\widetilde{O}(\frac{n\polylog(m)}{r\eps^2})$ multi-edges, and in total across all rounds, and all $\log(m)$ levels of sampling, we recover $\widetilde{O}(n \polylog(m)/ \eps^2)$ multi-edges. Because hyperedges are recovered only if a corresponding multi-edge is recovered, the final sparsity is therefore $\widetilde{O}(n \polylog(m)/ \eps^2)$ hyperedges. To summarize, we get the following lemma:

\begin{lemma}\label{lem:correctnessSparsityintro}
   Given as input a hypergraph $H$ and parameter $\eps \in (0,1)$,  \cref{alg:sparsifyintro} creates a $(1 \pm \eps)$ spectral sparsifier $\widetilde{H}$ of $H$ with only $\widetilde{O}(n \polylog(m) / \eps^2)$ hyperedges with probability $1 - 1/\poly(n,m)$.
\end{lemma}

In the next few sections,
we show how to implement our new sparsification framework in various modern models
of computation.

\subsection{Linear Sketching Hypergraph Spectral Sparsifiers}

The key observation is that \cref{alg:sparsifyintro} relies only on the ability to sample ordinary edges in proportion to their effective resistance in various vertex-sampled multi-graphs. The work of \cite{KLMMS14} initiated the study of linear sketches for creating spectral-sparsifiers, and in fact, this work already shows the existence of a linear sketch (in small space) which can sample edges at rates proportional to their effective resistance. 

Specifically, \cite{KLMMS14} showed that for a graph $G$ on $n$ vertices and a parameter $\eps > 0$, there is a linear sketch of size $\widetilde{O}(n / \eps^2)$ bits which can be used to recover a $(1 \pm \eps)$ spectral-sparsifier of $G$ with high probability. Extending this linear sketch to work for our multi-graphs is straightforward, and requires only blowing up the space by a factor of $\widetilde{O}(r\log(m))$ bits (a factor of roughly $r$ for the ``universe size'', i.e., number of distinct multi-edge slots, and a factor of roughly $\log(m)$ for the support size, i.e., the maximum number of multi-edges ever present in one multi-graph). Because this linear sketch already enables sampling in proportion to effective resistance, all that remains to implement \cref{alg:sparsifyintro} is to recover the indices of the corresponding hyperedges whenever we recover a multi-edge. However, this is essentially canonical: the hypergraph is represented by a vector in $\{0,1\}^{\binom{n}{r}}$, and the corresponding multi-graph is represented by a vector in $\{0,1\}^{\binom{n}{r} \cdot r^2}$ (where each hyperedge creates $O(r^2)$ multi-edge slots). Whenever the linear sketch recovers a multi-edge, this is reported as an index in $[\binom{n}{r} \cdot r^2]$. But, given this index, there is a single hyperedge which corresponds to this index, and therefore this hyperedge must be present.

The total space required for the linear sketch is just that of $\widetilde{O}(r \polylog(m,n))$ linear sketches for multi-graph effective resistance sampling, with each multi-graph defined on $O(\binom{n}{r}\cdot r^2)$ multi-edge slots, a support of $O(mr^2)$ multi-edges, and $O(n\polylog(n,m)/r)$ vertices, and each linear sketch using an over-sampling rate of $\widetilde{O}(\polylog(m,n)/\eps^2)$. For each multi-graph, the space required is $\widetilde{O}(\frac{n}{r} \cdot r\polylog(m) / \eps^2)$, and across all levels of sampling, and iterations of recovery, the number of multi-graphs is $O(r \polylog(m,n))$. Thus, the total space required by the linear sketch is $\widetilde{O}(n r \polylog(m) / \eps^2)$ bits. This yields \cref{thm:linearSketchintro}.

\subsection{Recursive Recovery Framework}

Unfortunately, re-using the same strategy for the fully-dynamic and online settings does not work. This is primarily due to the fact that while there are linear sketches which can accomplish effective-resistance sampling of multi-edges, there are no known fully-dynamic (or online) algorithms which can maintain effective-resistance samples of a graph as it undergoes insertions and deletions. In prior works on designing spectral sparsifiers of graphs in the fully-dynamic setting \cite{ADKKP16}, this is tackled by maintaining disjoint spanners at different levels of sampling.

More clearly, given a graph $G$, one can store a sequence $T = T_1 \cup \dots \cup T_{\polylog(n,m) / \eps^2}$ of disjoint spanners, where each $T_i$ is a $\log(n)$-spanner of $G - T_1 - \dots T_{i-1}$. In \cite{ADKKP16}, the authors showed via a simple combinatorial argument that any edge of effective resistance $\geq \frac{\eps^2}{\polylog(n)}$ \emph{must} be contained in this set $T$ of spanners. Immediately, this implies a simple algorithm for computing spectral sparsifiers whereby one recursively stores these disjoint spanners, and then subsamples the remaining edges at rate $1/2$. The correctness follows from the fact that the remaining edges have small effective resistance, whereby one can argue the concentration from \cite{ST11, SS11}, who showed that effective resistance sampling creates spectral sparsifiers. This leads to the following algorithm for creating spectral sparsfiers:

\begin{algorithm}[H]
    \caption{GraphSpannerSparsification$(G, n, m, \eps)$}\label{alg:iterativeRecoveryintro}
    Let $G_1 = G$. \\
    \For{$i \in [\log(m)]$}{
    Let $T^{(i)} = T^{(i)}_1 \cup \dots T^{(i)}_{\polylog(n,m)/\eps^2}$ be a sequence of $\polylog(n,m)/\eps^2$ disjoint spanners of $G_i$. \\
    Let $G_{i+1}$ be the result of sampling $G_i - T^{(i)}$ at rate $1/2$.
    }
    \Return{$T^{(1)} \cup2 \cdot T^{(2)} \cup \dots 2^{\log(m)} \cdot T^{(\log(m))}$}
\end{algorithm}

In \cite{ADKKP16}, the authors showed that this framework leads to fully-dynamic algorithms for spectral sparsifiers, as one can design decremental algorithms for maintaining these disjoint collections of spanners (and then bootstrap the decremental solution into a fully-dynamic one). Unfortunately for us however, our analysis still relies on being able to do \emph{effective-resistance sampling}, which is not directly in line with the above procedure.

In this direction, another one of our contributions is to show that the above procedure essentially simulates effective resistance sampling, up to some small degradations in the sampling probability. Specifically, we show the following:

\begin{lemma}\label{clm:ERSumRecoveryintro}
    Let $Q \subseteq G$ denote a set of multi-edges, and let $\eps > 0$ be a parameter. Then, if we let $S$ denote the set of multi-edges recovered in spanners as a result of running \cref{alg:iterativeRecoveryintro} on $G$, we have that 
    \[
    \Pr[S \cap Q \neq \emptyset] \geq \min \left (2/3, \sum_{e \in Q} R_{\mathrm{eff}, G}(e) / \eps^2 \right ) - \frac{\log(m)}{n^{20}}. 
    \]
\end{lemma}

In particular, for a hypergraph $H$, and for each hyperedge $e \in H$, we let the set $Q = K_e$, i.e., the set of multi-edges corresponding to $e$. Per \cref{prop:goalintro}, if the hyperedge $e$ is to be recovered by vertex sampling and effective-resistance sampling, then the constituent multi-edges in $K_e$ must at some have a significant effective resistance ($\geq \frac{\eps^2}{\polylog(n,m)}$). Then, by instead running our \emph{recursive} recovery procedure that stores spanners, we will be guaranteed that \emph{some multi-edge} in $K_e$ is recovered, which then suffices for recovering the hyperedge $e$ itself. Due to the degradation in error probability, we run the recursive recovery procedure some $\polylog(n,m)$ times to boost the probability of recovering all necessary hyperedges.

\subsection{Fully Dynamic Sparsification and Online Sparsification}

As a consequence of the above recursive recovery framework, we are able to construct fully-dynamic algorithms for building hypergraph sparsifiers and online algorithms for building hypergraph spectral sparsifiers. 
Roughly speaking, the key intuition here is that we have reduced the task of spectrally sparsifying a hypergraph to the task of maintaining a set of disjoint spanners of some corresponding multi-graphs. 

\paragraph{Fully-Dynamic} In the fully-dynamic setting, our starting point is existing constructions of \emph{decremental} spanners of simple graphs \cite{ADKKP16}. By developing some supplementary data structures, we are in fact able to extend these decremental spanners to arbitrary multi-graphs, and thus create a decremental implementation of hypergraph spectral sparsifiers. As in \cite{ADKKP16}, we are then able to leverage this decremental sparsifier data structure into a \emph{fully-dynamic} sparsifier data structure using a well-known reduction. All that remains then is to calculate the time complexity and space complexity of the fully-dynamic algorithm: it turns out that we store $r \polylog(m,n) / \eps^2$ vertex-sampled multi-graphs (each on $n/r$ vertices). For any hyperedge $e$ of arity $r$, this means that we expect there to be at most $r\polylog(m,n)/\eps^2$ corresponding multi-edges across \emph{all} multi-graphs. Upon removing a hyperedge, this leads to an amortized expected time-complexity of $O(r\polylog(m,n)/\eps^2)$, as removing a single multi-edge from a single spanner requires time $O(\polylog(n,m))$. The sparsity of the construction will be $\widetilde{O}(n \polylog(m) / \eps^2)$ hyperedges, as it is exactly in line with \cref{lem:correctnessSparsityintro}. This yields a proof of \cref{thm:fullyDynamicintro}.

\paragraph{Online} In the online setting, the algorithm is presented with a stream of hyperedge insertions and must decide immediately after seeing each hyperedge, whether or not to keep it, and decide on the corresponding weight to assign it. Fortunately, we have already reduced the task of hypergraph spectral sparsification to storing spanners, and creating spanners in an online manner is exceptionally straightforward: given a spanner constructed so far $T$, and the new edge $f$ to be inserted, we simply check if $f$ creates any cycles of length $\leq \log(n)$ in $T$. If not, we include $f$ in the spanner, and otherwise, we do not include $f$.

From a global perspective, whenever a hyperedge $e$ arrives, we try inserting all of the corresponding multi-edges to $e$ in each of the respective vertex-sampled multi-graphs. If, in any vertex-sampled multi-graph, a multi-edge $e$ is included in one of the spanners, then we know that we must keep the hyperedge $e$ with weight $1$. Otherwise, we flip a coin; with probability $1/2$ the hyperedge $e$ is deleted, and with probability $1/2$ we instead try inserting the hyperedge $e$ into the second level of spanners. We continue on in this manner for $\log(m)$ levels of spanners, until the hyperedge $e$ does not ``survive'' one of the coin flips. This reasoning yields the proof of \cref{thm:onlineintro}. Observe that the space of the online sparsifier follows from the fact that we store $r \polylog(m,n) / \eps^2$ $O(\log(n))$ spanners on graphs with $O(\frac{n}{r})$ vertices. This requires storing at most $O(n \polylog(m) / \eps^2)$ multi-edges, which stores at most $\widetilde{O}(n \polylog(m) / \eps^2)$ hyperedges. 

Thus, we have seen how the statement of \cref{prop:goalintro} is powerful, and yields a host of novel sublinear spectral sparsification results for hypergraphs. In the following section, we provide more insight into the proof of \cref{prop:goalintro}.

\subsection{Overview of Key Technical Theorem (\texorpdfstring{\cref{prop:goalintro})}{}}\label{sec:techOverviewAnalysis}

We now give an overview for our
key technical theorem (\cref{prop:goalintro}).
We first discuss our
\textit{Collective Energy Lemma} (\cref{lem:medium}) in \cref{sec:collectiveenergyintro},
and then discuss our
 \textit{Vertex Sampling Lemma} (\cref{lem:vse})
in
\cref{sec:vsintro}.
We refer the reader to our \hyperref[proof:thm1]{Proof of \cref*{prop:goal}}
for how we combine the two lemmas
to prove our key technical theorem.

Before diving into the discussion,
we first set up some definitions and notation, as well as present
the formal statements of the lemmas we are about to discuss.
For reasons that will be clear later in our discussion in \cref{sec:vsintro},
we restrict to only vertex potentials with entries in $[0,1]$ when presenting the statements
of the two lemmas.

\begin{restatable}[Vertex Potentials and Spanning Hyperedges]{definition}{defspan}
\label{def:span}
    A set of vertex potentials is a
    real-valued vector $x\in [0,1]^{n}$ supported
    on the $n$ vertices of a hypergraph $H$ (or its multi-graph $G$ which
    has the same set of vertices).
    We say a hyperedge $e\in H$ \textit{spans} $x$ if
    $\exists (u,v)\in e$ such that $x[u] = 1$ and $x[v] = 0$.
\end{restatable}
Recall that given vertex potentials $x_1,\ldots,x_k\in\mathbb{R}^{n}$,
    their collective energy is
    \begin{align*}
        \Ecal^H(x_1,\ldots,x_k) \defeq
        \sum_{e\in H}
        {\max_{(u,v)\in e} 
        \kh{ \sum_{i=1}^{k} \kh{x_i[u] - x_i[v]}^2 }
        }.
    \end{align*}
We also define
    for a set of hyperedges $F$ the following minimax resistance optimization:
    \begin{align*}
        \OPT_F \defeq
        \min_{W\in \Wcal} \max_{e\in F}
        \max_{(u,v)\in e} R^{G(W)}(u,v),
    \end{align*}
    where $\Wcal$ denotes the set of all valid weight assignments of the multi-graph
    $G=\Phi(H)$ of $H$.
    That is, $\OPT_F$ characterizes
    the minimum max sampling rates of the hyperedges in $F$
    over all valid weight assignments.

We present the formal statements of our two key lemmas
below.

\begin{restatable}[Collective Energy Lemma]{lemma}{lemmedium}
\label{lem:medium}
    For any $\theta\in(0,1)$,
    given a hypergraph
    $H$, 
    and a hyperedge set $F$ in $H$,
    if $\OPT_F \geq \theta$,
    then
    there exists a set of potentials $x_1,\ldots,x_k\in[0,1]^{n}$
    with $k\leq\poly(n)$
    such that
    the following two statements both hold:
    \begin{enumerate}
        \item Each $x_i$ is spanned by at least one hyperedge in $F$
        (cf. \cref{def:span}).
        \item $\Ecal^H(x_1,\ldots,x_k) \leq k \polylog(n,m) / \theta$.
    \end{enumerate}
\end{restatable}

\begin{restatable}[Vertex Sampling Lemma]{lemma}{lemvse}
\label{lem:vse}
Suppose we are given
a $\theta\in(0,1)$ satisfying $\theta \geq 1/\poly(n)$,
    a hypergraph
    $H$, 
    and a hyperedge set $F$ in $H$, along with sets of potentials $x_1,\ldots,x_k\in[0,1]^{n}$ such that
the following both hold:
   \begin{enumerate}
        \item Each $x_i$ is spanned by at least one hyperedge in $F$
        (cf. \cref{def:span}).
        \item $\Ecal^H(x_1,\ldots,x_k) \leq k \polylog(n,m) / \theta$.
    \end{enumerate}
    Then, there (deterministically) exists an $f\in F$ that spans
    at least one $x_i$ for which we have
    \begin{align*}
        \pr{f\in \VS(H,\theta^{-1}\polylog(n,m)}
        \geq 1 - 1/\poly(n,m),
    \end{align*}
     where the $\poly(n,m)$ in the success probability depends
        on the $\polylog(n,m)$ in the vertex-sampling scheme.
\end{restatable}

\subsubsection{Important Hyperedges and Collective Energy}\label{sec:collectiveenergyintro}

Our goal in this section is to connect important hyperedges
to a new notion of multi-way energy that we call
\textit{collective energy}, defined for
a set of potential vectors $x_1,\ldots,x_k$ as
\begin{align}\label{eq:collectiveenergyearly}
    \Ecal^H(x_1,\ldots,x_k) \defeq
    \sum_{e\in H}
    \max_{(a,b)\in e}
    \kh{
    \sum_{i=1}^{k} (x_i[a]-x_i[b])^2
    }.
\end{align}
We will eventually derive that
each important hyperedge is witnessed by a set of $k$ potential
vectors with small collective energy (in particular, $\leq k\polylog(n,m)$).
We refer the reader to the \textit{Collective Energy Lemma} (\cref{lem:medium})
for the formal statement.

This can be seen as the spectral analog of the work by Quanrud \cite{Qua23},
where they showed how to identify important hyperedges for cut sparsification
by looking at the \textit{ratio $k$-cuts}.
As we discuss in \cref{rmk:kcut} below,
their notion of ratio $k$-cuts can indeed be seen
as special cases of our collective energy.
However, we will take an entirely different approach,
as the cut counting bound derivation in \cite{Qua23} does not extend to spectral sparsification where we have to deal with real-valued vectors.

\begin{remark}[Collective Energy and $k$-Cuts]
    \label{rmk:kcut}
    We highlight an interesting connection between our notion
    of collective energy and the size of $k$-cuts in hypergraphs.
    
    Specifically, consider the case when
    $x_1,\ldots,x_k$ are the indicator vectors of
    a $k$-way partition $V_1,\ldots,V_k$ of the vertices.
    That is, $x_1,\ldots,x_k$ are all $0/1$ vectors with disjoint supports
    that form a partition of $V$.
    Then, one can verify that
    their collective energy
    $\Ecal^H(x_1,\ldots,x_k)$ equals \textit{exactly}
    twice the size of the corresponding $k$-cut in $H$.
    Consequently, in this special case,
    our small collective energy condition means that 
    this $k$-cut has size at most $k\polylog(n,m)$,
    which is exactly the criteria for determining important
    hyperedges for cut sparsification
    by Quanrud \cite{Qua23} based on their notion of ratio $k$-cuts.
    
\end{remark}

We start by discussing a conceptual idea for identifying the important hyperedges
that leads us to a certain optimization problem of effective resistances.

\paragraph{Minimax Optimization of Resistances.}

Consider the following idea of identifying the important hyperedges.
First, if we were able to choose a weight assignment $W$ such that
\textit{simultaneously for all hyperedges} $e$
we have
\begin{align*}
    \max_{(u,v)\in e} R^{G(W)}(u,v) < 1 / \polylog(n,m),
\end{align*}
then it would be great since we would have concluded that there are \textit{no} important
hyperedges that we need to keep in our sparsifier.
On the other hand, if no such weight assignment exists, we would like to find a certain
``certificate'' that certifies the absence of such weight assignments, which can in turn help us find
the ``bottleneck'' hyperedges preventing us from finding the desired $W$.
Ideally, the certificate should be oblivious to any weight assignment.
We will then classify the bottleneck hyperedges as important hyperedges.

Let us now rephrase the idea above from an optimization point of view.
Consider the following minimax optimization problem of effective resistances,
whose optimal value we denote by $\OPT_H$:
\begin{align}\label{eq:minimaxer}
    \OPT_H \defeq \min_{W\in \Wcal} \max_{e\in H} \max_{(u,v)\in e}
    R^{G(W)} (u,v),
\end{align}
where we write $\Wcal$ to denote the collection of all possible valid weight assignments.
If $\OPT_H < 1/\polylog(n,m)$, then we conclude that
there are \textit{no} important hyperedges that we are bound to keep in our sparsifier.
Otherwise,
we want to find some certificate certifying $\OPT_F \geq 1/\polylog(n,m)$,
which can in turn help us find the ``bottleneck'' hyperedges that prevent
$\OPT_F$ from ever going below $1/\polylog(n,m)$.
Since we want the certificate to be oblivious to any weight assignments,
a natural candidate is \textit{vertex potentials}, which are known
to characterize effective resistances through their \textit{energies}.

Indeed, we will exploit the following well-known connection between
effective resistances and energies of vertex potentials.
Specifically, for any $u,v\in V$, we have the following energy minimization
view of effective resistance:
\begin{align}\label{eq:erenergy}
    1/R^{G(W)}(u,v) =
    \min_{x\in\mathbb{R}^{n}: x[u]=1,x[v]=0}
    \Ecal^{W}(x),
\end{align}
where we call $x$ a set of \textit{vertex potentials},
and define $\Ecal^{W}(x) = \sum_{(a,b)\in G} w(a,b) (x[a] - x[b])^2$ to be the
\textit{energy} of $x$ in $G(W)$, with $w$ being the edge weight function specified by $W$.

Let us now define a few useful notations that will simplify the presentation below.
Let
$(u_1,v_1),\ldots,(u_t,v_t)$ be the set of \textit{all} vertex pairs such that
each $(u_i,v_i)\in e$ for some $e\in H$.
Furthermore, let $\Ecal_i^{W}$ be the minimum energy for the effective resistance
optimization problem (\ref{eq:erenergy}) for $R^{G(W)}(u_i,v_i)$, i.e.
\begin{align*}
    \Ecal_i^{W} \defeq \min_{x\in\mathbb{R}^{n}: x[u_i]=1,x[v_i]=0}
    \Ecal^{W}(x).
\end{align*}
Then $\OPT_H$ satisfies the following
identity:
\begin{align}\label{eq:maximinenergy}
    \frac{1}{\OPT_H} =
    \max_{W\in \Wcal}
    \min_{i} \Ecal^{W}_i.
\end{align}
Our goal then is to find a certificate that certifies
$(\ref{eq:maximinenergy})\leq \polylog(n,m)$.
One attempt to do so is to look at the
optimal $W^*$ of the outer maximization problem,
as well as the corresponding vertex potentials achieving the energies
$\Ecal_i^{W^*}$'s, hoping to use the optimality conditions of $W^*$ to connect
the energies of the latter to the value of (\ref{eq:maximinenergy}).
This however turns out to be difficult as
the inner minimization problem is not strictly convex, and in particular may
admit multiple minimizers.

To get around this, we will instead look at the \textit{dual} of
(\ref{eq:maximinenergy}).

\paragraph{Dual of (\ref{eq:maximinenergy}).}
In order to take the dual of (\ref{eq:maximinenergy}),
a first technicality we have to resolve is
to make
the inner minimization problem over a convex domain.
We do so by perhaps the most straightforward way where we extend the inner
domain to all convex combinations of $\Ecal_i^{W}$'s,
which does not change the optimal value:
\begin{align}\label{eq:primal00}
    \frac{1}{\OPT_H} =
    \max_{W\in \Wcal}\min_{\beta \in \Delta^{t-1}}
    \sum_{i\in [t]} \beta[i]\cdot \Ecal_i^{W},
\end{align}
where $\Delta^{t-1}$ denotes the simplex of dimension $t-1$, containing
all convex combination coefficients of $t$ numbers.
This reformulation has the following nice properties:
\begin{claim}[\cref{prop:convex} and \cref{prop:concave}]
    The objective function $\sum_{i\in [t]} \beta[i]\cdot \Ecal_i^{W}$
    is convex in $\beta$ and concave in $W$.
\end{claim}

The claim above allows us to invoke (generalizations of) von Neumann's Minimax Theorem
(see e.g., Corollary 37.3.2 of \cite{Rockafellar70})
and deduce strong duality for (\ref{eq:primal00}). In particular, we have
\begin{align}\label{eq:dual00}
    \frac{1}{\OPT_H} = \min_{\beta \in \Delta^{t-1}}
    \max_{W\in \Wcal}
    \sum_{i\in [t]} \beta[i]\cdot \Ecal_i^{W}.
\end{align}
Now, letting $\beta^*$ to be the minimizer of the outer minimization problem,
we can rewrite the dual (\ref{eq:dual00}) as
\begin{align}\label{eq:dual11}
    \frac{1}{\OPT_H} =
    \max_{W\in \Wcal}
    \sum_{i\in [t]} \beta^*[i]\cdot \Ecal_i^{W}.
\end{align}
That is, we have ended up with a single maximization problem
over all valid weight assignments,
with the objective function being a fixed convex combination
of the optimal energies between the $u_i,v_i$'s.
Moreover, the objective function is friendly, as its partial derivatives w.r.t. $W$
have simple forms:
\begin{claim}[\cref{prop:dr}]
     We have for any valid weight assignment $W\in \Wcal$,
    any $i\in[t]$, and any multi-edge $(a,b)\in G$ in the multi-graph,
    \begin{align}
        \frac{\partial \Ecal^W_{i}}{\partial w(a,b)} = \kh{x_{i}^W[a] - x_{i}^W[b]}^2,
    \end{align}
    where $x_i^W$ is defined to be the energy-minimizing potential vector for $u_i,v_i$
    in $G(W)$
    (cf. (\ref{eq:erenergy})).
    That is, the derivative of $\Ecal_i^W$ w.r.t. $w(a,b)$ is exactly the
energy contribution by the multi-edge $(a,b)$.
\end{claim}

Finally, by exploiting the KKT optimality conditions (see e.g., Theorem 28.2 of \cite{Rockafellar70}) on (\ref{eq:dual11}),
we show that
whenever $\OPT_H \geq 1/\polylog(n,m)$,
there are potential vectors
$\setof{x_1,\ldots,x_k} \subseteq \setof{x_1^{W^*},\ldots,x_t^{W^*}}$
satisfying
\begin{align}\label{eq:cert}
    \frac{1}{k}\cdot
    \Ecal^H(x_1,\ldots,x_k)
    \leq \polylog(n,m).
\end{align}
Here, we define
\begin{align}\label{eq:collectiveenergyoverview}
    \Ecal^H(x_1,\ldots,x_k) \defeq
    \sum_{e\in H}
    \max_{(a,b)\in e}
    \kh{
    \sum_{i=1}^{k} (x_i[a]-x_i[b])^2 
    }
\end{align}
to be the \textit{collective energy} of $x_1,\ldots,x_k$ in $H$,
$W^*$ to be the maximizer of (\ref{eq:dual11}),
and $x_i^{W^*}$ to be the energy-minimizing potential vector for $u_i,v_i$ in $G(W^*)$ (cf. (\ref{eq:erenergy})).
These potential vectors $x_1,\ldots,x_k$ thus together serve as certificate
for $\OPT_H \geq 1/\polylog(n,m)$,
and the hyperedges containing $(u_i,v_i)$'s corresponding
to $x_1,\ldots,x_k$ are considered the ``bottleneck'' hyperedges.

We refer the reader to the \textit{Collective Energy Lemma} (\cref{lem:medium}) for a formal statement. What we discussed above is of course a vast simplification of
the proof of the collective energy lemma. We refer the reader to \cref{sec:duality}
for the full derivation, in particular
\cref{prop:optimality} and
\cref{cor:opt} which show
how KKT conditions allow us to bypass weight assignments
completely.

\subsubsection{Recovering Important Hyperedges by Vertex Sampling}\label{sec:vsintro}

Our next step is then to show that the certificate $x_1,\ldots,x_k$ could help us recover
the important hyperedges. This can be seen as the spectral analog
of the work by Khanna, Putterman, and Sudan \cite{KPS24c}, where they showed
how to recover important hyperedges to cut sparsification
by carefully exploiting the (small) ratio
$k$-cuts, which are special cases of our certificate (cf. \cref{rmk:kcut}).
Although our task is more challenging as the
certificates are
arbitrarily real-valued
as opposed to $0/1$-valued,
we will nonetheless use a
drastically simpler algorithm, namely vertex sampling,
that readily reduces the task to effective resistance sampling in
\textit{ordinary graphs}, and relegate much of the complication to the analysis.
We refer the reader to the \textit{Vertex Sampling Lemma} (\cref{lem:vse})
for the formal statement.

From now on, we restrict our attention to potential
vectors in $[0,1]^n$.
This is without loss of generality since
we only deal with energy-minimizing potential vectors $x$
with the constraints that $x[u] = 1$ and $x[v]=0$ for some vertex pair $u,v$.
As such, any vector can be made in $[0,1]^n$ by setting all entries $>1$ to $1$
and all entries $<0$ to $0$ without increasing its energy or violating the constraints.

Let us now introduce the notion of \textit{potential-spanning hyperedges}
that will facilitate
our presentation.
For a given potential vector $x\in [0,1]^{n}$,
we say a hyperedge $e$ \textit{spans} $x$
if there exist $u,v\in e$ with $x[u] = 1, x[v]=0$.
For the vectors $x_1,\ldots,x_k$ serving as certificate in (\ref{eq:cert}),
each $x_i$ is bound to be spanned by at least one hyperedge,
in particular the hyperedge that contains the vertex pair $u_j,v_j$ for which
$x_i$ is energy-minimizing.

\paragraph{Extending $\OPT_H$ to Subsets of Hyperedges.}
Observe that in our definition of $\OPT_H$ in (\ref{eq:minimaxer}),
the inner maximization is over multiedges within
the \textit{entire} set of hyperedges in $H$.
However, there is nothing special about the entire set of hyperedges -
one can in fact change the inner maximization domain to any subset of hyperedges $F$
and obtain a new minimax optimization problem whose optimal value we denote by
$\OPT_F$:
\begin{align}\label{eq:minimaxer2}
    \OPT_F \defeq \min_{W\in \Wcal} \max_{e\in F} \max_{(u,v)\in e}
    R^{G(W)} (u,v).
\end{align}
Similarly,
when $\OPT_F\geq 1/\polylog(n,m)$,
by taking the dual and invoke the KKT optimality conditions,
we get a certificate $x_1,\ldots,x_k$ such that
each $x_i$ is spanned by a hyperedge $f\in F$,
and moreover
\begin{align}
    \frac{1}{k}\cdot \Ecal^H(x_1,\ldots,x_k) \leq \polylog(n,m).
\end{align}
Once again, we refer the reader to \cref{lem:medium} for the formal statement
and to \cref{sec:duality} for its proof.

\paragraph{Thought Process for Identifying Important Hyperedges.}

Now consider the following thought process for identifying important hyperedges:

\begin{enumerate}
    \item Initially, let $F_0$ contain all hyperedges in $H$.
    \item While $\OPT_{F_0} \geq 1/\polylog(n,m)$:
    \begin{enumerate}
        \item Find certificate $x_1,\ldots,x_k$ with collective energy
        $\leq k \polylog(n,m)$
        such that each $x_i$ is spanned by some $f_i\in F$.
    \item Remove a (deterministically chosen) hyperedge in $f_1,\ldots,f_k$
    from $F_0$. \label{step:rem}
    \end{enumerate}
    \item Identify all hyperedges \textit{not} in $F_0$ as important.
\end{enumerate}

No matter how we choose the hyperedge at Line \ref{step:rem} to remove,
by the condition of the while loop,
for the final $F_0$,
there exists a single weight assignment $W^*$ such that
\textit{all} multiedges within hyperedges in $F_0$ have effective resistances
at most $1/\polylog(n,m)$ in $G(W^*)$.
Thus we do not consider hyperedges in $F_0$ as important.
It then remains to show a good way to choose a hyperedge $f_i$ at Line \ref{step:rem}
to remove from $F_0$
so that (i) we are able to recover $f_i$ algorithmically, and
(ii) the number of important hyperedges is bounded by $n\polylog(n,m)$.
These will both follow by analyzing our vertex sampling algorithm $\VS(H,\lambda)$ presented in \cref{alg:VS}, which we restate below.

\begin{remark}
Crucially,
the thought process itself, along with the certificate
$x_1,\ldots,x_k$, are only used in our \textit{analysis} of the vertex sampling
algorithm. Algorithmically speaking, $\VS(H)$ is the only action we take to
recover important hyperedges.
\end{remark}

\vsov*

It is clear that the total number of hyperedges recovered by this process
is at most $n\lambda \polylog(n,m)$ due the resistance sampling.
We also show in \cref{sec:pflemvse} that
given any certificate $x_1,\ldots,x_k$ for $\OPT_F \geq 1/\polylog(n,m)$,
there \textit{deterministically exists} 
a hyperedge $f_i\in F$ spanning $x_i$ such that
\begin{align}
    \pr{f_i\in \VS(H)} \geq 1 - 1/\poly(n,m).
\end{align}
Thus this $f_i$ will be the one we remove from $F_0$
in the above thought process that identifies important hyperedges.
We refer the reader to the \textit{Vertex Sampling Lemma} (\cref{lem:vse})
for a formal statement.

Our proof of \cref{lem:vse} is rather delicate and contains several ideas carefully
pieced together, thus we do not intend to cover it fully in this technical overview.
Rather, in an effort to provide intuition behind the proof,
we will explain how to prove a (distinctly) weaker version of the claim.
Then
we will discuss the challenges we face in extending this warm-up solution to the full generality, and sketch what more we have to do to overcome them.

\paragraph{Warm-Up: Proving the $0/1$-Case.}
Specifically, recall that in \cref{rmk:kcut}, we noted that the ratio $k$-cuts
in \cite{Qua23} are special cases where a collection of $k$ $0/1$-valued vectors
with disjoint supports
have small collective energy.
In light of this connection, for illustration purpose,
we consider proving the following $0/1$-version of the vertex sampling lemma
as a warm-up.

\begin{lemma}[$0/1$-Vertex Sampling Lemma]\label{lem:svse}
Suppose we are given
$0/1$-valued vectors $x_1,\ldots,x_k\in \setof{0,1}^{n}$
with disjoint supports,
such that
each $x_i$ is spanned by at least one hyperedge in $H$,
and
\begin{align}
\frac{1}{k} \cdot \Ecal^H(x_1,\ldots,x_k) \leq \polylog(n,m).   
\end{align}
Then there (deterministically) exists an $f\in H$ spanning at least one $x_i$ such that
    \begin{align*}
        \pr{f\in \VS(H, \polylog(n,m))}
        \geq 1 - 1/\poly(n,m).
    \end{align*}
\end{lemma}

Note that we are doing effective resistance sampling in each vertex sampled multi-graph
in $\VS$. Thus intuitively, in order to recover a hyperedge $f\in H$,
it only helps if (i) at least one multi-edge within $f$ survives the vertex sampling,
and (ii) the surviving multi-edge(s) within $f$
has high resistance so they are likely to be sampled.
If, during one round of vertex sampling, a certain multi-edge $(u,v)\in f$
survives and it spans some $x_i$, we then hope to resort to the energy minimization
view (cf. (\ref{eq:erenergy})) of effective resistance to show $(u,v)$ has large resistance.

Specifically, let $H'$ be the vertex-sampled hypergraph that contains $(u,v)\in f$,
and let $G' = \Phi(H')$ be its multi-graph.
Suppose some $x_i$ satisfies $x_i[u] = 1$ and $x_i[v] = 0$.
Recall the energy minimization view of effective resistance:
\begin{align*}
    1/R^{G'}(u,v) =
    & \min_{x: x[u]=1, x[v]=0} \Ecal^{G'}(x) \\ =
    & \min_{x: x[u]=1, x[v]=0} \sum_{(a,b)\in G'} (x[a] - x[b])^2.
\end{align*}
This in particular means that the energy $\Ecal^{G'}(x_i)$ of $x_i$ in $G'$
is an upper bound on $1/R^{G'}(u,v)$.
As a result, it suffices to show that $\Ecal^{G'}(x_i)$ is small,
which will imply $R^{G'}(u,v)$ is large.

An immediate takeaway of the above discussion is that
only multi-edges that span at least one $x_i$ are useful.
We call these \textit{spanning multi-edges}.
Our first step of the analysis is to characterize these multi-edges by \textit{stars},
whose structure is friendly to analysis under vertex sampling.

\paragraph{Characterizing Spanning Multi-edges by Stars.}
Consider a potential vector $x_i$ and a hyperedge $f$ spanning it.
Then $f$ has at least one vertex with potential $0$ and at least one vertex with
potential $1$ in $x_i$. If the majority of them have potential $1$ in $x_i$,
then for each vertex $u\in f$ with potential $0$ in $x_i$,
we create a star with $u$ being the center
and all vertices in $e$ with potential $1$ in $x_i$ as leaves.
Otherwise, conversely, for each vertex $u\in f$ with potential $1$ in $x_i$,
we create a star with $u$ being the center
and all vertices in $e$ with potential $0$ in $x_i$ as leaves.
This way each star has $\Omega(r)$ multi-edges each has a potential difference
$1$ between its endpoints in $x_i$.
Let $\Scal(x_i)$ be the set of all stars we create from $x_i$ for all hyperedges.
Clearly, $\Scal(x_i)$ captures \textit{all} multi-edge energy contributions of $x_i$.

For ease of our analysis, we want to restrict
to a subset of $x_i$'s each with a roughly similar number of stars.
By a standard geometric grouping trick,
there exists $\setof{y_1,\ldots,y_{\ell}}\subseteq \setof{x_1,\ldots,x_k}$
such that $\ell \geq k/\polylog(n,m)$ and each $\sizeof{\Scal(y_j)}$ is
within $\polylog(n,m)$ of some number $h$.
We now consider the hyperedge $f^*\in H$ that contains the most number of stars
across all $y_1,\ldots,y_{\ell}$.
Note that $\ell \geq k/\polylog(n,m)$ implies that
$\Ecal(y_1,\ldots,y_{\ell}) \leq \ell \polylog(n,m)$,
which in turn yields that at most $\ell \polylog(n,m)$ hyperedges can contain stars,
as each of them contributes at least energy $1$ to their collective energy
$\Ecal^H(y_1,\ldots,y_{\ell})$ (cf. (\ref{eq:collectiveenergyoverview})).
Thus by an averaging argument, there exists one hyperedge $f^*$ that
contains at least $h/\polylog(n,m)$ stars across all $y_1,\ldots,y_{\ell}$.
Since $y_1,\ldots,y_{\ell}$ have disjoint supports,
the stars in $f^*$ must have at least $h/\polylog(n,m)$ different centers.
We next show that $f^*$ can be recovered by $\VS(H, \polylog(n,m))$ with high probability.

Within one round of vertex sampling at rate $1/r$,
the probability that one (out of $h/\polylog(n,m)$)
star center in $f^*$ survives is $\Omega(h/(r\polylog(n,m)))$.
Conditioned on a specific center being sampled, one of its star multi-edges
gets sampled with constant probability, as there are $\Omega(r)$ of them.
On the other hand, conditioned on a specific star center being sampled,
the total energy of the $y_j$ corresponding to the star center becomes
$\leq 1/r$ times smaller in expectation - this is because multi-edges incident on the center
survive with probability $1/r$, whereas other multi-edges survive with probability
$1/r^2 \leq 1/r$.
Since the $\leq h \polylog(n,m)$ stars in $\Scal(y_j)$ capture all energy of $y_j$,
we expect the total energy $\Ecal^{G'}(y_j)$ to be $\leq h \polylog(n,m)$.
Thus,
we expect the survival star multi-edge in $f^*$ to have effective resistance
at least $\Omega(1/(h\polylog(n,m)))$.
Together, in one round of vertex sampling, $f^*$ gets recovered with probability
at least
\begin{align*}
    \Omega(h/(r\polylog(n,m))) \cdot
    \Omega(1/(h\polylog(n,m)))
    = \Omega(1/(r\polylog(n,m))).
\end{align*}
As a result, $f^*$ gets recovered with high probability in
$r\polylog(n,m)$ rounds of vertex sampling.

\paragraph{Proof Sketch of the General Vertex Sampling Lemma.}
In the proof of the $0/1$-vertex sampling lemma above,
we are exploiting the fact all multi-edges with non-zero energy
contribution to the energy of the multi-graph can be useful in
recovering their parent hyperedges. This is however not true
when the vertex potentials have arbitrary real values.
In particular, multi-edges of small energy contribution
are not helpful in our recovery since the corresponding vertex potential vector
does not serve as a certificate of their resistances being high.
Yet, if their number is large, their total energy contribution may still
be large enough to overshadow the interesting multi-edges that do have large
energy contribution.

There is however a remedy for this, namely \textit{rounding} of the vertex
potential vector. For instance, if the multi-edges contributing small
energy have potentials between $[0,\eps]$ on their endpoints
for some small $\eps > 0$,
then we can \textit{round} the potential vector by setting
all potentials $\leq \eps$ to $\eps$.
This can only decrease the total energy, while also preserving
the multi-edges contributing large energy (e.g., $\Omega(1)$).

Making the rounding work in general turns out to require a more delicate analysis,
due the fact that we do not have this clear separation of the vertex potentials
in general. Our recipe for the general case is to build a
\textit{hierarchy} of star collections.
Specifically, we start from the star multi-edges we want to recover,
which we call the top-level stars.
If there is a way to round vertex potentials to make these multi-edges
stand out in terms of their energy contribution, then we are in good shape;
otherwise, we show that out of the multi-edges overshadowing the top-level
stars, we are able to create another level of stars of geometrically more
total energy, thus effectively expanding our star hierarchy by a new level.
We show that by choosing the potential rounding scheme carefully,
we can make sure (i) all stars we add to our hierarchy have large enough
energy contribution to make themselves useful in our recovery algorithm,
and (ii) the hierarchy can have at most $\polylog(n,m)$ levels.
Our recovery then proceeds in a bottom-up manner. Each level takes
$r\polylog(n,m)$ rounds of vertex sampling to recover.
So $\polylog(n,m)$ repetitions of these rounds can finally
recover the top-level star multi-edges we are interested in.

Once again, this is a vast simplification of the actual proof. We refer the reader to \cref{sec:pflemvse}
for the full proof detail.

\subsection{Organization}

\cref{sec:preli} establishes preliminary definitions and concepts that will be used throughout the paper. In \cref{sec:vsvsvs}, we present our main vertex sampling algorithm for spectral hypergraph sparsification and its analysis assuming the collective energy lemma and vertex sampling lemma,
which are then proved in \cref{sec:duality} and \cref{sec:pflemvse} respectively.
\cref{sec:linearSketching} shows how to implement our framework in the linear sketching model. \cref{sec:IterativeRecovery} develops our recursive recovery framework that forms the basis for both our online and dynamic algorithms.
In \cref{sec:9},
we generalize a result of \cite{ADKKP16} to recovering large effective resistance multi-edges in a multi-graph.
\cref{sec:OnlineSpectral} presents our algorithm for online spectral hypergraph sparsification and establishes its near-optimality via a matching lower bound in \cref{sec:ollb}. \cref{sec:FullyDynamic} extends our framework to the fully dynamic setting.

\section{Preliminaries}\label{sec:preli}

\subsection{Vertex Sampling}

In this paper we will make use of the notion of vertex-sampling.

\begin{definition}
    For a hypergraph $H = (V, E)$, a \textbf{vertex-sampling} at rate $p$ is the result of keeping every vertex (independently) with probability $p$. We denote the resulting vertex set by $V'$. The new hypergraph $H' = (V', E')$, where $E' = \{e \cap V': e \in E \}$.
\end{definition}

Note that if $e\intersect V' = \emptyset$, then the hyperedge disappears after
vertex-sampling. If $e \intersect V'\neq \emptyset$, we call $e \intersect V'$ the \textit{projected
hyperedge} of $e$.

We also make use of the corresponding multi-graph for any hypergraph.

\begin{definition}
    For a hypergraph $H = (V, E)$, the corresponding multi-graph for $H$, denoted by $\Phi(H)$ has the same vertex set $V$, and edge set $\bigcup_{e \in E} K_e$, where $K_e$ is the clique on vertices in $e$. 
\end{definition}

Note that if a hypergraph has arity bounded by $r$, then the number of multi-edges in the corresponding hypergraph is at most $|E| \cdot \binom{r}{2}$.

\subsection{Linear Sketches}

As an application of our results, we create \emph{linear sketches} for hypergraph spectral sparsification. We define linear sketches below:

\begin{definition}\label{def:graph-sketch}
    A \emph{linear sketch} of a hypergraph $G$ is given by a (randomized) matrix $S$ of dimension $s \times {2^n}$, chosen independently of the hypergraph. We associate the hypergraph $G$ with its indicator vector
    $\mathbf{1}_G \in\{0,1\}^{2^n}$, and then the \emph{sketch} of the graph is given by $M \cdot \mathbf{1}_G$. The space complexity of the sketch is given by the number of bits required to represent $M \cdot \mathbf{1}_G$.
\end{definition}

Note that linear sketches satisfy many convenient properties. For instance, when we wish to add or remove a hyperedge $e$ from $G$, we can simply add or delete the linear sketch of this hyperedge. That is,
\[
M \cdot \mathbf{1}_{G\pm e} = M \cdot \mathbf{1}_{G} \pm M \cdot \mathbf{1}_e.
\]

\subsection{Dynamic Algorithms}

In this paper we will be studying \emph{dynamic} algorithms for hypergraph problems. In general, for a graph (or hypergraph), a sparsifier may require $\Omega(n)$ (hyper)edges. Thus, it is infeasible to return the entire description of a hypergraph after each update to the underlying graph (i.e., each insertion or deletion of a (hyper)edge). Instead, in a \emph{dynamic} sparsifier, upon any update to the underlying graph, the algorithm responds with the necessary updates to the corresponding sparsifier. We make this more formal below:

\begin{definition}\label{def:dynamicHypergraphSparsifier}
    An algorithm $A$ is a $(1 \pm \eps)$ dynamic hypergraph sparsification algorithm if:
    \begin{enumerate}
        \item At every timestep $t$, $A$ receives a new update to the underlying hypergraph $H_{t-1}$. The update is either the insertion of a new hyperedge, or deletion of an existing hyperedge. We denote the new resulting hypergraph by $H_t$.
        \item At every timestep $t$, $A$ outputs a sequence of changes to update the previous sparsifier $\widetilde{H}_{t-1}$ to $\widetilde{H}_{t}$. This sequence of changes is either re-weighting of hyperedges, the addition of new hyperedges to the sparsifier, or the removal of hyperedges from the sparsifier. With probability $\geq 1 - 1 / \poly(n)$, $\widetilde{H}_{t}$ must be a $(1 \pm \eps)$ sparsifier to $H_t$.
    \end{enumerate}
\end{definition}

There are two key quantities that we will be interested in with respect to a dynamic hypergraph sparsification algorithm. First, we want to understand the \emph{sparsity} of the resulting hypergraph. It is trivial to see that if we do not care about the sparsity, then we can simply keep / delete every hyperedge as it is inserted / deleted. Second, we care about the \emph{update time} after each new insertion / deletion. If we allow the update time to be too large, we can simply re-calculate the entire sparsifier after each new insertion / deletion, therefore getting optimal sparsity, but going against the spirit of a dynamic algorithm. So, we add the following quantifiers to our description of a dynamic algorithm:

\begin{definition}
    We say that an algorithm $A$ is a $(1 \pm \eps)$ dynamic hypergraph sparsification algorithm of sparsity $s$ and update time $f$ if the algorithm satisfies the conditions of \cref{def:dynamicHypergraphSparsifier}, while also maintaining $\leq s$ hyperedges in each $\widetilde{H}_t$ and each update requiring amortized, expected time $\leq f$.
\end{definition}

We will use the following result from the dynamic \emph{graph} algorithms literature. 

\begin{theorem}\cite{ADKKP16}
For every $k \geq 2$ there is a decremental algorithm for maintaining a $2k-1$ spanner $H$ of expected size $O(k^2 n^{1 + 1 / k} \log(n))$ for an undirected graph $G$ on $n$ vertices with non-negative edge weights that has an expected total update time of $O(k^2 m \log(n))$. Additional $H$ has the following
property: Every time an edge is added to $H$, it stays in $H$ until it is deleted from $G$. 
\end{theorem}

\begin{remark}
Of particular note in this theorem is the final sentence. This \emph{monotonicity} (or what we call ``laziness'' later in the paper) property will be very useful for us as we design dynamic hypergraph sparsification algorithms, as it was in \cite{ADKKP16}.
\end{remark}

\section{Spectral Hypergraph Sparsification by Vertex Sampling}\label{sec:vsvsvs}

We give a meta algorithm for spectral hypergraph sparsification by vertex sampling which is for now stated without regard to any specific algorithmic implementation. We will discuss implementation details in various settings in the following sections.

\paragraph{Notation.}{We use $H$ to denote a hypergraph of arity $\in [r,2r]$ and
use $G = \Phi(H)$ to denote $H$'s underlying multi-graph where
we replace each hyperedge with a clique (of unit edge weights) on those vertices.
We write $W$ to denote a weight assignment
of the edges in $G$ where the edge weights of each hyperedge's clique sum up to exactly $1$,
and write $\Wcal$ to denote the collection of all such weight assignments.
We write $G(W)$ to denote the multi-graph weighted by $W$.}

For an ordinary multi-graph $G$, we write $R^G(u,v)$ to denote the effective resistance
between $u$ and $v$. We also write $R^G(e)$ to denote the effective resistance
between the endpoints of edge $e\in G$.

\subsection{Vertex-Sampling Algorithm}

We will use the following result from \cite{JambulapatiLS23} in a black-box manner.

\begin{theorem}[\cite{JambulapatiLS23}]\label{thm:JLS23}
    For any weight assignment $W\in \Wcal$ of the underlying multigraph
    $G$ of a hypergraph $H$, and $\eps\in(0,1)$,
    importance sampling each hypergraph $e\in H$ with probability
    \begin{align*}
        p_e \geq \max_{(u,v)\in e} R^{G(W)}(u,v) \polylog(n,m)/\eps^2
    \end{align*}
    gives a $(1+\eps)$-spectral sparsifier of $H$ with high probability
    in $n,m$.
\end{theorem}

Now consider the following vertex sampling algorithm.
Note that the output set $F$ contains at most
$O(n \lambda \polylog(n,m))$ hyperedges.

\quad

\vsov*
\quad

We will prove the following theorem, which essentially shows that
whenever a hyperedge $f$ cannot be recovered by vertex sampling
at oversampling rate $\theta^{-1}\polylog(n,m)$,
we must be able to sample $f$
with probability $\theta \polylog(n,m) / \eps^2$
while preserving the spectrum of the hypergraph.
Thus by setting $\theta$ to some $O((\eps/\log m)^2 / \polylog(n,m))$ and repeating
$O(\log m)$ times we get a $(1+\eps)$-spectral sparsifier
of $O(n\polylog(n,m)/\eps^2)$ hyperedges.

\keythmov*

To prove \cref{prop:goal}, we need to connect resistance to vertex potentials
and their (collective) energies.

\defspan*

\begin{definition}[Collective Energies]
    Given vertex potentials $x_1,\ldots,x_k\in\mathbb{R}^{n}$,
    their collective energy is
    \begin{align*}
        \Ecal^H(x_1,\ldots,x_k) \defeq
        \sum_{e\in H}
        \setof{\max_{(u,v)\in e} \left[
        \sum_{i=1}^{k} \kh{x_i[u] - x_i[v]}^2
        \right] }.
    \end{align*}
\end{definition}

\begin{definition}[Minimax Resistance Characterization]
    For a set of hyperedges $F$, define
    \begin{align*}
        \OPT_F \defeq
        \min_{W\in \Wcal} \max_{e\in F}
        \max_{(u,v)\in e} R^{G(W)}(u,v).
    \end{align*}
    That is, $\OPT_F$ characterizes
    the minimum max sampling rates of the hyperedges in $F$
    over all valid weight assignments.
\end{definition}

The following lemma will be proved in \cref{sec:duality}.

\lemmedium*

We prove the following lemma in \cref{sec:pflemvse}.

\lemvse*

Now, we show how to prove our main theorem using this lemma. Roughly, the intuition is that we start by considering $F = E$ (i.e., the entire hyperedge set). Now, either there is a hyperedge which is recovered with high probability from $F$, or every hyperedge in $F$ can afford to be sampled at probability $1/2$ while still preserving the spectrum of the hypergraph. In the second case, we are already done, and in the first case, we then simply define $F = F - \{e\}$, where $e$ is a hyperedge which is recovered with high probability. Then, we continue on inductively. Observe that the construction of this set $F$ is deterministic, and independent of the randomness used for vertex sampling. Thus, when we actually begin vertex sampling, we must only take a union bound over the $\leq m$ hyperedges that are explicitly recovered.

\begin{proof}[Proof of \cref{prop:goal}]
    \label{proof:thm1}{}
    For any given $\theta\in (0,1)$,
    we prove that there exists
    a hyperedge set $F_0$ such that
    $\OPT_{F_0} \leq \theta$
    and for each hyperedge $e\notin F_0$ we have
    \begin{align*}
            \pr{e\in \VS(H, \theta^{-1}\polylog(n,m))}
            \geq 1 - 1/\poly(n,m).
        \end{align*}
    This will immediately imply the proposition.
    To see why such an $F_0$ exists, we consider
    the following process for (implicitly) constructing it.
    \begin{enumerate}
        \item Initially, let $F_0$ contain all hyperedges in $H$.
        \item While $\OPT_{F_0} \geq \theta$:
        \begin{enumerate}
            \item By composing \cref{lem:medium} and \cref{lem:vse},
            $\exists f\in F_0$ such that
            \begin{align*}
            \pr{f\in \VS(H, \theta^{-1}\polylog(n,m))}
            \geq 1 - 1/\poly(n,m).
        \end{align*}
        \item Remove such an $f$ from $F_0$.
        \end{enumerate}
        \item Return $F_0$.
    \end{enumerate}
    Then it is clear that the while loop terminates in $\leq m$ iterations,
    and the returned $F_0$ has our desired property, completing the proof.
\end{proof}

In the following sections, we first describe the algorithmic implementations / consequences of vertex-sampling. Then, we provide the proofs of \cref{lem:medium}, \cref{lem:vse}.

\subsection{Sparsification via Vertex-Sampling}

In this section, we show how to use \cref{alg:VS} to create $(1 \pm \eps)$ spectral sparsifiers for hypergraphs. We present the algorithm below:

\sparsifyov*

The key claim is the following:

\begin{claim}\label{clm:oneRoundSparsifier}
Let $i \in [\log(m)]$, and let $H_i$ be as constructed by an iteration of \cref{alg:sparsify}. Then, with probability $1 - 1 / \poly(m)$, $F_i \cup 2 \cdot H_{i+1}$ is a $(1 \pm \eps)$ hypergraph spectral sparsifier for $H_i$.
\end{claim}

\begin{proof}
    Given the hypergraph $H_i$, \cref{alg:sparsify} recovers a set of hyperedges of $H_i$ via running \cref{alg:VS}. Now, by \cref{prop:goal}, there exists a single weight assignment $W^*_i$ of the multi-graph of $H_i$, such that every hyperedge satisfying 
    \[
    \max_{(u,v) \in e} R^{\Phi(H_i)(W^*_i)} \geq \frac{\eps^2}{\polylog(n,m)}
    \]
    is recovered with probability $\geq 1- 1 / \poly(m)$. Now, every hyperedge satisfying the above condition is essentially sampled with probability $1$, while all remaining hyperedges are sampled at rate $1/2$. In particular, for each hyperedge $e$, this means that the sampling rates we use satisfy
    \[
    p_e \geq \frac{C \log(n) \log(m)}{ \eps^2} \cdot \max_{(u,v) \in e} R_{\mathrm{eff}}^{\Phi(H_i)(W^*_i)}(u,v).
    \]
    By \cite{JambulapatiLS23}, this then means that the above sampling scheme (whereby we sample edges with probability $p_e$, and give weight $1 / p_e$), yields a $(1 \pm \eps)$ spectral-sparsifier with probability $1 -1 /\poly(m)$. $2 \cdot H_{i+1}$ is exactly the hyperedges who survive the sampling at rate $1/2$, and $F_i$ is exactly the hyperedges that we keep with probability $1$ (and hence weight $1$). This then yields the desired claim. 
\end{proof}

Now, we can conclude that \cref{alg:sparsify} produces spectral sparsifiers:

\begin{lemma}
    With probability $1 - 1 / \poly(m)$, the result of \cref{alg:sparsify} called on a hypergraph $H$ with parameter $\eps$ is a $(1 \pm \eps)$-spectral sparsifier of $H$. 
\end{lemma}

\begin{proof}
    Let us suppose by induction that $F_1 \cup 2 \cdot F_2 \cup \dots \cup 2^i \cdot F_i \cup 2^{i+1} \cdot H_{i+1}$ is a $(1 \pm 2i \eps)$ spectral sparsifier of $H$. Then, in the $i+1$st iteration of sparsification, we replace $H_{i+1}$ with $F_{i+1} \cup 2 \cdot H_{i+2}$. By composition of sparsifiers, if $F_{i+1} \cup 2 \cdot H_{i+2}$ is a $(1 \pm \eps)$ spectral sparsifier of $H_{i+1}$, then $F_1 \cup 2 \cdot F_2 \cup \dots \cup 2^{i+1} \cdot F_{i+1} \cup 2^{i+2} \cdot H_{i+2}$ is a $(1 \pm 2(i+1)\eps)$ spectral-sparsifier of $H$.

    By \cref{clm:oneRoundSparsifier}, and a union over all $\log(m)$ levels, this yields a $(1 \pm O(\eps \log(m)))$ sparsifier with probability $1 - 1 / \poly(m)$. Further, observe that after some $i = O(\log(m))$ iterations, every hyperedge in the hypergraph will be removed (i.e., sampled away) with probability $1 - 1 / \poly(m)$, and hence $H_i$ will be empty. Thus, all that remains is the $F_i$'s, and their union is the sparsifier.

    Finally, observe that we can use an error parameter $\eps' = \eps / \log(m)$ while only incurring a $\log^2(m)$ blow-up in the vertex-sampling parameter (which we absorb into the $\polylog(m)$). Thus, the result of \cref{alg:sparsify} on $H$ will be a $(1 \pm \eps)$ spectral sparsifier with probability $1 - 1 / \poly(m)$, as we desire.
\end{proof}

\subsection{Nearly-Linear Time Implementation}\label{sec:nltime}

\begin{restatable}[Nearly-Linear Time Implementation]{corollary}{nltime}
    \cref{alg:sparsifyintro} can be implemented on a static hypergraph $H$ with $\leq m$ hyperedges in time $\Otil((\sum_{e\in H} |e|) \polylog(m)/\eps^2)$ to get
    a $(1\pm\eps)$-spectral sparsifier of $H$ with $\Otil(n \polylog(m)/\eps^2)$ hyperedges,
    where $|e|$ denotes the arity of hyperedge $e$.
    The algorithm succeeds with probability $1-1/\poly(m)$.
\end{restatable}

\begin{proof}
Note that we can always group hyperedges by their arity into
$O(\log n)$ geometric groups such that hyperedges in the same group have arity within
a factor of $2$.
Thus, we focus on implementing
\cref{alg:sparsifyintro}
in $\Otil((mr + n)\polylog(m)/\eps^2)$ time,
where the input hypergraph only has arity $[r,2r]$.
Our implementation is fairly simple, using only linked list
data structures and effective resistance sampling in \cite{SS11}.

Note that all \cref{alg:sparsifyintro} does
is, for $O(\log m)$ adaptive iterations,
sampling/deleting hyperedges and repeatedly calling \cref{alg:VS}.
Thus it suffices to show that \cref{alg:VS} can be implemented
in $\Otil((mr+n)\polylog(m) /\eps^2)$ time.
We describe an implementation of \cref{alg:VS} as follows.

We first simulate the $N:= r\polylog(n,m)$ rounds of vertex-sampling upfront
as they are oblivious to other algorithmic actions.
Our goal is to obtain a linked list $\gamma_u$ for each vertex $u$ that
denotes the rounds in which $u$ gets sampled.
To do so,
for each vertex $u$, we first sample from Binomial distribution a number $K_u$ that corresponds to the number of successes when a coin with success prob. $p = 1/r$  is tossed $N$ times. We then repeatedly sample a random number between $1$ and $N$ until we have collected $K(u)$ distinct numbers. This corresponds to the rounds in which $u$ is selected, and we store them as a linked list $\gamma_u$.
This process takes $n\polylog(n,m)$ time in expectation, since $K_u \leq \polylog(n,m)$ with high probability in $m$.

We then do another preprocessing step that will be helpful for us to build each vertex-sampled
multi-graphs. Specifically, we go through each hyperedge one by one. For each hyperedge $e$,
we obtain linked lists $\ell_1^e,\ldots, \ell_{N}^e$ such that
$\ell_i^e$ contains all $e$'s endpoints that were sampled in the $i$-th round of vertex sampling.
This can be done by going through the endpoints of $e$ one by one and for each
endpoint $u$ appending $u$ to the linked lists $\ell_i^e$ for which $i\in \gamma_u$.
This preprocessing step takes running time $\Otil(m r \log(m))$ with high probability,
since with high probability in $m$ the arity of each hyperedge is bounded by $O(\log(m))$
in each round of vertex-sampling.

We then simulate the $r \polylog(n,m)$ adaptive rounds of effective resistance sampling. For each round $i$, we first create the vertex-sampled multigraph.
We do so by going through each remaining hyperedge $e$
that has not been deleted yet,
and creating a clique
supported on $e$'s vertices sampled in round $i$
which we have stored in linked list $\ell_i^e$. We also label each (multi-)edge
in the created clique by the identifier of the edge $e$.
Note that the creation of the vertex-sampled multigraph takes running time
$\Otil(m \log^2 m)$ since with high probability in $m$ each hyperedge's arity becomes
$O(\log(m))$ after vertex sampling.
Then we run
the nearly-linear time effective resistance sampling scheme in \cite{SS11}
to sample the multi-edges, which has a
running time of $\Otil(m \eps^{-2} \polylog(n,m))$ since with high probability in $m$
the vertex-sampled multi-graph has $\Otil(m \log(m)^2)$ edges,
and we over-sample the multi-edges by $\lambda = \polylog(n,m)$.
By looking at the labels
of the sampled multi-edges, we recover the corresponding
hyperedges as well. We then simply delete the recovered hyperedges from the graph and
proceed to the next round.

Our total running time adds up to $\Otil((mr+n) \polylog(m)/\eps^2)$ as desired.
\end{proof}

\section{From Minimax Resistance to Collective Energies} \label{sec:duality}

In this section, we will prove \cref{lem:medium}.

We consider a valid weight assignment
to be valid only if (i) the weights of the multi-edges within the clique of any single
hyperedge sums up to exactly $1$, and (ii) every multi-edge in the multi-graph has
weight at least\footnote{This is to make the multi-graph
always connected under any valid weight assignment, assuming the original hypergraph
is connected to start with.} $1/(8r^2)$. We denote the collection of valid weight assignments
by $\Wcal$.

Consider the following definition:
\begin{definition}[Minimax Resistance Characterization]
    For a set of hyperedges $F$, define
    \begin{align*}
        \OPT_F \defeq
        \min_{W\in \Wcal} \max_{e\in F}
        \max_{(u,v)\in e} R^{G(W)}(u,v).
    \end{align*}
    That is, $\OPT_F$ characterizes
    the minimum max sampling rates of the hyperedges in $F$
    over all valid weight assignments.
\end{definition}

Also consider the following collective energy lemma, which
we seek to prove in this section.

\begin{lemma}[Collective Energy Lemma]
    For any $\theta\in(0,1)$
    and hyperedge set $F$,
    if $\OPT_F \geq \theta$,
    then
    there exists sets of potentials $x_1,\ldots,x_k$
    with $k\leq\poly(n)$
    such that
    the following two statements both hold:
    \begin{enumerate}
        \item Each $x_i$ is spanned by at least one hyperedge in $F$.
        \item $\Ecal^H(x_1,\ldots,x_k) \leq k \polylog(n,m) / \theta$.
    \end{enumerate}
\end{lemma}

Let us fix a hyperedge set $F$ and a $\theta\in(0,1)$ such that
$\OPT_F \geq \theta$.
Let $P_F$ be the set of vertex pairs $(u,v)$'s such that $u,v$ both belong to a same
hyperedge in $F$.
Note that even if $(u,v)$ appears in more than one hyperedge in $F$, we only keep one copy
of it in $P_F$. As a result, $|P_F|\leq \binom{n}{2}$.
Then we can rewrite the minimax optimization problem as
\begin{align}
    \OPT_F = \min_{W\in \Wcal} \max_{(u,v)\in P_F} R^{G(W)}(u,v).
\end{align}
Our first step is to write the effective resistance in the optimization problem
as energy of vertex potentials.
To this end, let us set up some notation for vertex potentials and their energies.

Let $(u_1,v_1),\ldots,(u_{t},v_{t})$ where $t=|P_F|$ be the vertex pairs in $P_F$.

\paragraph{Vertex Potentials and Energies.}
Given a weight assignment $W$ of the underlying multigraph,
we write the energy of a vector $x\in\mathbb{R}^{n}$ in the multigraph
$G$ weighted by $W$ as
\begin{align}
    \Ecal^W(x) \defeq 
    \sum_{(a,b)\in G} w(a,b) (x[a] - x[b])^2.
\end{align}
It is known that the reciprocal of effective resistance of an edge $(u_i,v_i)$ is equal
to the minimum energy of any vertex potential spanned by $(u_i,v_i)$:
\begin{align*}
    \frac{1}{R^{G(W)}(u_i,v_i)}
    = \min_{x: x[u_i]=1, x[v_i]=0} \Ecal^W(x).
\end{align*}
We write $x_i^{W}$ to denote the optimal potentials for $R^{G(W)}(u_i,v_i)$,
and $\Ecal_{i}^{W}\defeq \Ecal^W(x_{i}^{W})$ to denote the corresponding minimum energy.
Then $\OPT_F$ can alternatively be written as
\begin{align}
    \frac{1}{\OPT_F} =
    & \max_{W\in\Wcal} \min_{i\in [t]} \min_{x: x[u_i]=1, x[v_i]=0} \Ecal^W(x) \notag \\ =
    & \max_{W\in\Wcal} \min_{i\in[t]} \Ecal^W_i. \label{eq:primal0}
\end{align}

\paragraph{Dual of $1/\OPT_F$.}
We will now look at the dual of $1/\OPT_F$.
First we make the minimization problem have a convex feasible solution space
by equivalently writing $1/\OPT_F$ as
\begin{align}\label{eq:primal}
    \frac{1}{\OPT_F} = & \max_{W\in\Wcal} \min_{\beta\in\Delta^{t-1}}
    \sum_{i\in[t]} \beta[i] \cdot \Ecal_i^{W}.
\end{align}
where $\Delta^{t-1}$ denotes the simplex of dimension $t-1$
(which contains all vectors denoting a distribution over $[t]$).
Note that now both domains $\Wcal$ and $\Delta^{t-1}$ are convex and compact.
We then prove that the objective function $\ip{\beta}{\ve^W_F}$ is convex
in $\beta$ and concave in $W$.
\begin{proposition}\label{prop:convex}
    For any fixed $W\in \Wcal$, 
    $\sum_{i\in[t]} \beta[i] \cdot \Ecal_i^{W}$ is convex
    in $\beta$ on $\Delta^{t-1}$.
\end{proposition}
\begin{proof}
    For any fixed $W\in \Wcal$,
    $\sum_{i\in[t]} \beta[i] \cdot \Ecal_i^{W}$ is a linear function of
    $\beta$, and thus is also convex.
\end{proof}
\begin{proposition}\label{prop:concave}
    For any fixed $\beta\in\Delta^{t-1}$,
    $\sum_{i\in[t]} \beta[i] \cdot \Ecal_i^{W}$ is concave in $W$ on $\Wcal$.
\end{proposition}
\begin{proof}
    It suffices to show that each $\Ecal^{W}_i$ is concave on $\Wcal$.
    Note that we have
    \begin{align*}
        \Ecal^W_{i} =
        \min_{x: x[u_i]=1, x[v_i]=0}
        \sum_{(a,b)\in G} w(a,b) (x[a] - x[b])^2.
    \end{align*}
    It suffices to show $\Ecal^W_{i}$ is concave in $W$.
    To this end,
    consider any weight assignments $U, V, W\in\Wcal$ such that
    $W = \lambda U + (1 - \lambda) V$ for $\lambda\in[0,1]$.
    We have
    \begin{align*}
        \Ecal_{i}^{W} = &
        \Ecal^{W}(x^{W}_{i}) \\ =
        & \lambda \Ecal^{U}(x^{W}_{i}) +
        (1 - \lambda) \Ecal^{V}(x^{W}_{i})
        \qquad \text{($\Ecal^W(x)$ is linear in $W$ when $x$ fixed)} \\
        \geq & \lambda \Ecal^{U}(x^{U}_{i}) +
        (1 - \lambda) \Ecal^{V}(x^{V}_{i})
        \qquad \text{($x^{U}_{i}$, $x^{V}_{i}$ minimize
        corresponding energies)} \\
        = & \lambda \Ecal^{U}_{i} + (1 - \lambda) \Ecal^{V}_{i}.
    \end{align*}
    This implies that $\Ecal_{i}^{W}$ is concave in $W$, as desired.
\end{proof}
The above propositions allow us to invoke (generalizations of) von Neumann's Minimax Theorem (see Corollary 37.3.2 of \cite{Rockafellar70}) and deduce
strong duality for (\ref{eq:primal}):
\begin{proposition}[Strong Duality for $1/\OPT_F$]
    We have
\begin{align}\label{eq:dual}
    \frac{1}{\OPT_F} = & \max_{W\in\Wcal} \min_{\beta\in\Delta^{t-1}}
    \sum_{i\in[t]} \beta[i] \cdot \Ecal_i^{W} \notag\\ = &
    \min_{\beta\in\Delta^{t-1}} \max_{W\in\Wcal} 
    \sum_{i\in[t]} \beta[i] \cdot \Ecal_i^{W}.
\end{align}
\end{proposition}

If we define $\beta^*$ to be the minimizer of (\ref{eq:dual}):
\begin{align*}
    \beta^* = \argmin _{\beta\in\Delta^{t-1}}
    \max_{W\in\Wcal} 
    \sum_{i\in[t]} \beta[i] \cdot \Ecal_i^{W}.
\end{align*}
then we have
\begin{align}\label{eq:dual1}
    \frac{1}{\OPT_F} =
    \max_{W\in\Wcal} 
    \sum_{i\in[t]} \beta^*[i] \cdot \Ecal_i^{W}.
\end{align}
We will then analyze the optimality conditions of (\ref{eq:dual1}) and
connect it to vertex potentials and their collective energy.

\paragraph{Optimality Conditions.}

We now analyze the optimality conditions of (\ref{eq:dual1}).
To this end, we first show that the derivative of the energy with respect to $W$
has the following simple form.

\begin{proposition}[Energy's Derivative with respect to Weight Assignment]
\label{prop:dr}
    We have for any valid weight assignment $W\in \Wcal$,
    any $i\in[t]$, and any multi-edge $(a,b)\in G$ in the multi-graph,
    \begin{align}
        \frac{\partial \Ecal^W_{i}}{\partial w(a,b)} = \kh{x_{i}^W[a] - x_{i}^W[b]}^2.
    \end{align}
\end{proposition}
\begin{proof}
We resort to the fact that
\begin{align*}
    \Ecal_i^{W} = \frac{1}{R^{G(W)}(u_i,v_i)} = 
    \frac{1}{\chi_i^T L_W^{\dag} \chi_i}.
\end{align*}
Here, we define $\chi_i$ to be the incidence vector of $(u_i,v_i)$, where
we have $\chi_i[u_i] = 1, \chi_i[v_i] = -1$ and $\chi_i[u] = 0,\forall u\neq u_i,v_i$.
We also define $L_W$ to be the Laplacian matrix of the weighted multi-graph $G(W)$,
and $L_W^{\dag}$ to denote its pseudo-inverse. Note that since any valid weight assignment ensures
that the corresponding weighted multi-graph is connected, the pseudo-inverse here always
means taking the inverse in the maximal subspace orthogonal to the all-one vector, with no
exceptions.

Using the chain rule, we get
\begin{align*}
    \frac{\partial \Ecal^W_{i}}{\partial w(a,b)} =
    - \frac{1}{\kh{\chi_i^T L_w^{\dag} \chi_i}^2}\cdot
    \frac{\partial \chi_i^T L_w^{\dag} \chi_i}{\partial w(a,b)}.
\end{align*}
Computing the derivative
via rank-one update formula for the pseudo-inverse \cite{ShermanM50,Meyer73}, we get
\begin{align*}
    \frac{\partial \chi_i^T L_w^{\dag} \chi_i}{\partial w(a,b)} =
    & -\chi_i^T \frac{\partial L_w^{\dag}}{\partial w(a,b)} \chi_i \\
    = &
    - \chi_i^T L_W^{\dag} \kh{ \chi_{ab} \chi_{ab}^T} L_w^{\dag} \chi_i \\
    = &
    - \kh{ \chi_i^T L_w^{\dag} \chi_{ab} }^2,
\end{align*}
where $\chi_{ab}$ denotes the incidence vector of $(a,b)$.
Plugging this back into the chain rule we get
\begin{align*}
    \frac{\partial \Ecal^W_{i}}{\partial w(a,b)} =
    & \kh{ \frac{\chi_i^T L_w^{\dag} \chi_{ab}}{\chi_i^T L_w^{\dag} \chi_{i}} }^2 \\
    = &
    \kh{ \chi_{ab}^T \frac{L_w^{\dag} \chi_i}{\chi_i^T L_w^{\dag} \chi_{i}} }^2.
\end{align*}
If we view $z := \frac{L_w^{\dag} \chi_i}{\chi_i^T L_w^{\dag} \chi_{i}}$ as a potential vector,
then the above equals exactly
the squared difference between $z[a]$ and $z[b]$.
Note that by definition, $z[u_i] - z[v_i] = 1$.
Thus it suffices to show $z$ minimizes the energy, which is indeed true since
\begin{align*}
    \Ecal^W(z) = z^T L_W z =
    \frac{\chi_i^T L_w^{\dag} L_w L_w^{\dag} \chi_i}{\kh{\chi_i^T L_w^{\dag} \chi_{i}}^2}
    = 
    \frac{\chi_i^T L_w^{\dag} \chi_i}{\kh{\chi_i^T L_w^{\dag} \chi_{i}}^2} =
    \frac{1}{\chi_i^T L_w^{\dag} \chi_{i}} = \Ecal_i^{W}.
\end{align*}
This finishes the proof.
\end{proof}

We will then use KKT conditions 
(see Theorem 28.2 of \cite{Rockafellar70})
to reason about optimality.
We first show that the optimization problem (\ref{eq:dual1}) satisfy some regularity conditions.

\begin{proposition}
    The value of (\ref{eq:dual1}) is finite and there is feasible point in the relative 
    interior of $\Wcal$ (i.e. Slater's condition is met).
\end{proposition}
\begin{proof}
    The value of (\ref{eq:dual1}) is bounded since each
    hyperedge contributes at most $\binom{n}{2}$ to the total energy,
    leading to a $m \binom{n}{2}$ upper bound on the value of (\ref{eq:dual1}).
    Moreover, the weight assignment that assigns uniform weight $1/\binom{|e|}{2}$
    to the multiedges of a hyperedge $e$ is feasible and in the relative interior
    of $\Wcal$ (since all weights are strictly bigger than $1/(8r^2)$ by
    $|e|\in [r,2r]$).
\end{proof}

Now we can invoke KKT conditions and get the following.
Here we let $W^*$ be the optimal weight assignment minimizing (\ref{eq:dual1}):
\begin{align*}
    W^* =
    \argmax_{W\in\Wcal} 
    \sum_{i\in[t]} \beta^*[i] \cdot \Ecal_i^{W},
\end{align*}
and $w^*$ be the weight function corresponding to $W^*$.
Also Define $x_i^* \defeq x_i^{W^*}$ and
$\Ecal_i^* \defeq \Ecal_i^{W^*}$.
That is, we have
\begin{align}\label{eq:dual2}
    \frac{1}{\OPT_F} = 
    \sum_{i\in[t]} \beta^*[i] \cdot \Ecal_i^*.
\end{align}

\begin{proposition}[Optimality Conditions for (\ref{eq:dual1})]
\label{prop:optimality}
For any hyperedge $f\in H$ and
any $(a,b),(c,d) \in f$, we have
\begin{enumerate}
    \item If $w^*(a,b) > 1/(8r^2)$ and $w^*(c,d) > 1/(8r^2)$, then
    \begin{align}
        & \sum_{i\in [t]} \beta^*[i] \kh{x_{i}^{*}[a] - x_{i}^{*}[b]}^2
        =
        \sum_{i\in[t]} \beta^*[i] \kh{x_{i}^{*}[c] - x_{i}^{*}[d]}^2.
    \end{align}
    \item If $w^*(a,b) > 1/(8r^2)$ and $w^*(c,d) = 1/(8r^2)$, then
    \begin{align}
        & \sum_{i\in[t]} \beta^*[i] \kh{x_{i}^{*}[a] - x_{i}^{*}[b]}^2 \geq
        \sum_{i\in[t]} \beta^*[i] \kh{x_{i}^{*}[c] - x_{i}^{*}[d]}^2.
    \end{align}
\end{enumerate}
\end{proposition}

\begin{proof}
    We first explicitly write (\ref{eq:dual1}) as a constrained maximization problem:
    \begin{align}
        &\mathrm{maximize}\quad
        \sum_{i\in[t]} \beta^*[i] \cdot \Ecal_i^{W} \notag \\
        & \mathrm{subject\ to} \notag \\
        & \qquad \forall (a,b)\in G,\ w(a,b) \geq 1/(8r^2) \notag \\
        & \qquad \forall f\in H,\ \sum_{(a,b)\in f} w(a,b) = 1.
        \label{eq:constraint}
    \end{align}
    The corresponding Lagrangian function is
    \begin{align}
        L(W; \mu,\lambda)=
        \sum_{i\in[t]} \beta^*[i] \cdot \Ecal_i^{W} +
        \sum_{(c,d)\in G} \mu_{ab} \cdot \kh{ w(c,d) - \frac{1}{8r^2}} +
        \sum_{f\in H}\lambda_f\cdot \kh{\sum_{(c,d)\in f} w(c,d) - 1},
    \end{align}
    where we have that the optimal value to (\ref{eq:constraint}) is equal to
    \begin{align*}
        \max_{W} \min_{\mu \geq 0, \lambda} L(W; \mu,\lambda).
    \end{align*}
    By KKT conditions (see Theorem 28.2 of \cite{Rockafellar70}), there exist $\mu^*\geq 0$ and $\lambda^*$ such that:
    \begin{enumerate}
        \item (Stationarity) For every hyperedge $f\in H$ and $(a,b)\in f$,
        \begin{align*}
            \frac{\partial}{\partial w(a,b)} L(W^*;\mu^*,\lambda^*) = 0.
        \end{align*}
        \item (Complementary Slackness) For every hyperedge $f\in H$ and $(a,b)\in f$,
        \begin{align*}
            \mu^*_{ab} > 0\ \Rightarrow\ w^*(a,b) = 0.
        \end{align*}
    \end{enumerate}
    By \cref{prop:dr},
    for a multi-edge $(a,b)$ in a hyperedge $f\in H$,
    we have
    \begin{align*}
        \frac{\partial}{\partial w(a,b)} \sum_{i\in[t]} \beta^*[i] \cdot \Ecal_i^{W} =
        \sum_{i\in [t]} \beta^*[i] \kh{x_i^{W}[a] - x_i^W[b]}^2.
    \end{align*}
    As a result, we have
    \begin{align*}
        \frac{\partial}{\partial w(a,b)} L(W;\mu,\lambda) =
        \sum_{i\in [t]} \beta^*[i] \kh{x_i^{W}[a] - x_i^W[b]}^2\ 
        +\ \mu_{ab}\  +
        \lambda_f.
    \end{align*}
    By the stationarity condition above,
    this partial derivative is zero at $W^*,\mu^*,\lambda*$, implying that
    \begin{align*}
        \sum_{i\in [t]} \beta^*[i] \kh{x_i^{*}[a] - x_i^*[b]}^2 = -\mu^*_{ab} - \lambda^*_f.
    \end{align*}
    If $w^*(a,b) > 1/(8r^2)$, then by the complementary slackness condition above,
    we have $\mu^*_{ab} = 0$, and thus
    \begin{align*}
        \sum_{i\in [t]} \beta^*[i] \kh{x_i^{*}[a] - x_i^*[b]}^2 = - \lambda^*_f,
    \end{align*}
    This is the same value for all $(a,b)\in f$ with $w^*(a,b) > 1/(8r^2)$,
    proving the first claim of the proposition.
    Otherwise ($w^*(a,b) = 1/(8r^2)$), using the fact that
    $\mu^*_{ab} \geq 0$, we get
    \begin{align*}
        \sum_{i\in [t]} \beta^*[i] \kh{x_i^{*}[a] - x_i^*[b]}^2 = -\mu^*_{ab} - \lambda^*_f
        \leq - \lambda^*_f,
    \end{align*}
    proving the second claim of the proposition.
\end{proof}

\paragraph{Getting Rid of Weight Assignments.}
We then prove the following useful inequality that characterizes $\OPT_F$ without
resorting to any weight assignment at all.

\begin{corollary}[of \cref{prop:optimality}]
\label{cor:opt}
    The vectors $x_{i}^{*}, (u_i,v_i)\in P_F$ satisfy that
    \begin{align}
        \sum_{f\in H}
        \max_{(a,b)\in f}
        \kh{ \sum_{i=1}^{t} \beta^*[i] \kh{x_{i}^{*}[a] - x_{i}^{*}[b]}^2 }
        \leq & \frac{2}{\OPT_F}.
    \end{align}
\end{corollary}
\begin{proof}
    Consider the clique of a hyperedge $f$
    and
    the multiedges $(a,b)$ therein satisfying
    \begin{align*}
        \sum_{i=1}^{t} \beta^*[i] \kh{x_{i}^{*}[a] - x_{i}^{*}[b]}^2
        <
        \max_{(c,d)\in f}
        \kh{ \sum_{i=1}^{t} \beta^*[i] \kh{x_{i}^{*}[c] - x_{i}^{*}[d]}^2 }.
    \end{align*}
    By \cref{prop:optimality},
    each such multi-edge $(a,b)$ has weight $1/(8r^2)$ with respect to $W^*$,
    and thus these multi-edges have total weight at most $1/2$ with respect to $W^*$.
    As a result we have
    \begin{align}
        & \sum_{f\in H}
        \max_{(a,b)\in f}
        \kh{ \sum_{i=1}^{t} \beta^*[i] \kh{x_{i}^{*}[a] - x_{i}^{*}[b]}^2 } \notag \\
        \leq & 2 \sum_{f\in H}
        \sum_{(a,b)\in f}
        w^*(a,b)
        \kh{ \sum_{i\in[t]} \beta^*[i] \kh{x_{i}^{*}[a] - x_{i}^{*}[b]}^2 }
        \qquad \text{\rm (\cref{prop:optimality})}
        \notag \\
        = & 2 \sum_{i\in [t]} \beta^*[i] \sum_{f\in H}
        \sum_{(a,b)\in f}
        w^*(a,b)
        \kh{x_{i}^{*}[a] - x_{i}^{*}[b]}^2
        \qquad \text{\rm (reordering summation)} \notag \\
        = & 2 \sum_{i\in[t]} \beta^*[i] \Ecal^{*}_{i} \notag \\
        = & \frac{2}{\OPT_F},
    \end{align}
    as desired.
\end{proof}

\paragraph{Proof of Collective Energy Lemma.}

We now prove the collective energy lemma, which we recall below.

\begin{lemma}[Collective Energy Lemma]
    For any $\theta\in(0,1)$
    and hyperedge set $F$,
    if $\OPT_F \geq \theta$,
    then
    there exists sets of potentials $x_1,\ldots,x_k$
    with $k\leq\poly(n)$
    such that
    the following two statements both hold:
    \begin{enumerate}
        \item Each $x_i$ is spanned by at least one hyperedge in $F$.
        \item $\Ecal^H(x_1,\ldots,x_k) \leq k \polylog(n,m) / \theta$.
    \end{enumerate}
\end{lemma}

\begin{proof}
    It suffices to exhibit vectors $x_1,\ldots,x_k$ such that
    \begin{align}\label{eq:goal1}
        \frac{1}{k}\cdot \Ecal^H(x_1,\ldots,x_k) \leq \frac{12\log_2 n}{\OPT_F}
    \end{align}
    and each $x_i$ is spanned by at least one hyperedge in $F$.
    To this end, let us consider the set of vectors
    $x_1^*,\ldots,x_t^*$, which are optimal vertex potentials
    for each vertex pair $(u_i,v_i)\in P_F$ when the multi-edges are weighted by $W^*$
    (optimal weight to (\ref{eq:dual1})).
    Clearly, each $x^{*}_{i}$ is spanned by at least one hyperedge in $F$
    which contains $u_i,v_i$ both.
    
    By \cref{cor:opt}, we know
        \begin{align}\label{eq:betaenergy2}
        \sum_{f\in H}
        \max_{(a,b)\in f}
        \kh{ \sum_{i\in [t]} \beta^*[i] \kh{x_{i}^{*}[a] - x_{i}^{*}[b]}^2 }
        \leq & \frac{2}{\OPT_F}.
    \end{align}
    The difference between the LHS of the above equation
    and the definition of collection energy
    is the parameters $\beta^*[i]$'s.
    In particular, if $\beta^*$  corresponded to the uniform distribution (i.e. $\beta^*[i]=1/t$ for all $i$),
    then the LHS would be the collective energy of $x_{i}^{*}$'s divided by $t$,
    and we would be done.
    To deal with the general case, consider doing geometric grouping of indices $i\in [t]$
    based on their $\beta^*[i]$ values.
    Specifically, for $j=0,1,\ldots,\log_2 n^3 - 1$,
    the $j$-th group $R_j$ contains indices $i$ with
    \begin{align*}
        \beta^*[i] \in (2^{-j-1}, 2^{-j}],
    \end{align*}
    whereas the last group $R_{\log_2 n^3}$ contains indices $i$ with
    \begin{align*}
        \beta^*[i] \in (-\infty, n^{-3}].
    \end{align*}
    Then we pick a group $R_{j^*}$ whose total $\beta^*$ values
    is at least $1/\log_2 r^3$. Note that it must be the case that
    $j^* < \log_2 n^3$, since there are $\binom{n}{2}$ pairs $(u_i,v_i)$ in total
    and thus the last group can have total $\beta^*$ values at most  $1/n$.
    Now restricting the inner summation of (\ref{eq:betaenergy2}) to
    group $R_{j^*}$, we have
    \begin{align*}
        \frac{2}{\OPT_F} \geq
        &\sum_{f\in H}
        \max_{(a,b)\in f}
        \kh{ \sum_{i\in[t]} \beta^*[i] \kh{x_{i}^{*}[a] - x_{i}^{*}[b]}^2 } 
        \qquad \text{(by (\ref{eq:betaenergy2}))}
        \\
         \geq &
        \sum_{f\in H}
        \max_{(a,b)\in f}
        \kh{ \sum_{i\in R_{j^*}} \beta^*[i] \kh{x_{i}^{*}[a] - x_{i}^{*}[b]}^2 }
        \qquad \text{(restricting inner sum to $R_{j^*}$)}
        \\
        \geq &
        \sum_{f\in H}
        \max_{(a,b)\in f}
        \kh{ \sum_{i\in R_{j^*}} 2^{-j^*-1} \kh{x_{i}^{*}[a] - x_{i}^{*}[b]}^2 }
        \qquad \text{($\beta^*[i] > 2^{-j^*-1}, \forall i \in R_{i^*}$)}
        \\
        \geq & \frac{1}{\log_2 n^3} \cdot \frac{1}{2 \sizeof{R_{j^*}}}
        \sum_{f\in G}
        \max_{(a,b)\in f}
        \kh{ \sum_{i\in R_{j^*}} \kh{x_{i}^{*}[a] - x_{i}^{*}[b]}^2 }
        \qquad \text{($R_{j^*}$ has total $\beta^*$ value $\geq 1/\log_2 n^3$)}
        \\
        = & \frac{1}{2 \log_2 n^3} \cdot \frac{1}{\sizeof{R_{j^*}}}
        \cdot \Ecal^H\kh{\setof{x_{i}^{*}}_{i\in R_{j^*}}}.
    \end{align*}
    This implies
    \begin{align*}
        \frac{1}{\sizeof{R_{j^*}}}
        \cdot \Ecal_G\kh{\setof{x_{i}^{*}}_{i\in R_{j^*}}} \leq
        \frac{12 \log_2 n}{\OPT_F},
    \end{align*}
    recovering (\ref{eq:goal1}) as desired.
\end{proof}

\section{Proof of the Vertex Sampling Lemma}\label{sec:pflemvse}

In this section, we will provide a complete proof of \cref{lem:vse}. To start, let us recall some definitions:

\defspan*

\begin{definition}[Collective Energies]
    Given vertex potentials $x_1,\ldots,x_k\in\mathbb{R}^{n}$,
    their collective energy is
    \begin{align*}
        \Ecal^H(x_1,\ldots,x_k) \defeq
        \sum_{e\in H}
        \setof{\max_{(u,v)\in e} \left[
        \sum_{i=1}^{k} \kh{x_i[u] - x_i[v]}^2
        \right] }.
    \end{align*}
\end{definition}

\begin{definition}[Minimax Resistance Characterization]
    For a set of hyperedges $F$, define
    \begin{align*}
        \OPT_F \defeq
        \min_{W\in \Wcal} \max_{e\in F}
        \max_{(u,v)\in e} R^{G(W)}(u,v).
    \end{align*}
    That is, $\OPT_F$ characterizes
    the minimum max sampling rates of the hyperedges in $F$
    over all valid weight assignments.
\end{definition}

Now, we let $x_1, \dots x_k$ be real-valued vectors with entries in $[0,1]$. We will show that there exist $\setof{y_1,\ldots,y_\ell}\subseteq \setof{x_1,\ldots,x_k}$
with $\ell = \Omega(k)$ such that for all $j\in [\ell]$,
the energy of $y_j$ in the multi-graph $G$ is bounded by
\begin{align}\label{eq:multiyj}
    \Ecal^G(y_j) \leq O(r^2\theta^{-1}\polylog(n,m)).
\end{align}
Recall that for a vector $x$, we define its energy in a multi-graph $G$ by
\begin{align*}
    \Ecal^G(x) = \sum_{e=(u,v) \in G} (x[u] - x[v])^2.
\end{align*}
In what follows, whenever we say ``energy'', we refer to the energy in a multi-graph;
we will explicitly say ``collective energy'' to refer to the collective energy in a hypergraph.

Note that because $y_1,\ldots,y_{\ell}$ is a subset of
$x_1,\ldots,x_k$ and $\ell=\Omega(k)$, together these imply that
the collective energy of the former is bounded by
\begin{align}\label{eq:colyjs}
    \Ecal^H(y_1,\ldots,y_{\ell}) \leq \ell \polylog(n,m) / \theta,
\end{align}
with the $\polylog(n,m)$ factor being potentially larger than the $\polylog(n,m)$
promised by the condition of the lemma.

We will from now on restrict our attention to $y_1,\ldots,y_{\ell}$,
with a goal to prove that $\exists f\in F$ that spans
at least one $y_j$ for which we have
\begin{align*}
    \pr{f\in \VS(H,\theta^{-1}\polylog(n,m)}
    \geq 1 - 1/\poly(n,m).
\end{align*}
We start by supplying the proof for (\ref{eq:multiyj}).
\begin{lemma}\label{lem:multienergyr}
    There exist $\setof{y_1,\ldots,y_{\ell}} \subseteq \setof{x_1,\ldots,x_k}$
    with $\ell = \Omega(k)$ such that for all $j\in[\ell]$
    \begin{align*}
        \Ecal^G(y_j) \leq O(r^2\theta^{-1}\polylog(n,m)).
    \end{align*}
\end{lemma}
\begin{proof}
    Note that by definition, we have
    \begin{align*}
        \sum_{i=1}^{k} \Ecal^G(x_i) =
        & \sum_{i=1}^{k} \sum_{e\in H}\sum_{(u,v)\in e}
        (x_i[u] - x_i[v])^2 \\
        = 
        & \sum_{e\in H} \sum_{(u,v)\in e} \sum_{i=1}^{k}
        (x_i[u] - x_i[v])^2 \qquad \text{(reordering summations)} \\
        \leq & \sum_{e\in H} \binom{r}{2}
         \setof{ \max_{(u,v)\in e} \left[ \sum_{i=1}^{k}
        (x_i[u] - x_i[v])^2 \right] } \\
        = & \binom{r}{2} \Ecal^H(x_1,\ldots,x_k) \\
        \leq & \binom{r}{2} k \polylog(n,m) / \theta.
    \end{align*}
    Then the lemma follows by a standard application of Markov's inequality.
\end{proof}

\paragraph{Characterization of Potentials by Stars.}
{
Now, let us consider a single hyperedge $e$, as well as a potential vector $y_j$. We will be interested in characterizing the energy contributed in the multi-edges corresponding to $e$. To this end, we introduce the notion of a \emph{star}. We establish a single vertex $c \in e$ as the \emph{center} of the star, and all other vertices are \emph{leaf} vertices of the star. Clearly, if we choose every single vertex to be a center, then our set of stars will capture all of the energy contributed by the multi-edges of the multi-graph. Our goal will be to carefully construct a set of stars which still captures the majority of the energy, yet also allows us to argue that certain hyperedges will be recovred with high probability. 

Hence, we now construct a hierarchy of collections of stars for each potential vector $y_j$.
We want that for each star center,
within its incident multi-edges in $G$,
the ones that appear in our star collections
contribute overwhelmingly more energy compared to the ones that do not.

We always create stars with the following
structure - each star contains multi-edges within a single hyperedge $e$, with one endpoint
as its center, and \textit{all} other endpoints as its leaves.
We will also ensure that the center has potential $0$,
and at least $|e|/2$ leaves have potential $\Omega(1)$.
For each star,
we call the leaf with the $\ceil{|e|/2}$th largest potential
the \textit{median} leaf.
We always guarantee that each star's median leaf has potential $\Omega(1)$.
Additionally,
for a \textit{single} potential vector,
all stars we create will have the \textit{same} center (though the stars will be for different hyperedges).

More specifically, consider a potential vector $y_j\in [0,1]^n$ that is spanned by
some hyperedge $f\in F$.
If $f$ has more endpoints with potentials in $[0,1/2]$ than in $(1/2,1]$, then
we modify $y_j$ by letting $y_j[u] \gets 1 - y_j[u]$ for all vertices $u$.
Note that this does not change the collective energy of $y_1,\ldots,y_{\ell}$
or the energy $\Ecal^G(y_j)$ at all.
Now
let $(c^*,d^*)\in f$ be such that $y_j[c^*] = 0$ and $y_j[d^*] = 1$.
For a $\eta\in [0,1]$, we write
$y_j^{\eta}$ to denote
the rounded version of $y_j$ where we set
all potentials below $\eta$ to $\eta$.
We use the following process to construct a
hierarchy
$\Scal_{y_j} = \Scal_{y_j}^1 \union \Scal_{y_j}^2\union\ldots \Scal_{y_j}^{O(\log n)}$
of collections of stars.
Crucially, we will ensure that all stars we create throughout the hierarchy
will have $c^*$ as their center.
\begin{enumerate}
    \item \textbf{Initializing $\Scal_{y_j}$:}
    We create the first level of stars, denoted $\Scal_{y_j}^{1}$, which
    contains the single star star with $c^*$ as the center,
    and $f$'s other endpoints as leaves.
    \item \label{step:goodstar} \textbf{Checking if the top level is good:}
    Let $\Scal_{y_j}^{p}$ be the current top level of stars in our hierarchy (i.e.
    $p$ is the number of levels which have been created so far).
    Among all median leaves in $\Scal_{y_j}^p$,
    let $b$ be the one with smallest potential,
    and define
    \begin{align*}
        \eta_p \defeq \kh{1 - \frac{1}{\log^2 n}} y_j[b].
    \end{align*}
    We then check, for the rounded potential vector $y_j^{\eta_p}$, whether
    the following inequality is true:
    \begin{align}\label{eq:check}
        \sum_{u:(c^*,u)\in G\setminus \Scal_{y_j}} \kh{y_j^{\eta_p}[u] - y_j^{\eta_p}[c^*]}^2
        \leq 16^{p+1} \cdot r.
    \end{align}
    If (\ref{eq:check}) is true,
    then we terminate with our current hierarchy; otherwise we proceed.

    \item \textbf{Expanding our hierarchy:}
    In the case that (\ref{eq:check}) is false,
    we next expand our hierarchy by adding
    another level $\Scal^{p+1}_{y_j}$ of stars that is initialized to be empty.
    
    Define $\eta'_p \defeq (1 - 1/\log^2 n) \eta_p$.
    Let us restrict our attention to the multi-edges $(c^*,u)\in G\setminus \Scal_{y_j}$
    satisfying $y_j[u] > \eta_p$,
    which
    contribute non-zero energy to the violation of (\ref{eq:check}).
    We say one such multi-edge $(c^*,u)$ is \textit{expandable} if, for
    the hyperedge $e$ that contains $(c^*,u)$, the majority of its vertices
    are assigned potentials $\geq \eta'_p$ by $y_j$, and \textit{non-expandable} otherwise
    (i.e. majority of $e$'s vertices have potentials $<\eta'_p$ in $y_j$).
    If a hyperedge $e$ contains one multi-edge $(c^*,u)\in G\setminus \Scal_{y_j}$
    that is expandable, then we add to the next level of collection $S^{p+1}_{y_j}$
    the star with $c^*$ being center and all of $e$'s other vertices as leaves.
    Otherwise, we ignore $e$ and the non-expandable multi-edges therein, if any. 

    \item \textbf{Repeat:} Go back to Step \ref{step:goodstar}
    with $\Scal_{y_j}^{p+1}$ being the new top level.
\end{enumerate}
}

At first glance, it may be unclear why we round the values in the potential vector, as this kinda of rounding can only decrease the energy contained in any particular multi-edge. However, the key point is that when it comes to sampling multi-edges and ultimately recovering a hyperedge, we only care about the \emph{relative} amount of energy contained in a specific multi-edge. Creating the star hierarchy as described above ensures that the energy of the stars contributes a large \emph{fraction} of the energy under the specific rounded potential vector that we create. Constructing the hierarchy is essential, as it allows us to define a new set of stars at each level, where each group of stars now uses a \emph{different} potential vector.

We now make some simply observations and remarks about the star hierarchy.

\begin{remark}\label{rmk:mlf}
    In each iteration,
    every star we add to $\Scal_{y_j}^{p+1}$
    has the same center $c^*$, and always has median leaf having
    potential at least $\eta'_p$ with respect to $y_j$.
\end{remark}

\begin{remark}
    We define expandability at \textit{multi-edge} level because this will make it easier
    to talk about their energy when we analyze vertex sampling later.
    However, expandability is in fact entirely determined by the \textit{hyperedge}
    that contains the multi-edge of interest - a hyperedge
    either only has expandable multi-edges or only has non-expandable multi-edges
    (or has neither),
    where ``neither'' happens when $c^*$ does not belong to the hyperedge,
    or all its vertices have potentials $\leq \eta_p$ in $y_j$.
\end{remark}

\begin{remark}
    Note that for each level $p$,
    the expandable and non-expandable multi-edges together \textit{entirely} capture
    the non-zero energy contribution to the energy of $y^{\eta_p}_j$ from multi-edges incident on $c^*$ but outside of
    $\Scal_{y_j}^{1}\union\ldots \Scal_{y_j}^{p}$.
\end{remark}

We show that the reason we can ignore non-expandable multi-edges is because
their number is small.
\begin{claim}\label{clm:fewbad}
    For the first $O(\log n)$ iterations of the above procedure,
    the number of non-expandable multi-edges is at most
    $O(r \theta^{-1}\polylog(n,m))$.
\end{claim}
\begin{proof}
    For each non-expandable multi-edge $(c,u)$ belonging to some hyperedge $e$,
    the multi-edges connecting $u$ to the $\Omega(r)$ (majority) vertices of $e$ with potentials
    $< \eta'$ in $y_j$ each has potential difference (in $y_j$) at least
    $$
    \eta - \eta' = \eta/\log^2 n \geq
    \kh{1 - \frac{1}{\log^2 n}}^{O(\log n)} / \log^2 n \geq
    \Omega(1/\log^2 n),
    $$
    resulting in a total energy contribution of $\Omega(r/\log^4 n)$
    to the energy of $y_j$.
    Moreover, if we were to sum up this contribution over all non-expandable
    multi-edges $(c,u)$,
    we would only count the contribution from each multi-edge at most twice,
    as it only has two endpoints.
    This coupled with the fact that $\Ecal^G(y_j) \leq O(r^2\theta^{-1}\polylog(n,m))$
    implies that the number of non-expandable multi-edges $(c,u)$ can be at most
    $O(r\theta^{-1}\polylog(n,m))$.
\end{proof}

We show that whenever we expand, we obtain geometrically more new stars in our collection.
\begin{claim}\label{clm:blowup}
    For the first $q = O(\log n)$ iterations with $q\geq 2$, the number
    of stars in $\Scal_{y_j}^{q}$ is at least
    \begin{align*}
        \sizeof{\Scal_{y_j}^{q}} \geq
        16^{q-1} - O(\theta^{-1} \polylog(n,m)).
    \end{align*}
\end{claim}
\begin{proof}
By the violation of (\ref{eq:check}), we know that, during the $q$-th iteration,
the incident multi-edges of $c$ not in $\Scal_{y_j}$ have energy
at least $16^{q}\cdot r$.
Thus the expandable multi-edges therein have energy 
$\geq 16^{q}\cdot r - O(r\theta^{-1}\polylog(n,m))$ by \cref{clm:fewbad}.
Finally, since each new star in $\Scal_{y_j}^{q}$ collects at most
$2r$ energy from the expandable multi-edges,
we have the desired lower bound on the number of stars.
\end{proof}

The above claim coupled with the energy upper bound of $y_j$ gives a total iteration
count of $O(\log n)$.
\begin{claim}\label{clm:itcount}
    Our process for constructing $\Scal_{y_j}$ terminates
    in $O(\log n)$ iterations.
\end{claim}
\begin{proof}
By \cref{rmk:mlf},
for $q=O(\log n)$,
each median leaf $b$ of our stars in $\Scal_{y_j}^{q+1}$ has potential in $y_j$ at least
\begin{align*}
\eta_{q}' \geq \kh{1 - \frac{1}{\log^2n}}^{O(\log n)} \geq \Omega(1).
\end{align*}
Thus by \cref{clm:blowup} and 
that $\Ecal^G(y_j) \leq O(r^2\theta^{-1}\polylog(n,m))$,
the process must terminate in $O(\log n)$ iterations before violating
the energy upper bound.
\end{proof}

\cref{clm:itcount} immediately implies the following:

\begin{claim}
    Each star in our star collection hierarchy has
    $\Omega(r)$ multi-edges each with $\Omega(1)$ energy in $y_j$.
\end{claim}

For each potential vector $y_j$, let $\levels{\Scal_{y_j}}$ be the number of levels
of the star collection hierarchy we created for $y_j$,
and let $\stars{\Scal_{y_j}}$ be the total number of stars
across all levels.

Consider the following process for recovering a hyperedge in $F$.
In the process, for the sake of analysis,
we will make changes to the set of potential vectors,
our star collection hierarchy,
and the hypergraph.
However, only Step \ref{step:vs} is what we actually do in our algorithm,
and the other steps are for analysis only.
\begin{enumerate}
        \item Let $\mu$ be the average of $\stars{\Scal_{y_j}}$ over
        the remaining potential vectors in $y_1,\ldots,y_{\ell}$. \label{stepmu}
        \item Discard all potential vectors $y_j$ with
        $\stars{\Scal_{y_j}} \geq \mu \log^2 n$ and all their stars. \label{step:discard}
        \item For each $y_j$:
        \begin{enumerate}
            \item While $p:=\levels{\Scal_{y_j}} \geq 2$
        and the current last level only has
        $|\Scal_{y_j}^p| \leq 16^p$ stars:
            \begin{enumerate}
                \item Delete the current last level of $\Scal^p_{y_j}$
                and all stars therein from our star collection hierarchy. 
                \item Repeat until the while condition is false.
            \end{enumerate}
        \end{enumerate}
        \item $F_1\gets \VS(H, \theta^{-1}\polylog(n,m))$ and delete hyperedges in $F_1$ from the hypergraph $H$. \label{step:vs}
        \item Delete the stars in hyperedges in $F_1$ from our star collection hierarchy.
        \item If $F_1$ contains a hyperedge also in $F$, then we terminate;
        otherwise, we go back to Step \ref{stepmu}.
\end{enumerate}

We first show that whenever a hyperedge contains sufficiently many stars
compared to the current upper bound we have on the total number of stars,
then we can recover it by vertex sampling.
We will next show that if no such hyperedges exist, then
we must be able to get a geometrically smaller upper bound
on the total number of stars, and can then make progress by analyzing the next level of stars. 

\begin{claim}\label{clm:vsr}
    Whenever we execute Step \ref{step:vs} in an iteration of the above process,
    for any hyperedge $e\in H$ containing
    $\Omega(\mu \theta/\polylog(n,m))$ stars
    of different centers, we have
    \begin{align*}
        \pr{e\in \VS(H, \theta^{-1}\polylog(n,m))} \geq
        1 - 1/\poly(n,m).
    \end{align*}
\end{claim}
\begin{proof}

This proof relies on several ingredients: First, we perform some simplifications to reduce our analysis to the case where only a single center $c_i \in e$ survives the vertex-sampling. Our goal is to show that when this center $c_i$ survives, then with relatively high probability, our effective resistance sampling will recover a multi-edge from $e$ which is incident on $c_i$. For this, we recall that the effective resistance of a multi-edge is equal to the maximum over all vertex-potentials of the fraction of energy contributed by this multi-edge. Naturally then, there are two quantities we want to bound: 1) an upper bound on the total energy of the entire vertex-sampled multi-graph, and 2) a lower bound on the energy contributed by the multi-edges corresponding to $e$. For the upper bound, we break the multi-edges into two parts, namely the multi-edges incident on $c_i$, and those not incident on $c_i$. For the lower bound, we simply argue that \emph{some} multi-edge in $e$ incident on $c_i$ will survive the vertex sampling. Now, we begin the rigorous proof:

Note that we can assume wlog that the $\mu\theta / \polylog(n,m)$ in the claim is
$O(r)$, since $e$ can only contain $\leq 2r$ stars of different centers (it only
has $\leq 2r$ vertices by our assumption on the arity of the hyperedge).
Let $N=\Omega(\mu \theta/\polylog(n,m))$ be the number of stars of $e$ of different centers.
Let $S_1,\ldots,S_{N}$ be stars of $e$ where
    each $S_i$ has a unique center $c_i\in e$.
    If $N > r/4$, we discard $N-r/4$ arbitrary stars and their respective centers
    from this collection, so that the total number of stars become exactly $r/4$
    (doing so only changes the number of stars by a constant factor),
    and update $N\gets r/4$.

Let $\EE_1$ denote the event that in one round of vertex sampling,
    at least one star center gets sampled.
    Since the stars have different centers,
    we have
    \begin{align*}
        \pr{\EE_1} =
        & 1 - \kh{1 - \frac{1}{r}}^{N} \\
        \geq & 1 - \exp\setof{-N/r} \\
        \geq & \frac{1}{e}\cdot N/r
        \qquad \text{(as $N \leq r/4$)}
        \\
        \geq & \Omega(\mu\theta r^{-1} / \polylog(n,m)).
    \end{align*}
    For an $i\in [N]$,
    let $\EE_2^i$ be the event that in a single round of vertex sampling,
    the center $c_i$ gets sampled.
    but all other star centers $c_j, j\neq i$ do not get sampled.
    Note that events $\EE_2^i$'s are mutually disjoint for different indices $i$'s,
    and their probabilities are all equal by symmetry.
    Moreover, letting $N'$ be the number of star centers sampled in $c_1,\ldots,c_N$,
    we have
    \begin{align*}
        \pr{\Union_{i=1}^{N} \EE_2^i | \EE_1} = & \pr{N'=1 | N' > 0} \\
        = & \frac{N (1/r) (1-1/r)^{N-1}}{1 - (1-1/r)^N}
        \qquad \text{(binomial distribution)}
        \\
        \geq & \frac{N(1/r) (1-(N-1)/r)}{1 - (1 - N/r)} \qquad
        \text{(first Bernoulli ineq. in
        \href{https://www.lkozma.net/inequalities_cheat_sheet/ineq.pdf}{here})} \\
        \geq & \frac{N/(4r)}{N/r} \qquad \text{($N\leq r/4$)} \\
        = & 1/4.
    \end{align*}
    As a result,
    \begin{align*}
        \pr{\Union_{i=1}^{N} \EE_2^i} =
        \pr{\Union_{i=1}^{N} \EE_2^i | \EE_1} \cdot \pr{\EE_1}
        = \Omega(\mu\theta r^{-1} / \polylog(n,m)).
    \end{align*}

    We divide the rest of the proof into
    the $\mu \geq \theta^{-1}$ case
    and the $\mu < \theta^{-1}$ case.\\[5pt]    
    
\noindent \textbf{The $\mu \geq \theta^{-1}$ Case:}{
    Consider conditioning on $\EE_2^i$ for a single arbitrary fixed $i\in[N]$.
    Let $y_j$ be the potential vector from which the star $S_i$ is created,
    let $q$ be the level where the star is created,
    and let $\eta_q$ be the threshold we used to check if level $q$ is good.
    Thus $S_i \in \Scal_{y_j}^{q}$.
    We then look at the rounded potential vector $y_j^{\eta_q}$.
    We consider how much $y_j^{\eta_q}$'s energy becomes after vertex sampling
    conditioned on $\EE_2^i$.
    For multi-edges not incident on $c_i$, their survival probability is
    either $1/r^2$ or $0$, depending on
    whether they are incident on another star center, so by (\ref{eq:multiyj})
    their expected total energy after vertex sampling conditioned on $\EE_2^i$ is
    $$O(1/r^2) \cdot \Ecal^G(y_j^\eta)
    \leq O(1/r^2) \cdot \Ecal^G(y_j) \leq O(\theta^{-1} \polylog(n,m)). $$
    We then restrict our attention to multi-edges incident on $c_i$ only.
    Specifically, we consider the following two cases:
    \begin{enumerate}
        \item \textbf{$q$ is the last level of $\Scal_{y_j}$ that hasn't been
        deleted: }
        In this case, either $q$ is the last level of $\Scal_{y_j}$ we have
        ever created, or there was a level $q+1$ in $\Scal_{y_j}$ but it has already been
        deleted.
        In the former case, we know the multi-edges incident on $c_i$
        but outside of $\Scal_{y_j}$
        contribute energy at most $16^{q+1} r$, which was the reason we stopped expanding
        $\Scal_{y_j}$. In the latter case,
        we know the multi-edges incident on $c_i$
        but outside of $\Scal_{y_j}$ contributing non-zero energy
        are the non-expandable multi-edges,
        plus those that were in $\Scal_{y_j}^{q+1}$, which were deleted because
        it has $\leq 16^{q+1}$ stars.
        Therefore,
        in either case, we know that the total energy of $y_j^{\eta_q}$ contributed
        by multi-edges incident on $c_i$ is at most
        \begin{align}\label{eq:contri1}
        O(\mu r \log^2 n) + O(16^{q+1} \cdot r) +
        O(r \theta^{-1} \polylog(n,m)),
        \end{align}
        with the first term contributed by multi-edges
        still remaining in $\Scal_{y_j}$, the second term contributed by
        the (expandable) multi-edges outside of $\Scal_{y_j}$, and the third contributed by
        the non-expandable multi-edges outside of $\Scal_{y_j}$.
        
        Note that
        $$\mu \geq |\Scal_{y_j}|/\log^2 n \geq |\Scal_{y_j}^{q}|/\log^2 n \geq 16^{q} / \log^2n,$$
        where the first inequality follows from Step \ref{step:discard},
        and the last inequality follows since level $q$ is the last level that hasn't
        been deleted.
        Also, recall that we are in the $\mu \geq \theta^{-1}$ case.
        So the total energy of $y_j^{\eta_q}$ of multi-edges incident on $c_i$
        is less than or equal to
        $\mu r \polylog(n,m)$.

        \item \textbf{There is a level $q+1$ in $\Scal_{y_j}$ that hasn't been deleted:}
        In this case, all multi-edges
        incident on $c_i$
        contributing nonzero energy to the energy of $y_j^{\eta_q}$
        are either in $\Scal_{y_j}^{q+1}$ or non-expandable. Thus the total energy of multi-edges
        incident on $c_i$ can be bounded by
        \begin{align}\label{eq:contri2}
        O(\mu r \log^2 n) +
        O(r \theta^{-1} \polylog(n,m)).
        \end{align}
        Recall once again that we are in the $\mu \geq \theta^{-1}$ case.
        So the total energy of $y_j^{\eta_q}$
        of multi-edges incident on $c_i$ can similarly bounded
        by $\mu r \polylog(n,m)$.
        \end{enumerate}
    }
    Conditioned on $\EE_2^i$, each multi-edge incident on $c_i$ survives vertex sampling
    with probability either $1/r$ or $0$, depending
    on whether they are incident on other star centers.
    Thus the expected total energy of these multi-edges becomes
    at most $\mu \polylog(n,m)$ after vertex sampling
    conditioned on $\EE_2^i$.

    Overall, in the case when $\mu \geq \theta^{-1}$,
    we can bound the expected total energy of $y_j^{\eta}$ of all multi-edges
    by $\mu \polylog(n,m)$
    after vertex sampling conditioned on $\EE_2^i$.

    On the other hand, we also have that with $\Omega(1)$ probability
    conditioned on $\EE_2^i$,
    we sample one of $S_i$'s star leaves with potential in $y_j$
    bigger than that of its median leaf.
    This is because there are at least $r/2$ such leaves,
    and at least $r/2 - N \geq r/4$ of them are not star centers,
    each of which survives vertex sampling with probability $1/r$ even conditioned
    on $\EE_2^i$.
    As a result, the corresponding star multi-edge to the sampled leaf
    contributes $\Omega(1/\log^4 n)$ energy
    to $y_j^{\eta_q}$
    via the corresponding star multi-edge.
    Thus with oversampling rate $\lambda = \theta^{-1}\polylog(n,m)$,
    conditioned on $\EE_2^i$,
    we recover hyperedge $e$ with probability
    $\Omega(\mu^{-1}\theta^{-1}/\polylog(n,m))$.
    Since this is the case if conditioned on any single $\EE_2^i$
    for all indices $i\in [N]$,
    the overall probability we recover $e$
    in a single round of vertex sampling
    is at least
    \begin{align*}
        & \Omega(\mu^{-1}\theta^{-1}/\polylog(n,m))
        \cdot \pr{\Union_{i=1}^{N} \EE_2^i} \\ =
        & \Omega(\mu^{-1}\theta^{-1}/\polylog(n,m))\cdot
        \Omega(\mu\theta r^{-1} / \polylog(n,m)) \\ =
        & \Omega(r^{-1}/\polylog(n,m)).
    \end{align*}
    Thus in $r\polylog(n,m)$ rounds, we recover $e$ with probability
    $1 - 1/\poly(n,m)$, as desired.\\[5pt]

\noindent \textbf{The $\mu < \theta^{-1}$ Case:}
    Finally we discuss the case when $\mu < \theta^{-1}$. In this case,
    the lower bound on the number of stars $e$ has becomes $1$ since it is an integral number.
    With $\mu < \theta^{-1}$, the dominating part in the energy
    in (\ref{eq:contri1}) and (\ref{eq:contri2}) is now $r\theta^{-1}\polylog(n,m)$.
    Thus, the expected energy of $y_j^{\eta_q}$
    after vertex sampling is now $O(\theta^{-1}\polylog(n,m))$.
    Fixing an arbitrary star center (which is potentially the only star center) of $e$,
    we know that with high probability, we will sample it in at least one
    round of vertex sampling, conditioned on which happening, we also
    sample a star multi-edge of $\Omega(1/\log^4 n)$ energy
    in $y_j^{\eta_q}$
    with $\Omega(1)$ probability.
    This coupled with the $\theta^{-1} \polylog(n,m)$ oversampling rate
    implies $e$ will be recovered with high probability.
\end{proof}

As promised, we show that
we get a geometrically smaller upper bound on the number of stars
after each iteration.
\begin{claim}\label{clm:starred}
At the end of each iteration, the total number of remaining stars satisfies
with probability $1-1/\poly(n,m)$
\begin{align*}
    \sum_{\text{remaining $y_j$}} \sizeof{\Scal_{y_j}} \leq
    \mu \ell / \log n.
\end{align*}
\end{claim}
\begin{proof}
Define each hyperedge $e$'s contribution to the collective energy as
\begin{align*}
    \Ecal^H_e(\setof{y_j}) \defeq
    \max_{(u,v)\in e}
    \left[ \sum_{\text{remaining $y_j$}} \kh{y_j[u] - y_j[v]}^2 \right].
\end{align*}
Note that we always have
\begin{align}
\sum_{e\in H} \Ecal^H_e(\setof{y_j}) =
\Ecal^H(\setof{y_j}) \leq \ell \polylog(n,m) / \theta,
\label{eq:colyjss}
\end{align}
where the inequality follows from (\ref{eq:colyjs}).
    Let $f(n)$ be the $\polylog(n,m)$ factor in (\ref{eq:colyjss}) (which is also
    the factor in (\ref{eq:colyjs})) multiplied with
    another $\log n$.
    Consider all hyperedges $e$ whose total number of stars across all remaining
    potential vectors is at least
    \begin{align}\label{eq:starup}
        \frac{\mu\theta}{f(n)}\cdot \Ecal^H_e(\setof{y_j}).
    \end{align}
    Since $\Ecal^H_e(\setof{y_j})$ captures the maximum multi-edge energy contribution,
    this implies that $e$ much have at least $\Omega(1)\cdot \frac{\mu\theta}{f(n)}$
    stars of different centers.
    By \cref{clm:vsr}, all those hyperedges can be recovered by vertex sampling
    with probability $1-1/\poly(n,m)$.
    Thus, all remaining hyperedges must contain at most (\ref{eq:starup}) stars.
    Summing this upper bound over all remaining hyperedges and plugging
    in the upper bound of collective energy in (\ref{eq:colyjss}), we get
    an upper bound on the total number of remaining stars as
    $\mu\ell / \log n$ as desired.
\end{proof}

We finally finish the proof of \cref{lem:vse} with the following claim.

\begin{claim}
    With probability $1 - 1/\poly(n,m)$,
    the above process terminates in $O(\log n / \log\log n)$ iterations,
    with at least one hyperedge in $F$ recovered.
\end{claim}
\begin{proof}
    Since each time we only discard potential vectors whose number of stars
    is more than $O(\log^2 n)$ times of the average,
    we discard $O(1/\log^2 n)$ fraction of the vectors by Markov's inequality.
    Thus for the first $O(\log n)$ iterations,
    we have $\Omega(1)$ fraction of the vectors remaining.
    Thus by \cref{clm:starred}, the total number of stars drops from $\Omega(\mu\ell)$
    to $O(\mu\ell/\log n)$,
    reducing by a $O(1/\log n)$ factor after each one of the
    first $O(\log n)$ iterations.
    Thus we only have $O(\log n)$ iterations before
    $\mu$ drops to $\theta^{-1}$ or below.

    If $\mu$ does not drop below $\theta^{-1}$,
    it must be the case that we have already recovered some hyperedge in $F$
    and thus terminated.
    If $\mu$ does drop below $\theta^{-1}$,
    then by \cref{clm:vsr}, we can recover all hyperedges
    that contain at least one star,
    among which one must be in $F$ since the first level star collection
    of every $y_j$ contains a single star in a hyperedge in $F$.
\end{proof}

\section{Linear Sketching Hypergraph Spectral Sparsifiers}\label{sec:linearSketching}

In this section, we will use the following linear sketch:

\begin{theorem}[Multi-graph Effective Resistance Sampler]\cite{KLMMS14}\label{thm:multigraphERSampling}
    Given any parameter $\phi\in \mathbb{R}$, there is a linear sketch $\mathcal{S}$ such that for any multi-graph $G$ on $\ell$ vertices with $\leq u$ potential edge slots, and $\leq \kappa$ edges, for any edge $e = (u,v) \in G$, $\mathcal{S}(G)$ recovers $e$ (independently) with probability at least $\phi \cdot r_{\text{eff}, G}(u,v)$. Further, $\mathcal{S}$ requires only $\widetilde{O}(\ell \phi \log(u) \polylog(\kappa) )$ bits of space to store. 
\end{theorem}

The above theorem follows from the linear sketch of \cite{KLMMS14} (among others). Although not explicit in their theorem statement, they independently sample each edge with probability at least as large as the leverage score (equal to the effective resistance for unweighted graphs), (see page 11, under ``{Correctness}'' section for more details). We discuss this theorem more completely in the appendix, see \Cref{sec:linearSketchingAppendix}.

Now, we will show that by storing copies of the above sketch for the vertex-sampled multi-graphs, we can create a linear sketch that can be used to recover a hypergraph sparsifier. We present an outline of the algorithm below:

\begin{algorithm}[H]
    \caption{LinearSketchSpectralSparsificationConstruction$(H, \eps)$}\label{alg:sketchConstruction}
    Let $H_0 = H$. \\
    \For{$i = 0,1 \dots 10\log(m)$}{
    \For{$j \in [r \polylog(n,m)]$}{
    Vertex sample $H_1$ at rate $1/r$ to get $H'_{(i,j)}$, with $\Phi(H'_{(i,j)})$ being its multi-graph.\\ 
    Store $\mathcal{S}(\Phi(H'_{(i,j)}), R_{i,j})$ with parameter $\phi = C \log(n) \log^2(m) / \eps^2$, and new randomness $R_{i,j}$.
    }
    }
\end{algorithm}

\begin{algorithm}[H]
    \caption{LinearSketchSpectralSparsificationRecovery$(H, \eps)$}\label{alg:sketchRecovery}
    Let $H_0 = H$. \\
    \For{$i = 0,1 \dots 10 \log(m)$}{
    $F_i = \emptyset$.\\
    \For{$j \in [r \polylog(n,m)]$}{
    Open $\mathcal{S}(\Phi(H'_{(i,j)} - \cup_{\ell \leq i} F_{\ell}, R_{i,j}))$ to yield a set of multi-edges $\widetilde{F}_i^{(j)}$. \\
    Let $F_i^{(j)}$ be the set of corresponding hyperedges for the multi-edges in $\widetilde{F}_i^{(j)}$. \\
    $F_i \leftarrow F_i \cup \widetilde{F}_i^{(j)}$.
    }
    }
    \Return{$\cup_i 2^i \cdot F_i$.}
\end{algorithm}

\begin{claim}\label{clm:linearSketchAccuracy}
    \Cref{alg:sketchRecovery} yields a $(1 \pm \eps)$ spectral-sparsifier for the hypergraph $H$ with probability $1 - 1 / \poly(m)$.
\end{claim}

\begin{proof}
First, observe that $\mathcal{S}(\Phi(H'_{(i,j)} - \cup_{\ell \leq i} F_{\ell}, R_{i,j}))$ is efficiently computable given our linear sketch. Indeed, because we store $\mathcal{S}(\Phi(H'_{(i,j)}), R_{i,j})$, and we have already explicitly recovered $\cup_{\ell \leq i} F_{\ell}$, it follows that $\mathcal{S}(\Phi(H'_{(i,j)} - \cup_{\ell \leq i} F_{\ell}), R_{i,j}) =\mathcal{S}(\Phi(H'_{(i,j)}), R_{i,j}) - \mathcal{S}(\Phi(\cup_{\ell \leq i} F_{\ell}), R_{i,j}) $. Here, we use our knowledge of the vertex-sampled partitions to compute the corresponding multi-edges that should be removed in correspondence with $\cup_{\ell \leq i} F_{\ell}$.

Next, observe that because $R_{i,j}$ is a random string independent of all other $R_{a, b}$, $a \neq i, b \neq j$, this means that the set of recovered edges $\cup_{\ell \leq i} F_{\ell}$ is independent of $R_{i,j}$. In particular, when we calculate $\mathcal{S}(\Phi(H'_{(i,j)} - \cup_{\ell \leq i} F_{\ell}), R_{i,j}) =\mathcal{S}(\Phi(H'_{(i,j)}), R_{i,j}) - \mathcal{S}(\Phi(\cup_{\ell \leq i} F_{\ell}), R_{i,j})$, this does not change the failure probability of the sketch, as we are simply performing an update operation which is independent of the randomness it is using (a similar fact is used in \cite{AGM12}). 

We must also show that we are able to recover $F_i^{(j)}$, which is the set of corresponding hyperedges for the multi-edges in $\widetilde{F}_i^{(j)}$. This is essentially a triviality: the linear sketch operates over a universe of size $O(\binom{n}{r} \cdot r^2)$, where there is one index in the universe for each multi-edge slot. We can view this universe as first choosing a hyperedge in $\binom{n}{r}$ ways, and then choosing a constituent multi-edge (in $\binom{r}{2}$ ways). Thus, whenever the linear sketch recovers an index $\in [O(\binom{n}{r} \cdot r^2)]$, it is explicit what the parent hyperedge that should be recovered is.

Finally, to conclude the correctness, we must only observe that \cref{alg:sketchRecovery} is exactly implementing \cref{alg:sparsify}, as the sketch $\mathcal{S}$ performs the multi-edge effective resistance sampling, and the hyperedge recovery is exactly  
\end{proof}

\begin{claim}\label{clm:linearSketchSpace}
    \Cref{alg:sketchRecovery} can be implemented with a (randomized) linear sketch of size $\widetilde{O}(n r \polylog(m) / \eps^2)$.
\end{claim}

\begin{proof}
    The space required by the above sketch is that of storing $r \polylog(m,n)$ multi-graph effective resistance samplers, each on a vertex set of size $\widetilde{O}(\frac{n}{r})$ with $\leq O(n^r \cdot r^2)$ multi-edge slots, $\leq O(m r^2)$ edges, and $\phi = \polylog(m,n) / \eps^2$. This translates to a space complexity of $\widetilde{O}(\frac{n}{r} \cdot r \polylog(m,n)/\eps^2)$ bits per multi-graph sketch (using \cref{thm:multigraphERSampling}), and thus a total space requirement of $\widetilde{O}(n r \polylog(m) / \eps^2)$ across all the sketches, as we desire.
\end{proof}

With this, we can state our final theorem for this section:

\begin{theorem}\label{thm:linearSketchmain}
    For hypergraphs of arity $\leq r$, with $\leq m$ hyperedges and an accuracy parameter $\eps \in (0,1)$, there is a linear sketch using $\widetilde{O}(n r \polylog(m) / \eps^2)$ bits that can be used to recover a $(1 \pm \eps)$-spectral sparsifier with probability $1 - 1 / \poly(n)$.
\end{theorem}

\begin{proof}
    The linear sketch itself is given by \Cref{alg:sketchConstruction}. The accuracy of the recovery procedure is given by \Cref{clm:linearSketchAccuracy}, and the size bound on the linear sketch is given by \Cref{clm:linearSketchSpace}.
\end{proof}

We also discuss the application of this linear sketch to the dynamic streaming setting:

\begin{corollary}\label{cor:dynamicStreaming}
    For any $\eps \in (0,1)$, there is a randomized dynamic streaming algorithm in $\widetilde{O}(nr \polylog(m) / \eps^2)$ bits of space that, for any sequence of insertions / deletions of hyperedges in an $n$-vertex unweighted hypergraph $H$ with at most $m$ edges of arity bounded by $r$, allows recovery of a $(1 \pm \eps)$ \emph{spectral}-sparsifier of $H$ with probability $1 - 1 / \poly(n,m)$ at the end of the stream.
\end{corollary}

\begin{proof}
    We use the linear sketch from \Cref{thm:linearSketchmain}. Let us denote this sketch by $\mathcal{S}$ and denote its randomness by $R$. Observe that for a hypergraph $H$ and a hyperedge $e$, we have \[
    \mathcal{S}(H \pm e, R) = \mathcal{S}(H, R)  \pm \mathcal{S}(e, R), 
    \]
    by the definition of a linear sketch. Thus, given a dynamic stream of a hypergraph, we must only store the randomness $R$, in addition to the current linear sketch $\mathcal{S}(H, R)$, and add / subtract the sketch of hyperedges that are inserted or deleted.

    The only remaining subtlety is to argue that the randomness $R$ need not be too large (namely, that it requires $\widetilde{O}(nr \polylog(m) / \eps^2)$ bits). For this, we refer the reader to Claim 7.1 in \cite{KPS24c}, which provides a general non-uniform derandomization result for hypergraph linear sketches. This then yields the corollary. 
\end{proof}

\section{Recursive Recovery Framework}
\label{sec:IterativeRecovery}
\subsection{General Framework}

From the previous sections, we have established that \cref{alg:VS} recovers all hyperedges whose sampling rate would be $\geq \frac{\eps^2}{\polylog(n,m)}$ in accordance with a weight assignment and \cref{thm:JLS23}. In this section, we will show how we can implement \cref{alg:VS} in an \emph{recursive} manner which will be amenable to both online and fully-dynamic algorithms for sparsification. 

Specifically, we will use the following algorithm, which operates on a multi-graph $G$ on $n$ vertices, with $\leq m$ multi-edges, and accuracy parameter $\eps$.

\begin{algorithm}[H]
    \caption{RecursiveRecovery$(G, n, m, \eps)$}\label{alg:iterativeRecovery}
    Let $G_1 = G$. \\
    \For{$i \in [\log(m)]$}{
    Let $F_i$ be such that $\{e \in G_i: R_{\mathrm{eff}, G_i}(e) \geq \frac{(1/2)^2}{1000 \log(n)\log^2(m)} \} \subseteq F_i$. \\
    Let $G_{i+1}$ be the result of sampling $G_i - F_i$ at rate $1/2$.
    }
    \Return{$S = F_1 \cup F_2 \cup \dots F_{\log(m)}$}
\end{algorithm}

First, we argue that the above procedure yields $(1 \pm \eps)$ spectral sparsifiers:

\begin{claim}
    For every $i \in  [\log(m)]$, $F_1 \cup \dots \cup 2^i \cdot F_i \cup 2^{i+1} \cdot G_{i+1}$ is a $(1 \pm O(i \cdot 1 / 2\log(m)))$-spectral sparsifier for $G$, with probability $1 - \log(m) / n^{20}$.
\end{claim}

\begin{proof}
    Let us assume inductively that $F_1 \cup \dots \cup 2^i \cdot F_i \cup 2^{i+1} \cdot G_{i+1}$ is a $(1 \pm 2(i \cdot 1 / 2\log(m)))$-spectral sparsifier for $G$ with probability $1 - i / n^{20}$. Then, in the next iteration, observe that $F_{i+1} \cup 2 \cdot G_{i+2}$ is a $(1 \pm 1 / 2\log(m))$-spectral sparsifier for $G_{i+1}$ with probability $1 - 1 / n^{20}$, as we are only sampling each edge $e$ independently with marginal probability $\geq \frac{1000 R_e \log^2(m) \log(n)}{\eps^2}$, where $\eps = 1/2$. Composing sparsifiers, and then taking a union bound over all $\log(m)$ levels then yields our claim. 
\end{proof}

Using the above, we can obtain simple bounds on the effective resistance of each edge as we go through levels of sampling. 

\begin{claim}
    With probability $\geq 1 - \log(m) / n^{20}$, for every $i \in [\log(m)]$, and for every $u, v \in \binom{V}{2}$, it is the case that 
    \[
    R_{\mathrm{eff}, G_i}(u,v) \geq 2^i\cdot 1/2 \cdot R_{\mathrm{eff}, G}(u,v).
    \]
\end{claim}

\begin{proof}
    First, from the previous claim, we know that with probability $\geq 1 - \log(m) / n^{20}$, for every $i \in [\log(m)]$, it is the case that $F_1 \cup \dots \cup 2^{i-1} \cdot F_{i-1} \cup 2^{i} \cdot G_{i}$ is a $(1 \pm 1/2)$-spectral sparsifier for $G$. In particular, this means that 
    \[
    R_{\mathrm{eff}, F_1 \cup \dots \cup 2^{i-1} \cdot F_{i-1} \cup 2^{i} \cdot G_{i}}(u,v) \geq 1/2 \cdot R_{\mathrm{eff}, G}(u,v),
    \]
    and therefore, 
    \[
     R_{2^{i} \cdot G_{i}}(u,v) \geq 1/2 \cdot R_{\mathrm{eff}, G}(u,v).
    \]
    Finally, when we remove the weight from $G^{i}$, we get that 
    \[
    R_{G_{i}}(u,v) \geq 2^i 1/2 \cdot R_{\mathrm{eff}, G}(u,v),
    \]
    as we desire. 
\end{proof}

Now, going forward, we will adopt a different perspective for the sub-sampling of edges. Indeed, instead of viewing the sub-sampling operation as taking place \emph{after} certain edges are recovered, let us instead fix a set of ``filter'' functions beforehand, where for each edge, a filter function keeps the edge with probability $1/2$, and deletes it otherwise. If edges are recovered in a round $i$, then we simply do not apply the filter functions for future levels.

\begin{claim}\label{clm:ERSumRecovery}
    Let $Q \subseteq G$ denote a set of multi-edges. Then, if we let $S$ denote the result of running \cref{alg:iterativeRecovery} on $G$, we have that 
    \[
    \Pr[S \cap Q \neq \emptyset] \geq \min \left (2/3, \sum_{e \in Q} R_{\mathrm{eff}, G}(e) / \eps^2 \right ) - \frac{\log(m)}{n^{20}}. 
    \]
\end{claim}

\begin{proof}
    Let $W = \sum_{e \in Q} R_{\mathrm{eff}, G}(e)$. First, observe that there must be some level of sampling $i \in [\log(m)]$ such that edges $e \in Q$ with effective resistance in $[\frac{1}{2^{i}},\frac{1}{2^{i-1}}]$ contribute a $\geq 1 / \log(m)$ fraction of $W$. Denoting this set by $Q^{(i)}$, we specifically are saying that 
    \[
    \sum_{e \in Q^{(i)}} R_{\mathrm{eff}, G}(e) \geq \frac{W}{\log(m)}.
    \]
    We remark that this immediately implies that there are $\geq \frac{W \cdot 2^i}{2\log(m)}$ edges in $Q^{(i)}$.

    Next, let us consider the graph after sampling at rate 
    \[
    \frac{1}{2^j} = \frac{100 \log(n)\log(m)}{2^i (1/2)^2}.
    \]
    By the previous claim, with probability $1 - \log(m) / n^{20}$, for every $(u,v) \in Q$, it will be the case that 
    \[
    R_{\mathrm{eff}, G_i}(u,v) \geq \frac{2^i (1/2)^2}{100 \log(n)\log(m)} (1 - \eps) \cdot \frac{1}{2^i} \geq \frac{(1/2)^2}{100 \log(n) \log(m)}.
    \]

    In particular, conditioned on this event happening, as long as \emph{some} edge in $Q^{(i)}$ survives the first $j$ filter functions, it will be the case that an edge in $Q^{(i)}$ is present and has large enough effective resistance, and therefore is recovered (and hence $S \cap Q \neq \emptyset$).

    Thus, it remains only to compute the probability that an edge in $Q^{(i)}$ survives the first $j$ filter functions. For this observe that when combined, the first $j$ filter functions are sampling at rate $1 /2^j = \frac{100 \log(n)\log(m)}{2^i (1/2)^2}$. Further, the number of edges in $Q^{(i)}$ is $\geq \frac{2^i \cdot W}{2 \log(m)}$. Thus,
    \[
    \Pr[\text{edge in }Q^{(i)} \text{ survives filters}] \geq 1 - (1 - \frac{100 \log(n)\log(m)}{2^i (1/2)^2})^{\frac{2^i \cdot W}{\log(m) 2}}
    \]
    \[
    \geq 1 - e^{-\frac{50 W \log(n)}{(1/2)^2}}.
    \]
    Now, we observe that if $e^{-x} \geq 1/3$, then $e^{-x} \leq 1 - x/2$. This means that either $\Pr[\text{edge in }Q^{(i)} \text{ survives filters}] \geq 2/3$, or 
    \[
    \Pr[\text{edge in }Q^{(i)} \text{ survives filters}] \geq \frac{50 W \log(n)}{2(1/2)^2},
    \]
    when $\frac{50 W \log(n)}{(1/2)^2} \leq \ln(3)$. Thus, we can conclude that 
    \[
    \Pr[\text{edge in }Q^{(i)} \text{ survives filters}] \geq \min(2/3, \frac{50 W \log(n)}{2(1/2)^2}).
    \]
    In particular, this means that 
    \[
    \Pr[\text{edge in }Q^{(i)} \text{ survives filters}] \geq \min(2/3, W).
    \]

    Finally, we observe that the only failure case for our recovering an edge from $Q$ is when \emph{either} the sequence of graphs we construct fails to be a spectral sparsifier, or an edge in $Q^{(i)}$ fails to survive the $j$ filter functions. Thus, 
    \[
    \Pr[S \cap Q = \emptyset] \leq \Pr[\text{Sparsifier fail}] + \Pr[\text{Filter fail}] \leq (1 - \min(2/3, W)) + \frac{\log(m)}{n^{20}}.
    \]
    This yields our stated claim. 
\end{proof}

\subsection{Re-Implementing Vertex-Sampling Recovery}

In this subsection, we will show that a variant of the above recursive recovery procedure can be used to implement the vertex-sampling recovery procedure. Specifically, for a fixed hyperedge $f$, let us set $Q$ to be the set of multi-edges corresponding to the clique-expansion of $f$. Observe that if $Q$ is to be recovered in the $r \polylog(m,n) \leq r^2 \leq n^2$ rounds of vertex-sampling with high probability, then in certain rounds of vertex-sampling, the multi-edges in $Q$ must have a lot of effective resistance (i.e., the sum of their effective resistances must be large). This is because multi-edges are sampled directly proportional to their effective resistance.

Finally, once we have established that the multi-edges have a lot of effective resistance, then we can use the previous section to argue that a multi-edge in $Q$ is recovered with high probability. Below, we make this intuition more formal. 

First, we introduce some notation: for a hyperedge $f$, and a round of vertex-sampling $\ell$, we let $Q^{(\ell, f)}$ denote the multi-edges correpsonding to $f$ in the $\ell$th round. We let $W^{(\ell, f)}$ denote the combined effective resistances for multi-edges in $Q^{(\ell, f)}$ with respect to the $\ell$th vertex-sampled multi-graph. 

\begin{claim}\label{clm:largeERSums}
    Suppose that a hyperedge $f$ is recovered with probability $\geq 1/2$ in $K \leq n^2$ rounds of vertex sampling (as defined in \cref{alg:VS}). Then, \[
    \sum_{\ell \in [K]: W^{(\ell, f)} \geq 1 / n^3} W^{(\ell, f)} \geq \frac{\eps^2}{\polylog(m,n)}.
    \]
\end{claim}

\begin{proof}
    First, observe that if a hyperedge is recovered with probability $\geq 1/2$, then this must mean across all rounds $\ell \in [K]$, 
    \[
     \sum_{\ell \in [K]} W^{(\ell, f)} \geq \frac{\eps^2}{\polylog(m,n)}.
    \]
    Indeed, if we suppose for the sake of contradiction that 
    \[
    \sum_{\ell \in [K]} W^{(\ell, f)} < \frac{\eps^2}{\polylog(m,n)}, 
    \]
    then 
    \[
    \Pr[\text{sample an edge from } Q] \leq \sum_{\ell \in [K]} \sum_{u,v \in Q^{(\ell, f)}} r_{\mathrm{eff}}(u,v) \cdot \polylog(m,n) / \eps^2 
    \]
    \[
    \leq \sum_{\ell \in [K]} W^{(\ell, f)}  \cdot \polylog(m,n) / \eps^2 < 1/2.
    \]

    Next, observe that 
\[
    \sum_{\ell \in [K]: W^{(\ell, f)} \leq 1 / n^3} W^{(\ell, f)} \leq 1/n,
    \]
    as there are at most $n^2$ rounds total, and under the above assumption, each such round is contributing at most $1/ n$ to the total sampling mass. 

    Finally, putting these together gives that 
    \[
    \sum_{\ell \in [K]: W^{(\ell, f)} \geq 1 / n^3} W^{(\ell, f)} 
    \]
    \[
    \geq \sum_{\ell \in [K]} W^{(\ell, f)}  - \sum_{\ell \in [K]: W^{(\ell, f)} \leq 1 / n^3} W^{(\ell, f)} \geq \frac{\eps^2}{\polylog(m,n)}.
    \]
\end{proof}

With this, we now consider the following algorithm:

\begin{algorithm}[H]
    \caption{RepeatedRecursiveRecovery$(G, n, m, \eps)$}\label{alg:RepeatedIterativeRecovery}
    Initialize $\widetilde{F} \gets \emptyset$ and $H_1 \gets H$.\\
    \For{$r \polylog(n,m)/ \eps^2$ rounds} {
    Vertex sample $H_1$ at rate $1/r$ to get $H'$, with $G$ being its multi-graph.\\ 
    Let $G_1 = G$. \\
    \For{$i \in [\log(m)]$}{
    Let $F_i$ be such that $\{e \in G_i: R_{\mathrm{eff}, G_i}(e) \geq \frac{(1/2)^2}{1000 \log(n)\log^2(m)} \} \subseteq F_i$. \\
    Let $G_{i+1}$ be the result of sampling $G_i - F_i$ at rate $1/2$.
    }
    Let $S =  F_1 \cup F_2 \cup \dots F_{\log(m)}$.\\
    Let $\widetilde{S}$
    contain all hyperedges of $H_1$ who have at least one
    clique multi-edge in $S$. \\
    Let $\widetilde{F} \gets \widetilde{F} \cup \widetilde{S}$ and delete $\widetilde{S}$ from $H_1$.
    }
    \Return{$\widetilde{F}$}
\end{algorithm}

\begin{lemma}\label{lem:RIRsame}
    Let $f$ be any hyperedge which is recovered by \cref{alg:VS} with probability $\geq 1/2$ (for a fixed set of vertex-samplings). Then, $f$ is recovered by \cref{alg:RepeatedIterativeRecovery} (with the same fixed set of vertex-samplings) with probability $\geq 1 - 1 / \poly(n,m)$.
\end{lemma}

\begin{proof}
    First, by \cref{clm:largeERSums}, we know that this means \[
    \sum_{\ell \in [K]: W^{(\ell, f)} \geq 1 / n^3} W^{(\ell, f)} \geq \frac{\eps^2}{\polylog(m,n)}.
    \]

    Now, by \cref{clm:ERSumRecovery}, in each iteration where $W^{(\ell, f)} \geq 1 / n^3$, we know that a multi-edge from $Q^{(\ell, f)}$ is recovered with probability $\geq \min(2/3, W^{(\ell, f)})/2$. Thus, because 
    \[
    \sum_{\ell \in [K]: W^{(\ell, f)} \geq 1 / n^3} W^{(\ell, f)} \geq \frac{\eps^2}{\polylog(m,n)},
    \]
    we also know that across these same rounds of vertex-sampling,
    \[
    \Pr[\text{edge from }Q \text{ is recovered}] \geq \frac{\eps^2}{\polylog(m,n)}.
    \]
    Finally, because we repeat this procedure some $\polylog(m,n) / \eps^2$ times, we get that any hyperedge which is recovered by \cref{alg:VS} with probability $\geq 1/2$ is also recovered by \cref{alg:RepeatedIterativeRecovery} with probability $1 - 1 / \poly(m,n)$.
\end{proof}

\subsection{Sparsification Algorithm}

The preceding subsections showed that we can use this recursive recovery procedure to recover the ``sensitive'' hyperedges (i.e., those with large sampling rate). Next, we show how we can turn this into an actual sparsification algorithm. The key observation is that we must only bootstrap the above algorithm $\log(m)$ times. In each iteration, we are sub-sampling the edges at rate $1/2$ and recovering the important hyperedges in the resulting hypergraph. 

Now, we re-analyze \cref{alg:sparsify} but with \cref{alg:RepeatedIterativeRecovery} as a sub-routine:

\begin{algorithm}[H]
\caption{HypergraphSparsify$(H, m, n , \eps)$}
\label{alg:finalSparsify}
Let $H_i = H$. \\
\For{$i  \in [\log(m)]$}{
Let $F_i = \mathrm{RepeatedIterativeRecovery}(H_{i}, \lambda = \frac{\polylog(m, n)}{\eps^2})$. \\
Let $H_{i+1}$ be the result of keeping each hyperedge in $H_i - F_i$ with probability $1/2$.
}
\Return{$\bigcup_{i \in [\log(m)]} 2^i \cdot F_i$.}
\end{algorithm}

\begin{claim}\label{clm:iterativeAccuracy}
    With probability $1 - 1 /\poly(n)$, for every $i \in [\log(m)]$, $2^{i+1} \cdot H_{i+1} \bigcup_{j \leq i} 2^i \cdot F_i$ as returned from \cref{alg:finalSparsify} is a $(1 \pm O(i \cdot \eps))$ hypergraph spectral sparsifier for $H$. 
\end{claim}

\begin{proof}
Recall that given the hypergraph $H_i$, \cref{alg:sparsify} recovers a set of hyperedges of $H_i$ via running \cref{alg:VS}. Now, by \cref{prop:goal}, there exists a single weight assignment $W^*_i$ of the multi-graph of $H_i$, such that every hyperedge satisfying 
    \[
    \max_{(u,v) \in e} R^{\Phi(H_i)(W^*_i)} \geq \frac{\eps^2}{\polylog(n,m)}
    \]
    is recovered with probability $\geq 1- 1 / \poly(m)$. Now, every hyperedge satisfying the above condition is essentially sampled with probability $1$, while all remaining hyperedges are sampled at rate $1/2$. The only new aspect of \cref{alg:finalSparsify} is that we have used \cref{alg:RepeatedIterativeRecovery} as a sub-routine. By \cref{lem:RIRsame} and taking a union bound over every hyperedge recovered by \cref{alg:sparsify}, we know that with probability $1 - 1 / \poly(n,m)$, every hyperedge recovered by \cref{alg:sparsify} is also recovered in \cref{alg:finalSparsify}.

    In particular, for each hyperedge $e$, this means that the sampling rates we use satisfy
    \[
    p_e \geq \frac{C \log(n) \log(m)}{ \eps^2} \cdot \max_{(u,v) \in e} R_{\mathrm{eff}}^{\Phi(H_i)(W^*_i)}(u,v).
    \]
    By \cref{thm:JLS23}\cite{JambulapatiLS23}, this then means that the above sampling scheme (whereby we sample edges with probability $p_e$, and give weight $1 / p_e$), yields a $(1 \pm \eps)$ spectral-sparsifier with probability $1 -1 /\poly(m)$. $2 \cdot H_{i+1}$ is exactly the hyperedges who survive the sampling at rate $1/2$, and $F_i$ is exactly the hyperedges that we keep with probability $1$ (and hence weight $1$). This then yields the desired claim. 
\end{proof}

\begin{lemma}
    With probability $1 - 1 / \poly(m)$, the result of \cref{alg:finalSparsify} called on a hypergraph $H$ with parameter $\eps$ is a $(1 \pm \eps)$-spectral sparsifier of $H$. 
\end{lemma}

\begin{proof}
    Let us suppose by induction that $F_1 \cup 2 \cdot F_2 \cup \dots \cup 2^i \cdot F_i \cup 2^{i+1} \cdot H_{i+1}$ is a $(1 \pm 2i \eps)$ spectral sparsifier of $H$. Then, in the $i+1$st iteration of sparsification, we replace $H_{i+1}$ with $F_{i+1} \cup 2 \cdot H_{i+2}$. By composition of sparsifiers, if $F_{i+1} \cup 2 \cdot H_{i+2}$ is a $(1 \pm \eps)$ spectral sparsifier of $H_{i+1}$, then $F_1 \cup 2 \cdot F_2 \cup \dots \cup 2^{i+1} \cdot F_{i+1} \cup 2^{i+2} \cdot H_{i+2}$ is a $(1 \pm 2(i+1)\eps)$ spectral-sparsifier of $H$.

    By \cref{clm:iterativeAccuracy}, and a union over all $\log(m)$ levels, this yields a $(1 \pm O(\eps \log(m)))$ sparsifier with probability $1 - 1 / \poly(m)$. Further, observe that after some $i = O(\log(m))$ iterations, every hyperedge in the hypergraph will be removed (i.e., sampled away) with probability $1 - 1 / \poly(m)$, and hence $H_i$ will be empty. Thus, all that remains is the $F_i$'s, and their union is the sparsifier.

    Finally, observe that we can use an error parameter $\eps' = \eps / \log(m)$ while only incurring a $\log^2(m)$ blow-up in the vertex-sampling parameter (which we absorb into the $\polylog(m)$). Thus, the result of \cref{alg:finalSparsify} on $H$ will be a $(1 \pm \eps)$ spectral sparsifier with probability $1 - 1 / \poly(m)$, as we desire.
\end{proof}

\section{Disjoint Spanners for Recovering High Effective Resistance Multi-Edges}\label{sec:9}

In this section, we will generalize a result of \cite{ADKKP16} to recovering large effective resistance multi-edges in a multi-graph. To this end, we introduce some definitions first:

\begin{definition}
    A subgraph $T \subseteq G$ is a $\log(n)$-spanner of $G$ if for every pair $u,v$, $d_{T}(u,v) \leq \log(n) \cdot d_G(u,v)$. Here, $d_G(u,v)$ is understood to be the distance from $u$ to $v$ in $G$.
\end{definition}

We generalize this to disjoint spanners as follows:

\begin{definition}
    We say that $T_1, \dots T_{\ell}$ form a disjoint collection of $\log(n)$-spanners if for every $i \in [\ell]$, $T_{i} \subseteq G - T_{1} - \dots T_{i-1}$, and $T_i$ is a $\log(n)$-spanner of $G - T_{1} - \dots T_{i-1}$.
\end{definition}

Our key claim (as in \cite{ADKKP16}) is the following:

\begin{claim}\label{clm:largeERdisjointSpanners}
    Let $T_1, \dots T_{\ell}$ be a disjoint collection of $\log(n)$-spanners of a multi-graph $G$. Then, every multi-edge $e = (u,v)$ for which $R_{\mathrm{eff}, G}(u,v) \geq \frac{\log(n)}{\ell}$ is contained in $T_1 \cup \dots \cup T_{\ell}$.
\end{claim}

\begin{proof}
    Our proof proceeds in the same manner as \cite{ADKKP16}. Indeed, consider any multi-edge $(u,v)$ which is not contained in $T_1 \cup \dots \cup T_{\ell}$. We will show every such multi-edge has effective resistance $\leq \frac{\log(n)}{\ell}$. Indeed, if a multi-edge $(u,v)$ is not recovered in any of $T_1, \dots T_{\ell}$, then it must be the case that $(u,v)$ is present in each graph $G - T_1 - \dots T_{i-1}$ when the $i$th spanner $T_i$ is being created. This means that there must be a path from $u$ to $v$ of length $\leq \log(n)$ in $T_i$. Because the $T_i$'s are disjoint, this means that across $T_1, \dots T_{\ell}$ there are $\ell$ disjoint paths of length $\leq \log(n)$ from $u$ to $v$. By the resistor network interpretation of graphs, this means that the effective resistance between $u$ and $v$ in $G$ must be $\leq\frac{\log(n)}{\ell} $, yielding our claim. 
\end{proof}

\section{Online Spectral Hypergraph Sparsification}\label{sec:OnlineSpectral}

In this section, we will show the following theorem:

\begin{theorem}
    Let $H$ be a hypergraph on $n$ vertices, $\leq m$ hyperedges, and arity $[r, 2r]$, whose hyperedges are revealed one by one. Then, there is an \emph{online} algorithm for computing a $(1 \pm \eps)$ spectral sparsifier for $H$ which retains only $\widetilde{O}(n \log(m) / \eps^2)$ hyperedges. 
\end{theorem}

Recall that in the online sparsification setting, the hyperedges are presented in a stream $e_1, \dots e_m$ (where we know the value $m$ before-hand). After each hyperedge $e_i$ is revealed, the algorithm must immediately decide whether or not to include $e_i$ in the sparsifier, and if included, must also decide on its weight. In particular, the sparsifier for $e_i$ must be decided without having seen the future hyperedges $e_{i+1}, \dots e_m$. 

To this end, we use the following claim:

\begin{claim}\label{clm:onlineSingleSpanner}
    Given a sequence of multi-edge insertions, there is an online algorithm which creates a $\log(n)$-spanner by keeping only $O(n)$ multi-edges.
\end{claim}

\begin{proof}
    The algorithm is very simple: the algorithm stores the spanner it has built so far, which is denoted by $T$. For the $i$th multi-edge $(u,v)$ that is revealed, the algorithm checks if $(u,v)$ are already connected by a path of length $\leq \log(n)$. If so, the algorithm does not include $(u,v)$, and otherwise the algorithm does include $(u,v)$. Clearly, this builds a $\log(n)$-spanner, as for any edge $(u,v)$ in $G$, $(u,v)$ are connected by a path of length $\leq \log(n)$ in $T$. 

    Next, we show that $T$ will never have a cycle of length $\leq \log(n)$. Indeed, in order for this to be the case, there must have been some time step $i$ where adding the edge $(u,v)$ created a cycle of length $\leq \log(n)$. This means that $(u,v)$ were connected by a path of length $\leq \log(n) - 1$, and the cycle was then completed by adding $(u,v)$. But, if there is already such a path of length $\leq \log(n) - 1$, then the edge $(u,v)$ would not be included. Thus, there must not be any cycles of length $\leq \log(n)$ in $T$. 

    Finally, it remains to bound the number of edges in $T$. This follows because any graph on $n$ vertices with no cycles of length $\leq k$ must have $n^{1 + O(1/k)}$ edges (see for instance, Proposition 2.3 in \cite{ABSHJKS20}). Plugging in $k = \log(n)$ then yields that $T$ must have $O(n)$ edges. 
\end{proof}

Now, by a simple bootstrapping of the above claim, we can also get the following:

\begin{claim}\label{clm:onlineDisjointSpanners}
    Given a sequence of multi-edge insertions, there is an online algorithm which creates a disjoint collection of $\ell$ $\log(n)$-spanners, keeping only $O(\ell n)$ multi-edges.
\end{claim}

\begin{proof}
    The algorithm again is very simple. The algorithm maintains the spanners $T_1, \dots T_{\ell}$ with the invariant that $T_i$ is a $\log(n)$-spanner of $G - T_1 - \dots - T_{i-1}$. 

    Upon receiving a multi-edge $(u,v)$, the algorithm checks if $(u,v)$ can be added to $T_1$ without creating a cycle of length $\log(n)$. If so, the $(u,v)$ is added to $T_1$, otherwise, the algorithm tries adding $(u,v)$ to $T_2$, and so on.

    Each $T_i$ has $O(n)$ edges by the same logic as the previous claim. Likewise, each $T_i$ is a $\log(n)$-spanner of $G - T_1 - \dots - T_{i-1}$, as we can simply view $T_i$ as the spanner that is build by applying the previous algorithm \cref{clm:onlineSingleSpanner} to the sequence of online edges with $T_1, \dots, T_{i-1}$ removed. Thus, we get a set of $\ell$ disjoint $\log(n)$-spanners, each with $O(n)$ edges, yielding our claim. 
\end{proof}

Now we are ready to present our online algorithm for spectral hypergraph sparsification. We adopt the mold of \cref{alg:RepeatedIterativeRecovery} and \cref{alg:finalSparsify}, but replace the black-box step of storing large-effective resistance multi-edges with storing a set of disjoint spanners:

\begin{algorithm}[H]
    \caption{RepeatedRecursiveRecoverySpanner$(G, n, m, \eps)$}\label{alg:RepeatedIterativeRecoverySpanner}
    Initialize $\widetilde{F} \gets \emptyset$ and $H_1 \gets H$.\\
    \For{$r \polylog(n,m)/ \eps^2$ rounds} {
    Vertex sample $H_1$ at rate $1/r$ to get $H'$, with $G$ being its multi-graph.\\ 
    Let $G_1 = G$. \\
    \For{$i \in [\log(m)]$}{
Store a set $T_1, \dots T_{4000 \log^2(n)\log^2(m)}$ of $4000 \log^2(n)\log^2(m)$ disjoint spanners of $G_i$.\\
    Let $F_i$ be the set of multi-edges $\{f: \exists j \in [4000 \log^2(n)\log^2(m)], f \in T_j \}$. \\
    Let $G_{i+1}$ be the result of sampling $G_i - F_i$ at rate $1/2$.
    }
    Let $S =  F_1 \cup F_2 \cup \dots F_{\log(m)}$.\\
    Let $\widetilde{S}$
    contain all hyperedges of $H_1$ who have at least one
    clique multi-edge in $S$. \\
    Let $\widetilde{F} \gets \widetilde{F} \cup \widetilde{S}$ and delete $\widetilde{S}$ from $H_1$.
    }
    \Return{$\widetilde{F}$}
\end{algorithm}

We can then adopt the following sparsification algorithm:

\begin{algorithm}[H]
\caption{HypergraphSparsifySpanner$(H, m, n , \eps)$}
\label{alg:finalSparsifySpanner}
Let $H_i = H$. \\
\For{$i  \in [\log(m)]$}{
Let $F_i = \mathrm{RepeatedIterativeRecoverySpanner}(H_{i}, \lambda = \frac{\polylog(m, n)}{\eps^2})$. \\
Let $H_{i+1}$ be the result of keeping each hyperedge in $H_i - F_i$ with probability $1/2$.
}
\Return{$\bigcup_{i \in [\log(m)]} 2^i \cdot F_i$.}
\end{algorithm}

\begin{claim}\label{clm:onlineCorrect}
    \cref{alg:finalSparsifySpanner} produces a $(1 \pm \eps)$ spectral hypergraph sparsifier of $H$ with probability $1 - 1 / \poly(m)$.
\end{claim}

\begin{proof}
    First, by \cref{clm:largeERdisjointSpanners}, it follows that in the multi-graph $G_i$, any multi-edge of effective resistance $\geq \frac{\log(n)}{4000 \log^2(n)\log^2(m)} = \frac{(1/2)^2}{1000 \log(n)\log^2(m)}$ is recovered in $F_i$, as the disjoint $\log(n)$-spanners $T_1, \dots T_{4000 \log^2(n)\log^2(m)}$ will include these multi-edges. Thus, \cref{alg:RepeatedIterativeRecoverySpanner} is a concrete instantiation of an algorithm which implements \cref{alg:RepeatedIterativeRecovery}.

    Then, we must only observe that \cref{alg:finalSparsifySpanner} is exactly implementing \cref{alg:finalSparsify}. The correctness directly follows. 
\end{proof}

\begin{claim}\label{clm:onlineSize}
    \cref{alg:finalSparsifySpanner} can be implemented as an online algorithm, storing only $\widetilde{O}(n \polylog(m) / \eps^2)$ hyperedges. 
\end{claim}

\begin{proof}
    First we show the bound on the number of hyperedges that are stored. The algorithm stores $O(r \polylog(m,n) / \eps^2)$ $\log(n)$-spanners, each on a vertex set of size $\widetilde{O}(n/r)$. Each spanner therefore stores $\widetilde{O}(n/r)$ multi-edges, and in total, the number of multi-edges stored is $\widetilde{O}(n \polylog(m) / \eps^2)$. Because each hyperedge that is recovered requires one of its corresponding multi-edges to be recovered, this means that the total number of recovered hyperedges is $\widetilde{O}(n \polylog(m) / \eps^2)$.

    Next, we show that the algorithm can be implemented in an online manner. This follows from the fact that we can construct the disjoint collections of spanners in an online manner (as per \cref{clm:onlineDisjointSpanners}) \emph{and} perform the sampling in an online manner. Before starting, the algorithm initializes all spanners to be empty, and uses public randomness to construct the vertex-sampled subsets of the original hypergraph. We will show that \cref{alg:RepeatedIterativeRecoverySpanner} can be implemented in an online manner. From there, observe that implementing \cref{alg:finalSparsifySpanner} online is straightforward, as we simply check if a hyperedge $e$ is stored by \cref{alg:RepeatedIterativeRecoverySpanner}. If so, the hyperedge is kept with weight $1$. Otherwise, we flip a coin: with probability $1/2$ we delete the hyperedge, and with probability $1/2$ we check if $e$ is stored by \cref{alg:RepeatedIterativeRecoverySpanner} at the second level of sampling. If it is recovered here, then we assign it weight $2$, and so on. 

    Finally, to implement \cref{alg:RepeatedIterativeRecoverySpanner} as an online algorithm, let us use $T^{(i,j)}_1, \dots T^{(i,j)}_{4000 \log^2(n)\log^2(m)}$ to denote the disjoint spanners for the $j$th round of vertex sampling (out of $r \polylog(n,m)/ \eps^2$ rounds), and $i \in [\log(m)]$ round of sub-sampling multi-edges. Whenever a new hyperedge $e$ arrives, we first calculate the vertex-sampled version of the hyperedge $e^{(1)}$, and attempt to insert the multi-edges corresponding to $K_{e^{(1)}}$ into $T^{(1,1)}_1, \dots T^{(1,1)}_{4000 \log^2(n)\log^2(m)}$. This can be done online as per \cref{clm:onlineDisjointSpanners}. Next, whichever multi-edges are not recovered, we sample at rate $1/2$ before attempting to insert into $T^{(2,1)}_1, \dots T^{(2,1)}_{4000 \log^2(n)\log^2(m)}$ (via \cref{clm:onlineDisjointSpanners}) and so for $i \in [\log(m)]$. After the $\log(m)$ levels, if any multi-edge from $K_{e^{(j)}}$ has been recovered, then we terminate. Otherwise, we continue on to the second vertex-sampled multi-graph and repeat trying to insert $K_{e^{(2)}}$ into $T^{(1,2)}_1, \dots T^{(1,2)}_{4000 \log^2(n)\log^2(m)}$, repeating the same logic. If no spanner ever recovers a multi-edge from $K_{e^{(j)}}$ across all rounds, then we do not store the hyperedge. Otherwise, we report it as being stored. This yields the online implementation. 
\end{proof}

Finally, we can conclude:

\begin{theorem}
    There is an online hypergraph spectral sparsification algorithm, which for hypergraphs $H$ on $n$ vertices with $\leq m$ hyperedges, and arity $\leq r$ undergoing a sequence of insertions maintains a $(1 \pm \eps)$-\emph{spectral}-sparsifier (and thus cut-sparsifier too) with probability $1 - 1 / \mathrm{poly}(m)$. The space complexity of the algorithm is $\widetilde{O}(nr\polylog(m) / \eps^2)$ bits and the sparsity of the sparsifier is $\widetilde{O}(n \polylog(m) / \eps^2)$ hyperedges.
\end{theorem}

\begin{proof}
    The algorithm and size of the sparsifier follows from \cref{clm:onlineSize} (the bit complexity follows because there are $\widetilde{O}(n \polylog(m) / \eps^2)$ stored multi-edges, but each multi-edge has a label of size $O(r)$ bits to its parent hyperedge). The correctness of the sparsifier follows from \cref{clm:onlineCorrect}. 
\end{proof}

\subsection{Online Sparsification Lower Bound}\label{sec:ollb}

In this section, we detail a lower bound for online hypergraph (cut) sparsification in the model where we can re-weigh hyperedges that have already been included in the sparsifier. Note that this is a strictly easier setting than the setting discussed above, where required spectral sparsification (stronger than cut), and did not allow the sparsifier to alter the weights of the hyperedges it kept. 

Further, we restrict the model to only allow insertions of one copy of each hyperedge, and that inserted hyperedges must be unweighted.

\subsubsection{Lower Bound Set-up}

Now, we explain the set-up for the lower bound. We will let $k$ denote the number of rounds of insertions that we will perform. We partition the vertex set into a left side $L = [n]$ and a right side $R = [n]$. For each round $j \in [k]$, we randomly partition the left side $L$ into two pieces $L_j^1, L_j^2$, and the right side into two pieces $R_j^1, R_j^2$. Then, for each left vertex $u \in L_j^1$, we add $n^{2j}$ random hyperedges of arity $n/4$ from to $R_j^1$, and likewise for every $u \in L_j^2$ we add $n^{2j}$ random hyperedges from to $R_j^2$.

\subsection{Lower Bound Analysis}

First, we establish that if the sparsifier is small, then necessarily all of its edges must be crossing between the partition established in $R$ with high probability. 

\begin{claim}
    Suppose that after round $j-1$, the sparsifier contains $\leq n^2$ hyperedges. Then, with probability $1 - n^2 / 2^{\Omega(n)}$, each hyperedge in the sparsifier crosses from $R_j^1$ to $R_j^2$.
\end{claim}

\begin{proof}
    Indeed, consider any hyperedge $e$ in the sparsifier. Note that the partition $R_j^1, R_j^2$ is chosen independently of the sparsifier. So, the probability that $e$ is completely contained in exactly one of $R_j^1, R_j^2$ is bounded by $(1/2)^{n/4-1} = 1 / 2^{\Omega(n)}$. Taking the union bound over all edges $e$ in the sparsifier yields our claim. 
\end{proof}

Next, we show that with high probability, there is no way to re-weigh existing hyperedges to match certain cut sizes without necessarily violating other cuts. 

\begin{claim}
    Suppose that after round $j-1$, the sparsifier contains $\leq n^2$ hyperedges. Then, in round $j$, the sparsifier must include $\Omega(n)$ new hyperedges in the sparsifier. 
\end{claim}

\begin{proof}
    Indeed, consider a single vertex $u \in L$, and let us assume WLOG that $u \in L_j^1$. Now, in the $j$th round, we insert $n^j$ hyperedges that are incident upon $u$. For the sake of contradiction, let us assume that the sparsifier does not include any of these hyperedges, and instead re-weighs existing hyperedges. Because the degree of $u$ must be preserved in order to be a $(1 \pm \eps)$ cut-sparsifier, this means that the hyperedges incident to $u$ in the sparsifier must be re-weighed by an average factor of $\Omega(n)$ to compensate. However, now let us consider cut defined by $S = L_j^1 \cup R_j^1$. In the actual hypergraph (not the sparsifier), the size of this cut is bounded by $\sum_{i \leq j-1} n^{2i+1} = O(n^{2j-1})$. However, in the sparsifier, with probability $1 - 1 / 2^{\Omega(n)}$ (from the previous claim), every hyperedge incident to $u$ is now crossing from $R_j^1 $ to $R_j^2$. The sum of the weights of these hyperedges is $\Omega(n^{2j})$, and therefore the cut is exceeded by a factor of $n$, and therefore not a sparsifier. 

    This means that the sparsifier \emph{must} select a new hyperedge among those incident on $u$ to be included. Hence, $n$ new hyperedges must be included, as there is one for each left-vertex. 
\end{proof}

\begin{claim}
    After $k \leq n/10$ rounds, the size of the sparsifier is $\Omega(nk).$
\end{claim}

\begin{proof}
    For each of the $k$ rounds, the sparsifier adds $\Omega(n)$ hyperedges. 
\end{proof}

\begin{theorem}
    Online hypergraph cut-sparsifiers with re-weighing allowed require $\Omega(n \log(m) / \log(n))$ hyperedges .
\end{theorem}

\begin{proof}
    Given an allotment of $m$ hyperedges, this allows us to perform the above procedure for $\Omega(\log_n(m)) = \Omega(\log(m) / \log(n))$ rounds. By the previous claim, this leads to a total of $\Omega(n \log(m) / \log(n))$ hyperedges that must be in the sparsifier. 
\end{proof}

\section{Fully Dynamic Algorithms for Spectral Hypergraph Sparsification}\label{sec:FullyDynamic}

In this section, we explain how to implement a fully-dynamic algorithm for spectral hypergraph sparsification. We will use \cref{alg:finalSparsifySpanner} as a starting point, and show how we can leverage decremental algorithms for maintain spanners in multi-graphs to create a fully-dynamic sparsification algorithm. To this end, we start by explaining the well-known reduction from fully-dynamic sparsification algorithms to decremental sparsification algorithms.

\subsection{Reducing Fully-Dynamic to Decremental}

We use the well-known reduction for turning decremental sparsifiers into fully dynamic sparsifiers (see for instance, \cite{ADKKP16}). Note that the previous reductions were specifically done for graphs, hence our motivation here is to re-produce it for hypergraphs. We follow the exposition and statement of \cite{ADKKP16} closely. 

We show the following lemma:

\begin{lemma}\label{lem:decToDynamic}\cite{ADKKP16}
    Given a decremental algorithm for maintaining a $(1 \pm \eps)$ spectral-sparsifier of size $S(m, n, \eps)$ for an undirected hypergraph with total update time $m \cdot T(m, n, \eps)$, there is a fully dynamic algorithm for maintaining a $(1 \pm \eps)$-cut sparsifier of size $O(S(m, n, \eps) \log(m))$ with amortized update time $O(T(m, n , \eps)\log(m))$.
\end{lemma}

As in \cite{ADKKP16}, this lemma relies on a notion of decomposability. 

\begin{claim}
    Let $H = (V, E)$ be a hypergraph, and let $E_1, \dots E_k$ be a partition of the edge-set into $k$ pieces. Let $\widetilde{H}_i$ be a $(1 \pm \eps)$-spectral sparsifier for $H_i = (V, E_i)$. Then, $\widetilde{H} = \cup_{i = 1}^k \widetilde{H}_i$ is a $(1 \pm \eps)$-spectral sparsifier for $H$. 
\end{claim}

\begin{proof}
    Consider any vector $x \in \R^V$. We have that $Q_H(S) = \sum_{i = 1}^k Q_{H_i}(S)$. Thus, because $Q_{\widetilde{H}_i}(S) \in (1 \pm \eps) Q_{H_i}(S)$, we see that 
    \[
    Q_{\widetilde{H}}(S)  = \sum_{i = 1}^k Q_{\widetilde{H}_i}(S) \in (1 \pm \eps) Q_H(S),
    \]
    as we desire.
\end{proof}

Now, we are ready to prove the overall lemma.

\begin{proof}[Proof of \cref{lem:decToDynamic}]
    Set $k = \lceil \log(m) \rceil$. For each $1 \leq i \leq k$, we maintain a set $E_i \subseteq E$ of edges, and an instance $A_i$ of the decremental algorithm running on the hypergraph $H_i = (V, E_i)$. We also keep a binary counter $C$ that counts the number of insertions $\mod m$ with the least significant bit in $C$ being the right most one. 

    A deletion of an edge $e$ is carried out by simply deleting $e$ from the set $E_i$ it is contained in, and propagating the deletion to instance $A_i$ of the decremental algorithm. An insertion of an edge $e$ is more involved: we let $j$ be the highest (i.e., left-most) bit that gets flipped in the counter when increasing the number of insertions. Thus, in the updated counter, the $j$th bit is $1$, and all lower bits are $0$. We first add the edge $e$, as well as all edges in $\cup_{i = 1}^{j-1} E_i$ to $E_j$. Then, we set $E_i = \emptyset$ for all $1 \leq i \leq j-1$. Finally, we re-initialize the instance $A_j$ on the hypergraph $H_j = (V, E_j)$.

    Now, we can bound the total update time for each instance $A_i$ of the decremental algorithm. First, we observe that the $i$th bit of the binary counter is reset after every $2^i$ edge insertions. A simple induction then shows that at any time $|E_i| \leq 2^i$, for all $1 \leq j \leq k$. Now, consider an arbitrary sequence of updates of length $\ell$. The instance $A_i$ is re-initialized after every $2^i$ insertions. It therefore is re-initialized at most $\ell / 2^i$ times. For every re-initialization, we pay a total update time of $|E_i| \cdot T(|E_i|, n, \eps) \leq 2^i \cdot T(m, n, \eps)$. For the entire sequence of $\ell$ updates, the total time spent for instance $A_i$ is therefore $(\ell/2^i) \cdot 2^i \cdot T(m, n \eps) = \ell \cdot T(m ,n , \eps)$. Thus, we spend total time $O(\ell \cdot T(m ,n , \eps) \log(m))$ for the entire algorithm, which gives an amortized update time of $O(T(m, n, \eps)\log(m))$.
\end{proof}

\subsection{Building Decremental Disjoint Multi-graph Spanners}

Recall that the key building block in \cref{alg:finalSparsifySpanner} and \cref{alg:RepeatedIterativeRecoverySpanner} is the construction of disjoint spanners in a multi-graph. The work of \cite{ADKKP16} already studied constructing these in \emph{ordinary} (simple) graphs. In this section, we will show how we can extend their result to work for multi-graphs. Roughly speaking, the difficulty comes from the fact that whenever we recover a multi-edge, we must also recover the corresponding \emph{label} (i.e., parent hyperedge) corresponding to it. 

We start with the spanners of \cite{ADKKP16}:

\begin{theorem}\cite{ADKKP16}\label{thm:decrementalSpanners}
For every $k \geq 2$ there is a decremental algorithm for maintaining a $\log(n)$ spanner $H$ of expected size $O( n \log^3(n))$ for an undirected graph $G$ on $n$ vertices with non-negative edge weights that has an expected total update time of $O( m \log^3(n))$. Additionally, $H$ has the following
property: Every time an edge is added to $H$, it stays in $H$ until it is deleted from $G$. 
\end{theorem}

Recall that we refer to this additional property as the ``lazy'' property. Next, recall our notion of disjoint spanners:

\begin{definition}
    Let $G = (V, E)$ be a multi-graph. We say that a set of spanners $T_1, \dots T_{\ell}$ is an $\ell$-bundle of $\log(n)$-spanners for $G$ if $T_i$ is a $\log(n)$-spanner of $G - T_1 - T_2 - \dots - T_{i-1}$. 
\end{definition}

Notationally, for $i \in [\ell]$, we will let $G_i$ denote the (simple) graph for which $T_i$ is a spanner. That is, $G_i$ is the result of taking the multi-graph $G - T_1 - \dots - T_{i-1}$, and then putting only a single copy of any present multi-edge. Note that the same edge $(u,v)$ may be used in different spanners if there are multiple copies of the edge present. Now, as a warm-up, we show that there is a decremental data structure for maintaining $\ell$-bundles of spanners in multi-graphs. 

\begin{claim}\label{clm:dataStructureWithoutLabels}
    There is a decremental algorithm for maintaining an $\ell$-bundle of $\log(n)$-spanners of an undirected, unweighted multi-graph $G$ with amortized update time $O(\ell \log(\ell) \log^3(n) )$. Moreover, the algorithm satisfies the ``lazy'' property, in the following sense:
    Every time an edge $e$ is inserted, the only potential change is the insertion of $e$ to the bundle of spanners. Whenever an edge $e$ is \emph{deleted}, the only potential changes are the removal of $e$ from the bundle of spanners, and the addition of a single new edge to the bundle of spanners. 
\end{claim}

\begin{proof}
    In addition to maintaining the $\ell$ disjoint spanners (using \cref{thm:decrementalSpanners}), we maintain a simple table data structure. This table data structure maps each possible edge $e = (u,v)$ to a tuple with three elements. The first element in the tuple is the number of copies of the multi-edge, denoted  $w_e$, the second is the number of spanners which contain the multi-edge $t_e$, and the final element is a self-balancing binary search tree (e.g., AVL tree) containing the indices $\subseteq [\ell]$ of all the spanners which contain this multi-edge, as well as a pointer to the final index $f_e \in [\ell]$ in the AVL tree.
Note that since different multi-edges connecting the same pair of vertices are
    interchangeable for the purpose of building spanners,
    we only keep a counter for the number of spanners using $e$ rather than
    track which copies they are using.

    Now, let us consider what happens when an edge $e$ is inserted. First, if $w_e > t_e$, then the multi-edge is already present in every copy of the graph (i.e., every $G_i$), and all we do is increment $w_e$. So, we consider the case when $w_e = t_e$, that is, every copy of the multi-edge is already being used in spanners. We first increment $w_e$ by $1$. Then, we access $f_e$. Starting $i = f_e + 1$, we insert the edge $e$ into $G_i$. If $T_{i}$ updates to include the edge $(u,v)$, then increment $t_e$, and set $f_e = i$. Otherwise, we continue on, incrementing $i$. Eventually, either every graph $G_i$ has $e$ inserted, or some spanner $T_i$ adds the edge, in which case we increment $t_i$ and set $f_e = i$, and add $i$ to the AVL tree and stop the process. Thus, the total update time for an insertion is bounded by the table updates, which are constant time,
    insertion of $i$ into an AVL tree that takes $O(\log \ell)$ time,
    and the $\leq \ell$ insertions into the graphs $G_i$ for which we are maintaining the spanners (amortized time $O(\ell \log^2(n))$). Note that we trivially satisfy the laziness condition, i.e., either one tree adds the edge $e$, or no tree adds the edge $e$. 

    Next, we consider deletions. Indeed, let some multi-edge $e = (u,v)$ be deleted. First, we access the table. If $w_e \geq t_e + 1$, then this means that not every copy of the edge is being used. So, we simply decrement $w_e$ by $1$. If $w_e = t_e + 1$, then we must also go to $f_e$, and starting with $i = f_e + 1$, remove the edge $e$ from $G_i, G_{i+1}, \dots$. However, since $e \notin T_i, T_{i+1}, \dots$, this leads to no cascading changes, as we are removing an edge not present in the spanners. 
    
    So, let us consider the remaining case, when $w_e = t_e$, which means that every copy of the multi-edge is being used in the spanners.
As we mentioned at the beginning of the proof, all copies of $e$ are interchangeable, so we can always assume without loss of generality that the $f_e$th spanner is using the copy of the multi-edge that is being deleted.
    When we remove the edge $e$, we first decrement $w_e$. Then, we store the index $f_e$, and also delete $f_e$ from the AVL tree. We then access the \emph{new} index of the final spanner that contains the edge $e$, which we denote by $f'_e$ (and update this in the table to be the new final spanner index). For all the indices $i \in [f'_e + 1, f_e - 1]$, we remove the edge $e$ from $G_i$. Note that this causes no cascading changes, as $e$ was not used in any of these spanners. Then, we remove the edge $e$ from $G_{f_e}$. Because $e$ was also used in $T_{f_e}$, this does cause a change, as $T_{f_e}$ now potentially adds a new edge. Indeed, if $T_{f_e}$ does not add a new edge, then nothing changes, so the only interesting case is when $T_{f_e}$ does add a new edge, let us call this $e'$. We insert $f_e$ into the AVL tree of spanners containing $e'$. But, because we inserted a copy of $e'$ into $T_{f_e}$, we must also ensure that there are not too many copies of $e'$ present in the spanners. So, we find $w_{e'}$ and compare this with $t_{e'}$. If $w_{e'} = t_{e'}$ (i.e., every copy of $e'$ was being used), then we simply go to $f_{e'}$, and remove $e'$ from $T_{f_{e'}}$, repeating the above procedure. Note that the number of deletions is bounded by $\ell$, as each time a new edge is inserted (i.e., a spanner deletes $e$ and adds $e'$), the indices we consider go from $f_e \rightarrow f_{e'} \rightarrow f_{e''}$. So, the indices of the spanners we update are monotonically increasing.
    Thus we have total amortized update time $O(\ell \log(\ell) \log^2(n))$ for the deletion.
\end{proof}

However, this still is not enough for our purposes. Recall that in \cref{alg:finalSparsifySpanner}, our target is not only to store spanners, but to have spanners \emph{report} the corresponding hyperedges for each edge. So, to capture this, we introduce the notion of a \emph{labelling} for each multi-edge. 

\begin{definition}
    Let $G = (V, E)$ be a multi-graph, where for each multi-edge there is a corresponding label $\in \mathcal{U}$ where $\mathcal{U}$ admits a total ordering. We say that a set of spanners $T = T_1, \dots T_{\ell}$ is a label-respecting $\ell$-bundle of spanners for $G$ if $T_i$ is a spanner of $G - T_1 - T_2 - \dots - T_{i-1}$, and for any update to $T$, $T$ reports the change in edge $e = (u,v)$ \emph{as well as} the corresponding label for the edge $(u,v)$ that was changed.
\end{definition}

We now show how to implement such a label-respecting $\ell$-bundle of spanners in the \emph{decremental} model.

\begin{claim}\label{clm:decrementalLabelledSpanning}
    There is a decremental algorithm for maintaining a label-respecting $\ell$-bundle of spanners of an undirected, unweighted multi-graph $G$ with label set $\mathcal{U}$, with $\leq m$ edges total. The amortized update time is $O(\ell
    \log(\ell) \log^2(n) + \log(m) \log(\mathcal{U}))$. Moreover, the algorithm satisfies the ``lazy'' property, in the following sense. Whenever an edge $e$ with corresponding label $x$ is \emph{deleted}, the only potential changes are the removal of $e$ with label $x$ from the bundle of spanners, and the addition of a single new edge (with label) to the bundle of spanners. Further, the data structure can be initialized on an instance with $m$ edges in time $O(m \ell \log(\ell) \log^2(n) + m \log(m) \log(|\mathcal{U}|))$.
\end{claim}

\begin{proof}
    In addition to maintaining the spanners (using \cref{thm:decrementalSpanners}), we maintain a simple table data structure. This table data structure maps each possible edge $e = (u,v)$ to a tuple with four elements. The first element in the tuple is the number of copies of the multi-edge $w_e$, the second is the number of spanners which contain the multi-edge $t_e$, the third element is an AVL tree containing the indices $\subseteq [\ell]$ of all the spanners which contain this multi-edge, as well as a pointer to the final index $f_e \in [\ell]$ in the AVL tree, and the final element is an AVL tree of all valid labels for edge $(u,v)$. After the initialization of the data structure,
and throughout the entire process of processing the (decremental) updates,
    we always associate the labels for edge $e = (u,v)$
    in the $t_e$ spanners using $e$
    with the first $t_e$ labels (i.e., the first $t_e$ labels in the sorted order). 
    
    Now we show how to handle deletions. Let some multi-edge $e = (u,v)$ with corresponding label $x$ be deleted. First, we access the table. If $w_e \geq t_e + 1$, then this means that not every copy of the edge is being used. So, we simply decrement $w_e$ by $1$. If $w_e = t_e + 1$, then we must also go to $f_e$, and starting with $i = f_e + 1$, remove the edge $e$ from $G_i, G_{i+1}, \dots$. We also search in the AVL tree of valid labels for $(u,v)$ to find the label $x$, and remove $x$. If $x$ is among the first $t_e$ labels, we also access the new $t_e$th label, $x'$. The data structure reports a change that edge $(u,v)$ has changed from label $x$ to label $x'$.

    Now, let us consider the remaining case, when $w_e = t_e$, which means that every copy of the multi-edge is being used. When we remove the edge $e$, we first decrement $w_e$. Then, we store the index $f_e$, and also delete $f_e$ from the AVL tree. We then access the \emph{new} index of the final spanner to contain the edge $e$, which we denote by $f'_e$ (and update this in the table to be the new final spanner index). We also remove the label $x$ from the AVL tree. The data structure reports that edge $e$ with label $x$ has been removed from the spanner. 
    
    For all the indices $i \in [f'_e + 1, f_e - 1]$, we remove the edge $e$ from $G_i$. Note that this causes no cascading changes, as $e$ was not used in any of these spanners. Then, we remove the edge $e$ from $G_{f_e}$. Because $e$ was also used in $T_{f_e}$, this does cause a change, as $T_{f_e}$ now potentially adds a new edge. Indeed, if $T_{f_e}$ does not add a new edge, then nothing changes, so the only interesting case is when $T_{f_e}$ does add a new edge, let us call this $e'$, so we focus on this case: First, we insert $f_e$ into the AVL tree of the indices of which spanners contain $e'$. But, because we inserted a copy of $e'$ into $T_{f_e}$, we must also ensure that there are not too many copies of $e'$ present in the spanners. So, we find $w_{e'}$ and compare this with $t_{e'}$:
    \begin{enumerate}
        \item If $w_{e'} = t_{e'}$ (i.e., every copy of $e'$ was being used), then we simply go to $f_{e'}$, and remove $e'$ from $T_{f_{e'}}$, repeating the above procedure.
        Note that this cascading effect can only happen $\ell$ times at most, as each time we go from $f_e \rightarrow f_{e'}$, the value increases by at least one. 
        \item If $w_{e'} > t_{e'}$ (i.e. there were available copies of $e'$), the data structure increments $t_{e'}$ by $1$, and sets $f_{e'} = \max(f_{e'}, f_e)$. If $w_{e'}$ now equals $t_{e'}$, then we additionally remove $e'$ from the final $\ell - f_{e'}$ graphs (but this causes no cascading changes, as none of these graphs were using $e'$). The data structure outputs that edge $e'$ has been added to the bundle of spanners with label $x'$, where $x'$ is the $t_{e'}$st label among the sorted labels for edge $e'$.
    \end{enumerate}
    
    The total run-time thus sums up to $O(\ell \log(\ell) \log^2(n))$. For maintaining the labels, observe that there are $\leq m$ labels present in any edge slot. Assuming that the label universe is $\mathcal{U}$ and supports comparison operations in time $O(\log(|\mathcal{U}|))$, we can insert / delete the labels in time $O(\log(m) \log(|\mathcal{U}|))$. Since at most $O(1)$ labels are updated, this yields our final update-time. 

    Finally, to see the initialization time, observe that given a list of multi-edges / labels, we can construct the data structure of \cref{clm:dataStructureWithoutLabels} in total time $O(m \ell \log(\ell) \log^2(n))$. From this point, all that remains is to create the AVL tree of labels. This can be done by (for instance) inserting the labels of each $(u,v)$ into the AVL tree one by one. This requires time $O(m \log(m) \log(|\mathcal{U}|))$ to construct, yielding our run-time. 
\end{proof}

Finally, we require one more tool before being able to present our decremental sparsifier. Recall that we repeatedly vertex-sample the hypergraph $H$ to get a smaller vertex set $V'$, and so for any hyperedge $e$, we must insert the multi-edges in $K_{V' \cap e}$ into the graph. However, because there are $r \polylog(m,n)$ different vertex sampled graphs, if we just naively compute this intersection separately for each graph, we will be doomed to have a running time of $\Omega(r^2)$ (since each intersection requires time $\Omega(r)$ to calculate). Thus, we now show that just using a table suffices for computing all of these intersections in time $\widetilde{O}(r \polylog(m,n))$.

\begin{claim}\label{clm:computingIntersections}
    Let $V^{(1)}, \dots V^{r \polylog(m,n)}$ denote the result of independently vertex sampling $V$ at rate $\frac{1}{r}$. There exists a data structure such that given any hyperedge $e \subseteq V$ of arity $r$, we can compute $e \cap V^{(i)}, \forall i \in [r\polylog(m,n)]$ in time $O(r \polylog(m,n))$.
\end{claim}

\begin{proof}
    We store a table that for every vertex $v \in [n]$, maps it to a list of the corresponding indices $i \in [r\polylog(m,n)]$ for which $v \in V^{(i)}$. 
    
    Now, for any hyperedge $e$, we first initialize $r\polylog(m,n)$ empty lists (one for each $V^{(i)}$), and denote these by $L_1, \dots L_{r\polylog(m,n)}$. Then, we iterate through $e$, and for each $v\in e$, we access the table to find which $V^{(i)}$ contain $v$. For any such $i$, we then add $v$ to corresponding list $L_i$. Clearly then, at the end of this procedure $L_i$ contains exactly $e \cap V^{(i)}$.

    Next, we bound the run-time. Initializing the lists takes time $O(r\polylog(m,n))$. For each $v \in e$, we access the table to see which $V^{(i)}$ contain $v$. The time for this is bounded by the number of $V^{(i)}$ that contain $v$. For this, observe that we are vertex sampling at rate $\frac{1}{r}$ for $r\polylog(m,n)$ rounds. Hence, the number of such $V^{(i)}$ is bounded by $O(\polylog(m,n))$ in expectation. Repeating this for all $v \in e$ (of which there are $\leq 2r$), yields an expected running time of $O(r\polylog(m,n))$ (or also a worst-case run-time with probability $1 - 2^{- \Omega(\log(n) \log(m))}$).
\end{proof}

\subsection{Decremental Recovery Algorithm}

Finally, we are now able to present a decremental data structure for \cref{alg:RepeatedIterativeRecoverySpanner}. 

\begin{claim}\label{clm:decrementalRIRSpanner}
    There is a decremental data structure which, given parameters $r, \eps$, maintains a set of hyperedges $F$ in accordance with \cref{alg:RepeatedIterativeRecoverySpanner}. The data structure can be implemented with amortized update time $O(r\polylog(m,n, 1/\eps) \cdot (1 / \eps)^2)$.
\end{claim}

\begin{proof}
    First, we describe the data structure. We store the data structure of \cref{clm:computingIntersections} so that we can efficiently compute the intersections of each hyperedge with the vertex-sampled subsets. We then additionally store the data structure of \cref{clm:decrementalLabelledSpanning}, with parameter $\ell = \polylog(m,n) / \eps^2$ on each vertex-sampled multi-graph. 
    
    However, it remains to show how we can implement Line 11 of \cref{alg:RepeatedIterativeRecoverySpanner}. In particular, we need to show how, given only the (labelled) spanners, we can also exactly recover the corresponding set of hyperedges. 
    
    For this, we first explain how to implement the step using a naive method. Indeed, if we simply let the label of each multi-edge denote the corresponding hyperedge, then in addition to the spanners, we can maintain a sorted set of all the labels for which there is a multi-edge in the bundle of spanners using that label, along with a counter for how many occurrences there are for that label. After any deletion of an edge / label pair, the data structure of \cref{clm:decrementalLabelledSpanning} either does nothing (i.e., no changes to the data structure) or reports the inclusion of a new edge / changing of a label, and we simply increment / decrement the corresponding counters in our set of labels. The only overhead here beyond that of \cref{clm:decrementalLabelledSpanning} is in updating the counters of labels (which takes an additional time $O(r \log(n))$ to binary search, since $|\mathcal{U}| \approx n^r$). Note that whenever a new label is introduced for the first time, we also update all down-stream spanners (i.e.,  to remove the edges corresponding to this newly recovered hyperedge from the later multi-graphs $\Phi(H^{(i+1)}), \dots$).

    Now, we can bound the run-time. The key claim is the following, which is a consequence of the laziness of our data structure.

    \begin{claim}
        Let $F^{(i)}$ be as defined in \cref{alg:finalSparsifySpanner} when the disjoint spanners are implemented with the above data structure. If no hyperedges are inserted into $H$, then this also holds for $H - \bigcup_{j < i} F^{(j)}$ for every $i \in [16r\log(m)\log(n)^2]$.
    \end{claim}

    \begin{proof}
        We can prove this by induction on $i$. The claim is trivially true for $i = 1$. Let us now assume it holds by induction up to $i-1$, and we will show it holds true for $i$. Indeed, suppose that if no hyperedges are inserted into $H$, then no hyperedges are inserted in to $H - \bigcup_{j < i-1} F^{(j)}$. Thus, the only way for a hyperedge to be inserted into $H - F^{(i)}$, would be to remove a hyperedge from the $i-1$st set of recovered hyperedges ($F^{(i)} -F^{(i-1)} $). However, this will never happen, as the only time a hyperedge is removed from a recovered set of hyperedges is when the hyperedge itself is deleted, by the laziness of our data structure. 
    \end{proof}

Note that there might be cascading deletion happening in the vertex-sampled
    multi-graphs - if in a multi-graph we recover a new hyperedge due to
    deletion of another previously recovered hyperedge, then we shall remove this hyperedge from
    all subsequent multi-graphs, potentially causing $\polylog(n,m)$ multi-edge deletions
    in those multi-graphs. And the latter might trigger further edge deletions for the same
    reason. However, as we argue below, this does not blow up our amortized runtime since
    each hyperedge (and their corresponding multi-edges) can only be deleted once
    from each multi-graph.
    
    Now, the remainder of our proof follows from observing that the hyperedges (and their corresponding multi-edges) can \emph{only be deleted once} from each vertex-sampled multi-graph. Thus, we finish by bounding the total run-time of deleting all the hyperedges once. Indeed, consider a single hyperedge $e$. To delete the hyperedge, we first compute all of the corresponding multi-edges for each sub-sampled hypergraph. This can be done via \cref{clm:computingIntersections} in time $O(r\polylog(m,n))$ total. Next, each hyperedge becomes $\leq r \polylog(m,n)$ multi-edges. For each multi-edge, deleting it from each label-respecting spanner and reporting the changes can be done in time $O(\polylog(m,n) \cdot (1/\eps)^2 \cdot \log(1/\eps) \log^2(n) + \log(m) \log(\mathcal{U})) = O(\polylog(m,n)  \cdot (1/\eps)^2 \log(1/\eps)  + \log(m) \log(\mathcal{U}))$ via \cref{clm:decrementalLabelledSpanning}. Thus, because there are $\leq m \cdot r \polylog(m,n)$ total multi-edges, the total-time spent by the data structure in removing \emph{every} hyperedge is bounded by 
    \[
    O(m \cdot r \log^2(m) \log^3(n)) + O(m \cdot r \polylog(m,n) \cdot (1/\eps)^2 \cdot \log(1/\eps)  \log^2(n) + m \cdot r \log(m) \log(\mathcal{U}))
    \]
    \[
    = O(mr\polylog(m,n) \cdot (1/\eps)^2 \log(1/\eps)  + mr \log(m)\log(\mathcal{U}) ).
    \]
    Thus, the amortized update time, per hyperedge deletion, is $O(r\polylog(m,n) \cdot (1/\eps)^2 \log(1/\eps)  + r \log(m)\log(|\mathcal{U}|))$. Note however, that this is not quite the bound that we would like. Indeed, because we are simply using $\mathcal{U}$ to be the entire set of hyperedges of arity $[r, 2r]$, $\log(|\mathcal{U}|) = \Omega(r\log(n))$. Thus, the amortized update time would be $\Omega(r^2)$, when ideally we would only have a linear dependence on $r$. 
    
    To fix this, we introduce our final trick. Instead of associating each multi-edge of a hyperedge $e$ with itself as the label, we instead associate $e$ with a random $C \cdot \log(m)$ bit label, which we call $x_e$. We include a table which maps $x_e$ to $e$ (and like-wise one which maps $e$ to $x_e$), and additionally, for the data structure \cref{clm:computingIntersections}, instead of accessing this table with the hyperedge $e$, we instead access it with the label $x_e$. The main insight is that with overwhelmingly high probability, we have no collisions within the new (smaller) labels, because the hypergraph itself never has more than $m$ hyperedges. Additionally, for each internal operation, we can use $\log(|\mathcal{U}|)= O(\log(m))$. Thus, the new amortized update (i.e., deletion) time is $O(r\polylog(m,n) \cdot (1 / \eps)^2\log(1/\eps) )$.
\end{proof}

We also show that the data structure of \cref{clm:decrementalRIRSpanner} can be initialized efficiently.

\begin{claim}\label{clm:decrementalLowStrengthInit}
    The data structure of \cref{clm:decrementalRIRSpanner} can be initialized on a hypergraph $H$ on $n$ vertices, $m$ hyperedges, arity $[r, 2r]$, and error parameter $\eps$ in time $O(m r \polylog(m,n) (1/\eps)^2 \log(1/\eps) )$.
\end{claim}

\begin{proof}
    Let $E$ denote the edge set with $m$ hyperedges. First, we choose the vertex-sample subsets which requires time $\widetilde{O}(n\log(m))$ (for each vertex $v$, choose which rounds of vertex sampling $v$ will be present for). Then, we initialize the data structure of \cref{clm:computingIntersections}, which takes time $O(r \polylog(m,n))$. Then, we choose the random labels we assign to each hyperedge to decrease the time for binary searching. This takes time $O(r +  \log(m))$ per hyperedge to store the label, hyperedge pair in the table. We also create a binary search tree of the hyperedges, such that for each hyperedge one can find the corresponding label assigned. This takes time $O(m \log(m) \cdot r \log(n))$ to compute ($m$ edges to sort, time $r \log(n)$ per comparison). 

    Next, we also store a list of all the \emph{active} hyperedge labels (i.e., labels for which the corresponding hyperedge has not already been recovered in some spanner). Initially, this is just the set of all hyperedge labels. Now, for the first vertex sampled hypergraph, we insert all of the corresponding multi-edges (with their corresponding labels) into the data structure of \cref{clm:decrementalLabelledSpanning}, by using \cref{clm:computingIntersections} to find the multi-edges to insert. This takes time $O(m \ell \log(\ell) \log^2(n) + m \log(m) \log(|\mathcal{U}|)) = O(m \polylog(m,n)(1/\eps)^2\log(1/\eps))$, using the determined values of $\ell$, $\log(|\mathcal{U}|)$. Note that this data structure will report some labels as being used, because some hyperedges will have multi-edges stored in the spanner. Hence, for each of these labels which is used, we remove them from our list of \emph{active} hyperedge labels. If we let $\kappa_1$ denote the number of hyperedges recovered in this round, this takes time $\kappa_1 \log^2(m)$ to remove ($\kappa_1$ times, binary search in the list which requires $\log(m)$ comparisons, each in time $\log(m)$). Now, we simply repeat this procedure for the second vertex-sampled graph, and so on up until the $r\polylog(m,n)$th round. It must be the case that $\kappa_1 + \kappa_2 + \dots + \kappa_{r\polylog(m,n)} \leq m$, because no hyperedge can be stored in different rounds of vertex sampling. Thus, the total insertion time across all the rounds of vertex sampling is bounded by $O(m r \polylog(m,n) (1/\eps)^2\log(1/\eps))$ Finally, to report the hyperedges which have been recovered, we simply take the labels of hyperedges which were recovered, and use our table to recover the corresponding hypergraph labels. This takes time $\leq m \cdot r\log(n)$ ($m$ hyperedges to search, recovering $r \log(n)$ bits each). Thus, the total time to build this data structure is 
    \[
    O(mr \polylog(n,m)) + O(m r \polylog(m,n) (1/\eps)^2\log(1/\eps))  =  O(m r \polylog(m,n) (1/\eps)^2\log(1/\eps)).
    \]
\end{proof}

Finally, we can use the data structure from the preceding section to implement our overall sparsifier data structure as in \cref{alg:finalSparsifySpanner}. In particular, we simply store the data structure of \cref{clm:decrementalRIRSpanner} for each of $\log(m)$ levels of sparsification. We formalize the claim below:

\begin{claim}\label{clm:decrementalSparsifierUpdateTime}
    There is a decremental data structure maintaining a $(1 \pm \eps)$-spectral sparsifier of a hypergraph $H$ of arity $[r, 2r]$, $\leq m$ hypredges (with probability $1 - 1 / \mathrm{poly}(m)$) with amortized update time $O(r\polylog(m,n) \cdot (1/\eps)^2\log(1/\eps))$. Further, the intialization time of the data structure is 
    \[O(m r \polylog(m,n) (1/\eps)^2\log(1/\eps)).\]
\end{claim}

\begin{proof}
    We avoid repeating the details from the previous proof. However, the key insight remains the same. Because the data structures are lazy, hyperedges can only be removed a \emph{single} time. In order to implement the $\log(m)$ levels of sampling, we associate each new hyperedge with $\log(m)$ random bits, where the $i$th bit dictates whether the hyperedge survives the $i$th level of sampling. Note that, as before, this property does not alter the laziness of the data structure.
    Thus, it suffices to bound the update time for the operations involving only a single hyperedge, ignoring any cascading effects. The update time for any single hyperedge deletion is simply $\log(m)$ times that of \cref{clm:decrementalRIRSpanner} (for the at most $\log(m)$ rounds)
    thus yielding the above claim. Note that the correctness of the data structure follows from \cref{clm:onlineCorrect} (and probability of correctness), as we are exactly implementing the same algorithm (only with a different spanner implementation). 

    To see the initialization time, we simply use the initialization time bound of \cref{clm:decrementalLowStrengthInit}. Note that each time we store this data structure, it reports some hyperedges as being recovered. Thus, we maintain an ``active list'' of hyperedges as in the proof of \cref{clm:decrementalLowStrengthInit}, when initializing the data structure at different levels. The initialization is thus at most $\log(m)$ times that of \cref{clm:decrementalLowStrengthInit}.
\end{proof}

We are thus ready to conclude:

\begin{theorem}
    For a hypergraph on $n$ vertices and $\leq m$ hyperedges, there is a fully dynamic algorithm for maintaining a $(1 \pm \eps)$-spectral sparsifier of size $\widetilde{O}(n \polylog(m) / \eps^2)$ hyperedges with an amortized, expected update time of $\widetilde{O}(r \polylog(m) / \eps^2)$.
\end{theorem}

\begin{proof}
    This follows from \cref{lem:decToDynamic}, \cref{clm:decrementalSparsifierUpdateTime}, and \cref{clm:decrementalLowStrengthInit}. The bound on the number of hyperedges follows from the same bound as in the online setting (since the number of spanners that are stored is small) \cref{clm:onlineSize}.
\end{proof}

\bibliographystyle{alpha}
\bibliography{ref}

\newpage 
\appendix

\section{A Naive Linear Sketching Algorithm}\label{sec:lsnaive}

As before, we assume the input hypergraph $H$ has hyperedges of arity within $[r,2r]$ only,
which is without loss of generality up to a $\log r$ factor in the sparsity of the sparisifer.
As we mentioned in our technical overview, a balanced weight assignment of the
multi-graph is hard to compute in the linear sketching setting as that requires repeat
and complete access to the underlying hypergraph.
However, one can naively sparsify the hypergraph using the uniform weight assignment,
where for each multi-edge associated with a hyperedge $e$, we assign a weight
$1/\binom{|e|}{2}$.
We then resort to the sampling rate
$p_e = \eps^{-2} \polylog(n)\cdot \max_{(u,v) \in e} R^{G(W)}(u,v)$
by \cite{KKTY21b,Lee23,JambulapatiLS23} with $W$ being the uniform weight assignment.
By the analysis in \cite{BansalST19} (see Section 5.2), we know that
\begin{align}\label{eq:bst19}
p_e \leq \eps^{-2} \polylog(n) \frac{1}{r} \sum_{(u,v)\in e} R^{G(W)}(u,v).
\end{align}

Note that in the uniform weight assignment $W$, all edge weights are within a constant factor
of $1/r^2$,
as the input hypergraph $H$ has hyperedges of arity within $[r,2r]$ only.
Therefore we can in fact work with the unweighted multi-graph $\Phi(H)$ of $H$,
and do effective resistance sampling with an oversampling rate
of $r\polylog(n)\eps^{-2}$, and recover a hyperedge $e$ if it has an associated multi-edge
that got sampled.
This is guaranteed to be an oversampling compared to $p_e$ by (\ref{eq:bst19}).
Moreover, by \cref{thm:multigraphERSampling},
this can be done by a linear sketch of $\Otil(nr^2 \polylog(m) / \eps^2)$ bits
using the algorithm in \cite{KLMMS14}.

\section{Space Bounds for \texorpdfstring{\cite{KLMMS14}}{KLMMS14}}\label{sec:linearSketchingAppendix}

\subsection{\texorpdfstring{$\ell_2$}{l2} Heavy-Hitters}

In this section, we explain in more detail how to derive the space bounds of \cite{KLMMS14} when dealing with multi-graphs. We will let $u$ denote the universe size (i.e., the total number of possible multi-edge slots), and let $m$ denote the support size (i.e., the actual maximum number of present multi-edges). 

Recall that \cite{KLMMS14} require the following notion of an $\ell_2$ heavy-hitter:

\begin{definition}
    For a parameter $\eta \in (0,1)$ and a vector $x \in \R^u$, we say that a vector $w$ is an $\ell_2$ heavy-hitter for $x$ with parameter $\eta$ if 
    \[
    \left \Vert x - w \right \Vert_{\infty} \leq \eta \Vert x \Vert_2. 
    \]
\end{definition}

In \cite{KLMMS14}, they show the existence of a linear sketch implementing an $\eta$ $\ell_2$ heavy-hitter that requires $O(\log^2(u) / \eta^2)$ bits to store. Here, we show that this bound can more tightly be written as $O(\log(u)\log(m) / \eta^2)$. 

\begin{claim}\label{clm:heavyhitter}
    For vectors $x \in \R^u$ with support size bounded by $m$, and all entries bounded in magnitude by $C$, there is a linear sketch (using public randomness) implementing an $\eta$ $\ell_2$ heavy-hitter that requires $O(\log(u)\log(mC) / \eta^2 + \log(u) \log(1 / \delta))$ bits to represent, returns a vector of sparsity $O(1 / \eta^2)$ and has success probability $1 - 1 / \text{poly}(u) - \delta$.
\end{claim}

\begin{proof}
    We simply follow the standard CountSketch protocol. In each copy of CountSketch, we initialize a uniformly random hash function $h$ which maps $[u] \rightarrow [100 / \eta^2]$, and for each $i \in [u]$, we also associate a random sign $\sigma_i \in \pm 1$. For each of the $j \in [100 / \eta^2]$ buckets, we store $c_j = \sum_{i: h(i) = j} \sigma_i x_i$. From prior work (\cite{JW18} for instance, or \url{https://www.cs.cmu.edu/afs/cs/user/dwoodruf/www/teaching/15859-fall17/weekEight.pdf}), it is known that if we simply store $O(\log(u))$ rounds of this CountSketch (initialized independently), and let $c_j^{(p)}$, $h^{(p)}$ denote the sketch values, hash function respectively in the $p$th round, then we can acquire a good estimator for each value $x_i$. Indeed, we get an unbiased estimator for $x_i$ by looking at $\sigma_i^{(j)} \cdot c_{h^{(j)}(i)}^{(j)}$ and taking the median over these values across $j = 1, \dots, O(\log(u))$. In these prior works, it is established that this yields estimates to each $x_i$ with additive error at most $\frac{\eta \Vert x \Vert_2}{10}$ with probability $1 - 1 / \text{poly}(u)$. Note then, by additionally storing a linear sketch for estimating the $\ell_2$ norm of $x$ (which can be done to accuracy $\eta / 1000$ with probability $1 - \delta$ in space $O(\log(u)\log(1 / \delta) / \eta^2 )$ via \cite{AMS96}), we can then return every element $x_i$ which attains a value of at least $\frac{8 \eta \hat{\ell_2(x)}}{10}$, where $\hat{\ell_2}(x)$ is our estimate of the $\ell_2$ norm of $x$. Because we overestimate $\Vert x \Vert_2$ by at most $(\eta / 1000)$, and we have additive error at most $\eta \Vert x \Vert_2 / 10$, any element $x_i$ of size at least $\eta \Vert x \Vert_2$ will report a size which is $\geq \frac{9 \eta \Vert x \Vert_2}{10} \geq \frac{8 \eta (1  + \eta / 1000)\Vert x \Vert_2 }{10}$, and thus be recovered. 

    Further, note that the sparsity of the returned vector will be $O(1 / \eta)$, as at most $\frac{10}{7 \eta}$ elements can contribute $\geq 7 \eta \Vert x \Vert_2 / 10$ and thus be recovered. 

    Finally, assuming that the randomness is provided to us via a re-readable random tape, we can see that the total space required by this linear sketch is $O(\frac{1}{\eta^2} \cdot \log(u) \cdot \log(mC) + \log(u) \log(1 / \delta))$, where the final $\log(mC)$ term is for storing the running sum $c_j$ in each bucket, and $\delta$ is the failure probability of the \cite{AMS96} sketch.
\end{proof}

\begin{claim}
    For a multi-graph with $u$ possible edge slots, and at most $m$ multi-edges inserted, one can implement the sketch of \cite{KLMMS14} in space $\widetilde{O}(n \log(u) \polylog(m) / \eps^2)$ with failure probability at most $1 - 1 / \poly(n)$, where $n$ is the number of vertices in the multi-graph. 
\end{claim}

\begin{proof}
    Recall that \cite{KLMMS14} simply stores heavy-hitter sketches of the down-sampled incidence matrices. Letting $B$ denote the incidence matrix, there are at most $O(\log(m))$ levels of sampling we must do until the matrix is empty. For each level, we store a heavy-hitter sketch with parameter $\eta = \frac{\eps}{c_1 \cdot \sqrt{\log(m)}}$. We initialize the failure probability $\delta = 1/\poly(n)$, where $n$ is the number of vertices in the multi-graph. It follows then that the total space required is bounded by $O((n \log(m) ) \cdot (\log(u) \polylog(m) / \eps^2 + \log(u) \log(n))) = \widetilde{O}(n \log(u) \polylog(m) / \eps^2)$ (here we are also using that the entries of the vector we sketch with $\ell_2$ heavy hitters is bounded by $\poly(m)$ as well). Correctness follows exactly from the work of \cite{KLMMS14}, as they use \Cref{clm:heavyhitter} as a black-box in their sketch.
\end{proof}
\end{document}